\documentclass[english,notitlepage,nofootinbib]{revtex4-2}
\usepackage[T1]{fontenc}
\usepackage[latin9]{inputenc}
\usepackage{color}
\usepackage{babel}
\usepackage{amsmath}
\usepackage{amssymb}
\usepackage{graphicx}
\PassOptionsToPackage{normalem}{ulem}
\usepackage{ulem}
\usepackage[bookmarks=false,
 breaklinks=false,pdfborder={0 0 1},backref=false,colorlinks=false]
 {hyperref}

\makeatletter

\providecommand{\tabularnewline}{\\}

\usepackage{babel}

\allowdisplaybreaks

\makeatother

\begin{document}
\title{Exclusive photoproduction of $\chi_{c}\gamma$ pairs}
\author{Marat Siddikov$^{1}$, Ivan Zemlyakov$^{1,2}$}
\affiliation{$^{1}$Departamento de Física, Universidad Técnica Federico Santa
María,~~~~\\
 y Centro Científico - Tecnológico de Valparaíso, Casilla 110-V, Valparaíso,
Chile}
\affiliation{$^{2}$Instituto de Física, Pontificia Universidad Católica de Valparaíso,
Av. Brasil 2950, Valparaíso, Chile}
\begin{abstract}
In this manuscript we study the exclusive photoproduction of $\chi_{c}\gamma$
pairs in the collinear factorization framework. We found the coefficient
functions for all possible spins and helicity projections of $\chi_{c}$
mesons and final-state photons in the leading order in the strong
coupling $\alpha_{s}$. In our analysis we focused on the contribution
of the leading twist chiral even GPDs, and found that for all spin
states of $\chi_{c}$ the suggested process is determined by the behavior
of the gluon GPDs $H_{g}$ in the ERBL kinematics. Using the phenomenological
parametrizations of the GPDs from the literature, we estimated numerically
the photoproduction cross-sections and the expected counting rates
in the kinematics of middle-energy photoproduction experiments that
could be realized at the Electron-Ion Collider. We observed that the
$\chi_{c1}$, $\chi_{c2}$ mesons are produced predominantly with
the same polarization as the incoming photon, and the expected counting
rates of $\chi_{c1}\gamma$ and $\chi_{c2}\gamma$ pairs are sufficiently
large for their experimental study. We also analyzed the angular distribution
of $\chi_{c}$ mesons in \emph{electro}production experiments, and
noticed that for some helicity components there are sizable angular
asymmetries, which can be used as complementary observables for experimental
study. Finally, we estimated the role of this process as a potential
background to $\chi_{c}$ photoproduction, which has been recently
suggested as a tool for studies of odderons. We found that the contribution
of $\chi_{c}\gamma$ (with undetected photon) is on par with expected
contributions of odderons and Primakoff mechanisms in the kinematics
of small momentum transfer $|t|\lesssim1$ GeV$^{2}$, but becomes
negligible at larger $|t|$. 
\end{abstract}
\maketitle

\section{Introduction}

At present the Generalized Parton Distributions (GPDs) are being used
as standard tools to describe the nonperturbative dynamics of partons
in the hadronic target~\cite{Diehl:2000xz,Goeke:2001tz,Diehl:2003ny,Guidal:2013rya,Boer:2011fh,Dutrieux:2021nlz,Burkert:2022hjz,AbdulKhalek:2021gbh,Accardi:2012qut}.
Nowadays \textit{ab initio} computations of the GPDs are not feasible,
and for this reason phenomenological extractions from experimental
data remain the main way to access these objects. At present the available
parametrizations of the GPDs largely rely on experimental data for
$2\to2$ processes, such as Deeply Virtual Meson Production or Compton
Scattering (DVMP and DVCS respectively). However, these channels have
low sensitivity to some kinematic domains, and for this reason do
not fix uniquely the GPDs of the target (the so-called ``\emph{shadow
GPD problem}'')~~\cite{Kumericki:2016ehc,Bertone:2021yyz,Moffat:2023svr}.
Although the experimental data provide reasonable constraints for
some of the GPDs, currently, these constraints remain quite weak for
the transversity GPDs and the gluon GPDs in the Efremov-Radyushkin-Brodsky-Lepage
(ERBL) kinematics~\cite{Guo:2023ahv}. This stimulated a search
of new channels that could complement existing $2\to2$ data, and
several novel $2\to3$ processes have been proposed as possible instruments
for phenomenological extractions of the GPDs~\cite{GPD2x3:9,GPD2x3:8,GPD2x3:7,GPD2x3:6,GPD2x3:5,GPD2x3:4,GPD2x3:3,GPD2x3:2,GPD2x3:1,Duplancic:2022wqn,Qiu:2024mny,Qiu:2023mrm,Deja:2023ahc,Siddikov:2022bku,Siddikov:2023qbd}.
All these channels inherently complement one another due to sensitivity
to a unique flavor combination of the GPD flavors. The conditions
for applicability of the factorization in such processes are usually
satisfied in presence of hard scales which guarantee a good kinematic
separation of the final-state hadrons~\cite{GPD2x3:10,GPD2x3:11}.
The aforementioned studies largely focused on the production
of light meson or photon pairs, with their amplitudes primarily controlled
by quark GPDs in both chiral-odd and chiral-even sectors. While gluon
GPDs can also contribute in some of these channels, their contribution
is typically intertwined with that of quarks, making the phenomenological
analysis more complex. Furthermore, as was discussed in~\cite{Duplancic:2024opc,Nabeebaccus:2023},
the presence of massless quarks and gluons in the theory may lead
to nontrivial overlapping singularities which challenge the applicability
of the collinear factorization unless the distribution amplitude of
produced light meson vanishes rapidly enough at the endpoints.

Recently in~\cite{Siddikov:2024blb} we suggested that the gluonic
GPDs can be studied in exclusive photoproduction processes which include
a quarkonium and a photon in the final state. The process $\gamma p\to J/\psi\gamma p$
is strongly suppressed, since it requires $C$-odd echanges in $t$-channel. For this raeson, in our previous study we focused on $\eta_{c}\gamma$ production. In this manuscript we extend those studies and analyze $\chi_{c}\gamma$
photoproduction, which also proceeds without $C$-odd exchanges in
$t$-channel. In contrast to $\eta_{c}$ mesons, the $\chi_{c}$ mesons
are the lightest $P$-wave quarkonia, and thus can be used as independent
tools for study of GPDs. Due to large branching fractions of the radiative
decays $\chi_{cJ}\to J/\psi\,\gamma\,\,(J=1,2)$ and very clean experimental
signature of $J/\psi$ mesons (an $\ell^{+}\ell^{-}$ lepton pair
with appropriate mass), we expect that it will be possible to distinguish
experimentally the events which correspond to this process. The sufficiently
heavy mass of $\chi_{c}$ and the fact that it is an excited state
implies that feed-down contributions (from decays of heavier quarkonia)
should be negligible. The mass of this quarkonium also serves as a
natural hard scale, which justifies the perturbative treatment even
for the photoproduction~\cite{Korner:1991kf,Neubert:1993mb}, and
allows to avoid many collinear and soft singularities which may appear
in diagrams with massless quarks. Finally, the heavy quark and quarkonia
masses justify the application of NRQCD approach for description of
the $\bar{Q}Q$ pair hadronization into final state quarkonium~\cite{Bodwin:1994jh,Maltoni:1997pt,Brambilla:2008zg,Feng:2015cba,Brambilla:2010cs,Cho:1995ce,Cho:1995vh}.
We expect that the suggested process can be studied in low- and middle-energy
photon-proton collisions in ultraperipheral kinematics at LHC, as
well as in electron-proton collisions at the future Electron Ion Collider
(EIC)~\cite{Accardi:2012qut,AbdulKhalek:2021gbh,Burkert:2022hjz}.

The proposed process $\gamma p\to\chi_{c}\gamma\,p$ (with unobserved
photon) also deserves attention as a possible background to the exclusive
$\chi_{c}$ photoproduction $\gamma p\to\chi_{c}p.$ The latter process
has been proposed recently in~\cite{Benic:2024} for study of odderons.~
Since the $\gamma p\to\chi_{c}\gamma p$ process does not require
$C$-odd exchanges in $t$-channel, its cross-section in certain kinematics
is sufficiently large and can lead to a substantial background for
odderon searches via $\chi_{c}$ photoproduction.

The subsequent sections are organized as follows. In the following
Section~\ref{sec:Formalism} we explain in detail the kinematics
of the process and describe the theoretical approach used for evaluation
of the partonic amplitudes (coefficient functions) of the process.
However, due to a large number of similarly looking expressions for
helicity components, which occupy a lot of space, the final expressions
for these objects are relocated to Appendix~\ref{sec:CoefFunction}
 The Section~\ref{sec:Numer} includes our numerical estimates for
the cross-sections and counting rates obtained with some commonly
used parametrizations of the GPDs. In its subsection~\ref{subsec:odderon}
we compare the cross-sections of the odderon-mediated $\gamma p\to\chi_{c}p$
and the proposed $\gamma p\to\chi_{c}p\gamma$ with unobserved photon.
In the final Section~\ref{sec:Conclusions} we summarize our findings
and draw conclusions.

\section{Theoretical framework}

\label{sec:Formalism} The choice of the reference frame and light-cone
parametrization of the particles' momenta for $\chi_{c}\gamma$ -photoproduction
largely coincide with what was done in~~\cite{Siddikov:2024blb}
for analysis of $\eta_{c}\gamma$ production. In the limit of massless
quarkonium (but fixed invariant mass of quarkonium) this parametrization
reduces to the expressions obtained in~\cite{GPD2x3:8,GPD2x3:9}
for exclusive photoproduction of\emph{ light} meson-photon pairs $\gamma\pi,\,\gamma\rho$
with large invariant mass. In what follows we will consider that the
mass of the quarkonium $M_{\chi_{c}}$ is a large parameter, parametrically
of the same order as the invariant mass $M_{\chi_{c}\gamma}$. At
the same time, we will explicitly exclude the kinematics where the
emitted photon is soft, and thus $M_{\chi_{c}\gamma}\approx M_{\chi_{c}}$.
Since the spectrum of equivalent photons in $ep$ and $pp$ collisions
is dominated by quasi-real photons with small virtuality $Q\approx0$,
in what follows we will focus only on the photoproduction by transversely
polarized real photons. Below in the subsections~\ref{subsec:Kinematics},
~\ref{subsec:Amplitudes} we describe the kinematic variables used
to describe the 4-momenta of all particles and explain the main steps
used for evaluation of the partonic amplitudes .

\subsection{Kinematics of the process}

\label{subsec:Kinematics} For our evaluations we will work in the
photon-proton collision frame, where the colliding proton and the
incoming quasireal photon move along the axis $z$ in opposite directions.
In what follows we will use a notation $q$ for the 4-momentum of
the incoming photon, $P_{{\rm in}},P_{{\rm out}}$ for the momenta
of the target before and after interaction, $k$ for the momentum
of the final-state photon, and $p_{\chi_{c}}$ for the momentum of
produced $\chi_{c}$ meson. For the light-cone components of all the
vectors, we will use the Kogut-Soper convention~\cite{Brodsky:1997de}
, and bold-face letters with subindex $\perp$ for the transverse
(1,2)-components, namely
\begin{equation}
v^{\mu}=\left(v^{+},\quad v^{-},\quad\boldsymbol{v}_{\perp}\right)=v^{+}n_{+}+v^{-}n_{-}+\boldsymbol{v}_{\perp},\qquad v^{\pm}=\frac{v^{0}\pm v^{3}}{\sqrt{2}},\qquad\boldsymbol{v}_{\perp}=v_{x}\hat{\boldsymbol{x}}+v_{y}\hat{\boldsymbol{y}},
\end{equation}
where the light-cone vectors $n_{\pm}$ are defined as $n_{+}=\left(1,\,0,\,\boldsymbol{0}_{\perp}\right)$
and $n_{-}=\left(0,\,1,\,\boldsymbol{0}_{\perp}\right)$. The square
of the vector in this convention is given by 
\begin{equation}
v^{2}\equiv v^{\mu}v_{\mu}=2v^{+}v^{-}-\boldsymbol{v}_{\perp}^{2},
\end{equation}
and the convolution with Dirac $\gamma$-matrix reads as 
\begin{equation}
\hat{v}=\gamma^{\mu}v_{\mu}=\gamma^{+}v^{-}+\gamma^{-}v^{+}-\boldsymbol{\gamma}_{\perp}\cdot\boldsymbol{v}_{\perp}.\label{eq:gammaSlash}
\end{equation}
 The light-cone decomposition of the particles' momenta may be written
as~\cite{GPD2x3:8,GPD2x3:9}
\begin{align}
q^{\mu} & =\left(0,\,\sqrt{\frac{s}{2}},\,\boldsymbol{0}_{\perp}\right)\label{eq:q}\\
P_{{\rm in}}^{\mu} & =\left(\left(1+\xi\right)\sqrt{\frac{s}{2}},\,\frac{m_{N}^{2}}{\sqrt{2s}\left(1+\xi\right)},\,\,\boldsymbol{0}_{\perp}\right),\\
P_{{\rm out}}^{\mu} & =\left(\left(1-\xi\right)\sqrt{\frac{s}{2}},\,\frac{m_{N}^{2}+\boldsymbol{\Delta}_{\perp}^{2}}{\sqrt{2s}\left(1-\xi\right)},\,\boldsymbol{\Delta}_{\perp}\right),\\
\Delta^{\mu} & =P_{{\rm out}}^{\mu}-P_{{\rm in}}^{\mu}=\left(-2\xi\,\sqrt{\frac{s}{2}},\,\frac{2\xi m_{N}^{2}+\left(1+\xi\right)\boldsymbol{\Delta}_{\perp}^{2}}{\sqrt{2s}\left(1-\xi^{2}\right)},\,\boldsymbol{\Delta}_{\perp}\right)\\
p_{\chi_{c}}^{\mu} & =\left(\frac{\left(\boldsymbol{p}_{\perp}^{\chi_{c}}\right)^{2}+M_{\chi_{c}}^{2}}{\alpha_{\chi_{c}}\sqrt{2s}},\,\alpha_{\chi_{c}}\sqrt{\frac{s}{2}},\,\boldsymbol{p}_{\perp}^{\chi_{c}}\right),\qquad\boldsymbol{p}_{\perp}^{\chi_{c}}:=-\boldsymbol{p}_{\perp}-\frac{\boldsymbol{\Delta}_{\perp}}{2}\\
k^{\mu} & =\left(\frac{\left(\boldsymbol{k}_{\perp}^{\gamma}\right)^{2}}{\left(1-\alpha_{\chi_{c}}\right)\sqrt{2s}},\,\left(1-\alpha_{\chi_{c}}\right)\sqrt{\frac{s}{2}},\,\boldsymbol{k}_{\perp}^{\gamma}\right),\qquad\boldsymbol{k}_{\perp}^{\gamma}:=\boldsymbol{p}_{\perp}-\frac{\boldsymbol{\Delta}_{\perp}}{2}\label{eq:k}\\
 & \boldsymbol{p}_{\perp}\equiv\left(\boldsymbol{k}_{\perp}^{\gamma}-\boldsymbol{p}_{\perp}^{\chi_{c}}\right)/2
\end{align}

We may rewrite the invariant Mandelstam variables $S_{\gamma N},t$
in terms of these variables as
\begin{align}
S_{\gamma N} & \equiv W^{2}=\left(q+P_{{\rm in}}\right)^{2}=s\left(1+\xi\right)+m_{N}^{2},\\
t & =\left(P_{{\rm out}}-P_{{\rm in}}\right)^{2}=-\frac{1+\xi}{1-\xi}\Delta_{\perp}^{2}-\frac{4\xi^{2}m_{N}^{2}}{1-\xi^{2}}.\label{eq:tDep}
\end{align}
From Eq.~(\ref{eq:tDep}) we can see that at given $\xi$, the invariant
momentum transfer $t$ is bound by 
\[
t\le t_{{\rm min}}=-\frac{4\xi^{2}m_{N}^{2}}{1-\xi^{2}}.
\]
In order to characterize the kinematics of the final-state quarkonium
and photon, we also define the invariants 
\begin{align}
 & u'=\left(p_{\chi_{c}}-q\right)^{2},\qquad t'=\left(k-q\right)^{2},\qquad M_{\gamma\chi_{c}}^{2}=\left(k+p_{\chi_{c}}\right)^{2},\label{eq:uPrimetPrime}
\end{align}
which correspond to invariant momentum transfer from the incoming
photon to the $\chi_{c}$ meson, momentum transfer to the final-state
(outgoing) photon, and the invariant mass of the $\gamma\chi_{c}$
pair. These variables are related with each other as
\begin{align}
 & -u'-t'=M_{\gamma\chi_{c}}^{2}-M_{\chi_{c}}^{2}-t.\label{eq:Constr}
\end{align}
Rewriting the variable $u'$ in the rest frame of the quarkonium,
we may obtain an additional constraint 
\begin{equation}
u'=M_{\chi_{c}}^{2}-2E_{\gamma}M_{\chi_{c}}\le M_{\chi_{c}}^{2}\qquad\Rightarrow\qquad-t'\le M_{\gamma\chi_{c}}^{2}-t.
\end{equation}

The polarization vectors of the incoming and outgoing real photons
with momenta $\boldsymbol{k}$ are chosen in the light-cone gauge
as
\begin{equation}
\varepsilon_{T}^{(\lambda=\pm1)}(\boldsymbol{k})=\left(\frac{\boldsymbol{e}_{\lambda}^{\perp}\cdot\boldsymbol{k}_{\perp}}{k^{-}},\,0,\boldsymbol{e}_{\lambda}^{\perp}\right),\qquad\boldsymbol{e}_{\lambda}^{\perp}=\frac{1}{\sqrt{2}}\left(\begin{array}{c}
1\\
i\lambda
\end{array}\right),\label{eq:PolVector-1}
\end{equation}
namely they satisfy the gauge condition $A^{-}=0$. As we discuss
in Appendix~\ref{subsec:WF}, this choice is preferable since it
allows to avoid evaluation of additional contributions which stem
from coupling of the external photons to the gauge link in the $\chi_{c}$
meson.

\begin{figure}

\includegraphics[scale=0.4,height=8.5cm]{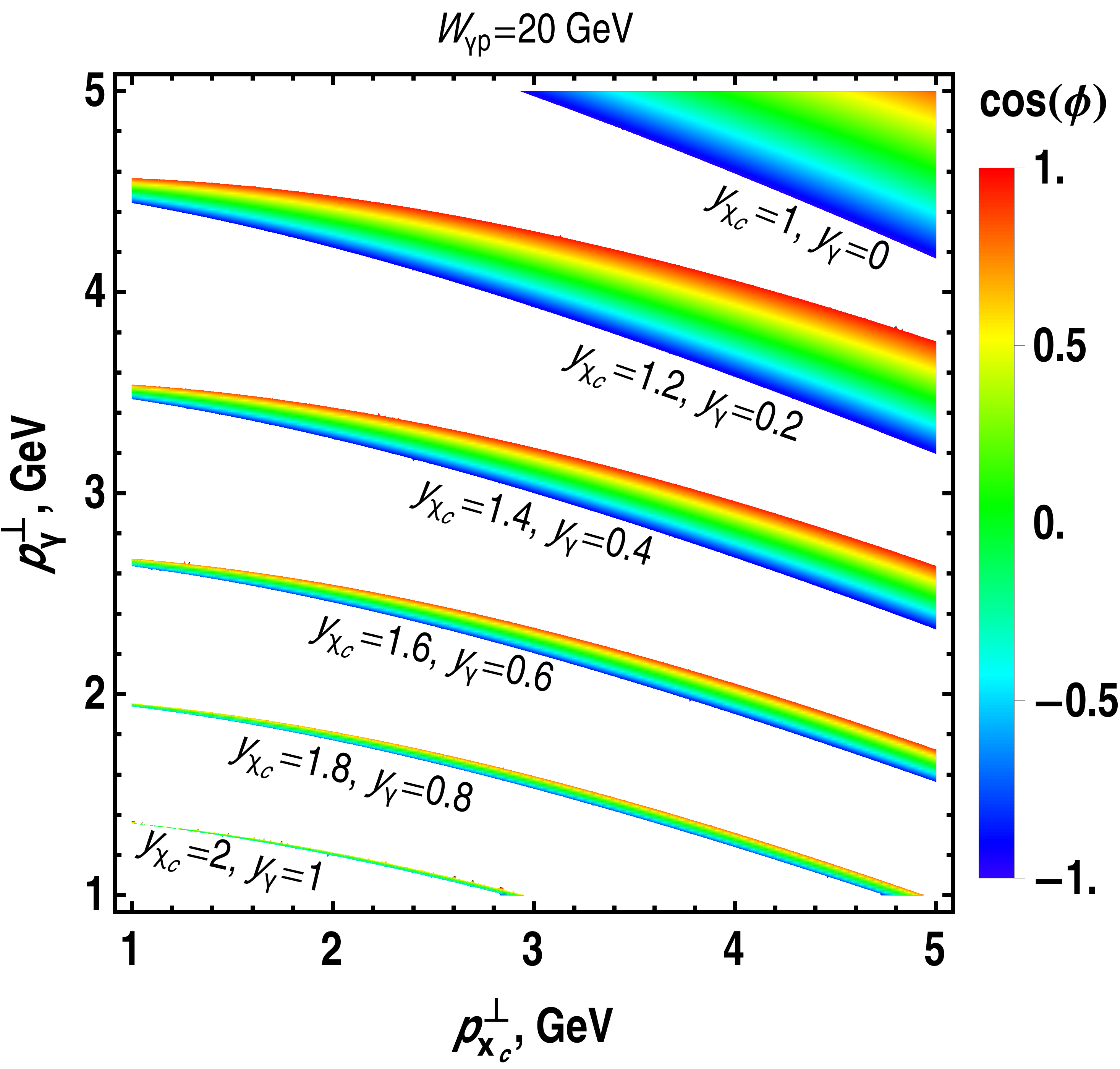}\includegraphics[scale=0.4,height=8.5cm]{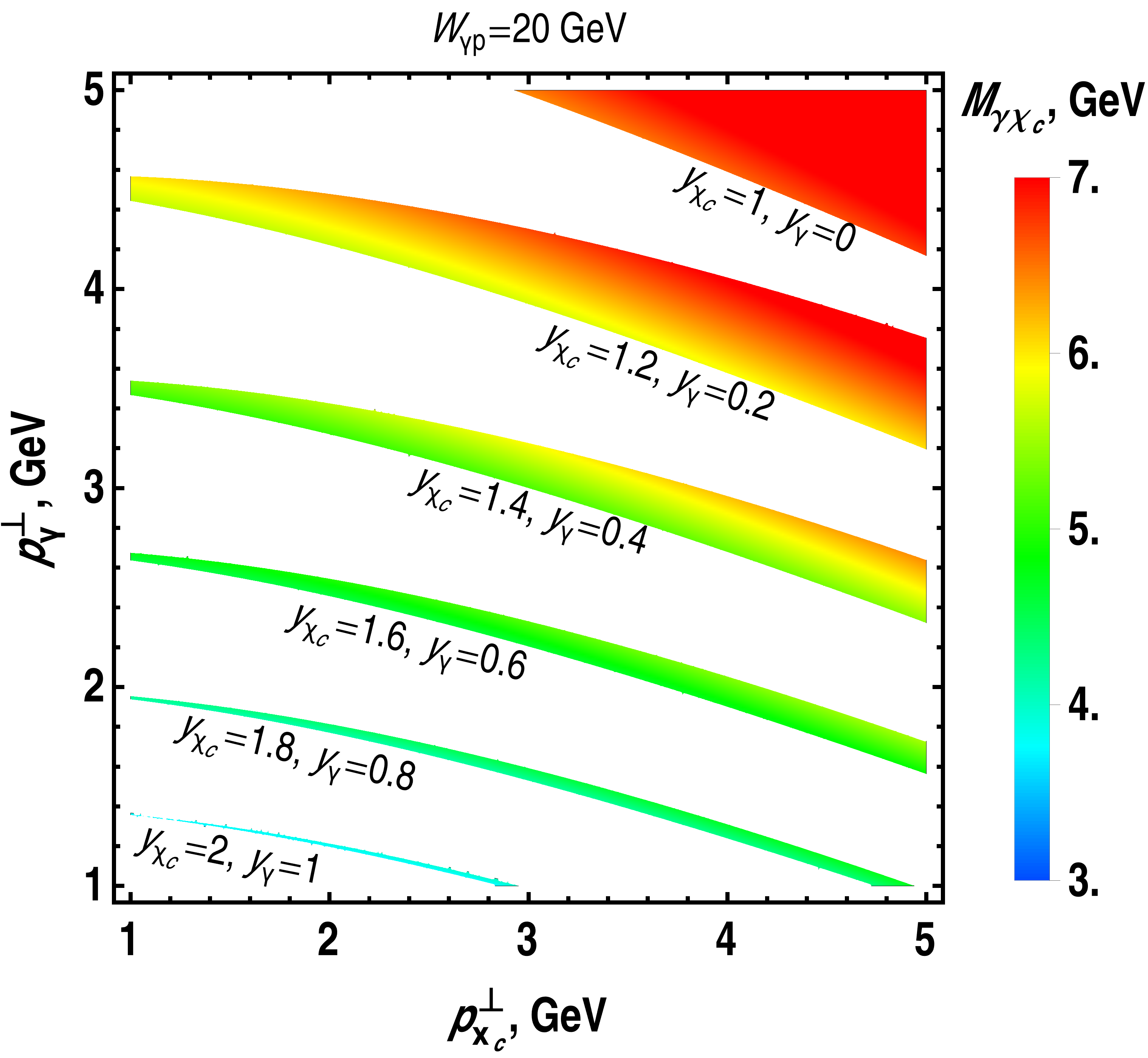}

\caption{\protect\label{fig:Domain}(Color online) Colored bands represent
kinematically allowed region for $\chi_{c}\gamma$ pair production
\uline{at fixed invariant energy} $W$ and fixed rapidities $y_{\chi_{c}},\,y_{\gamma}$.
For both particles we chose positive sign of rapidity in direction
of the photon. Since photon moves in \textquotedblleft minus\textquotedblright{}
direction in~(\ref{eq:q}), this corresponds to nonstandard definition
of rapidity $y_{a}=\frac{1}{2}\ln\left(k_{a}^{-}/k_{a}^{+}\right),\qquad a=\chi_{c},\gamma$.
The increase of rapidities implies increase of the longitudinal momenta,
and thus in view of energy conservation corresponds to smaller transverse
momenta. In the left plot the color of each pixel reflects the azimuthal
angle $\phi$ between the transverse momenta $\boldsymbol{p}_{\chi_{c}}^{\perp}=-\boldsymbol{p}_{\perp}-\boldsymbol{\Delta}_{\perp}/2$
and $\boldsymbol{p}_{\gamma}^{\perp}=\boldsymbol{p}_{\perp}-\boldsymbol{\Delta}_{\perp}/2$.
Similarly, in the right plot the color reflects the value of invariant
mass $M_{\gamma\chi_{c}}$ in the chosen kinematics.}
\end{figure}

\begin{figure}
\includegraphics[totalheight=6cm]{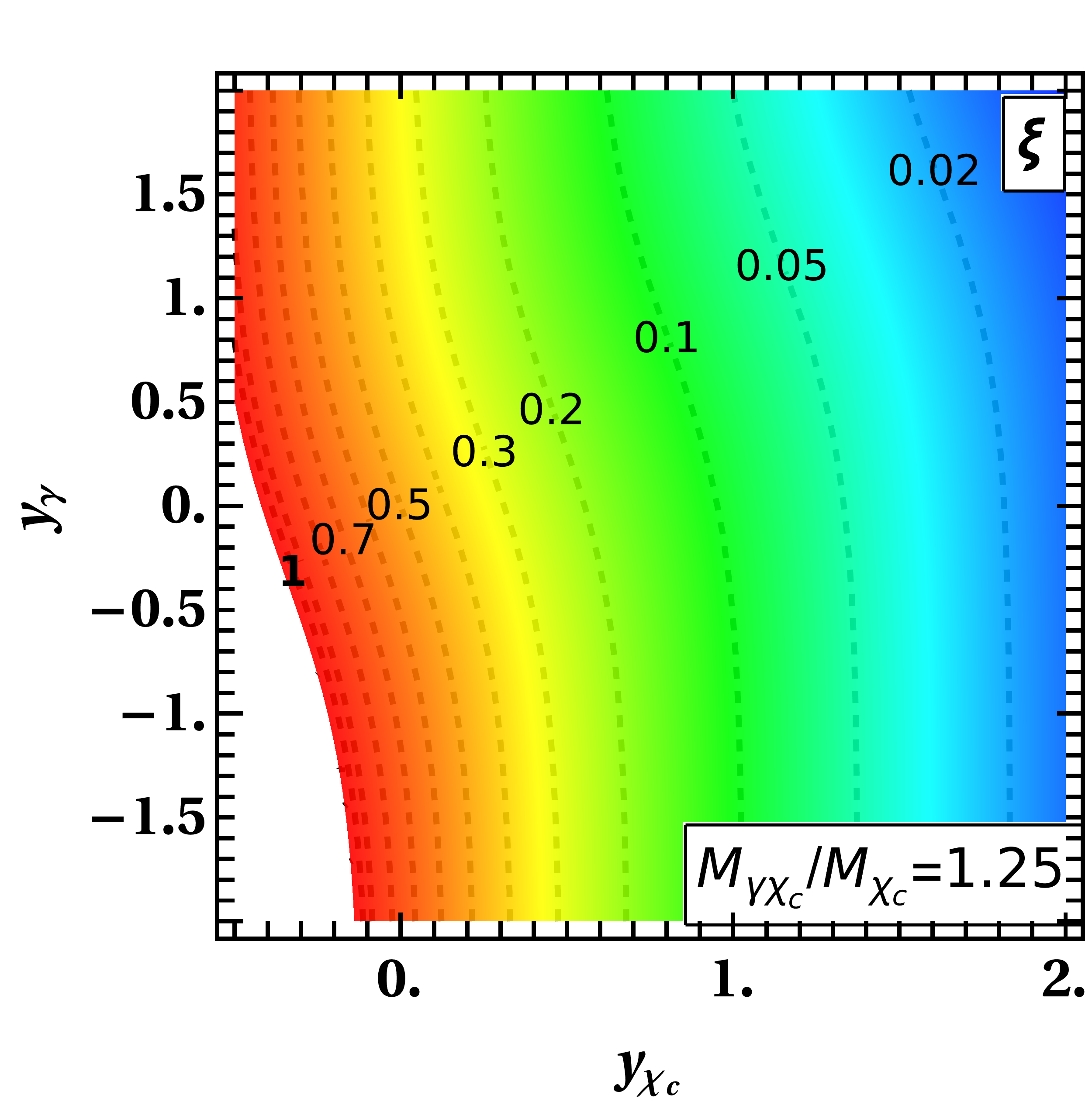}\includegraphics[totalheight=6cm]{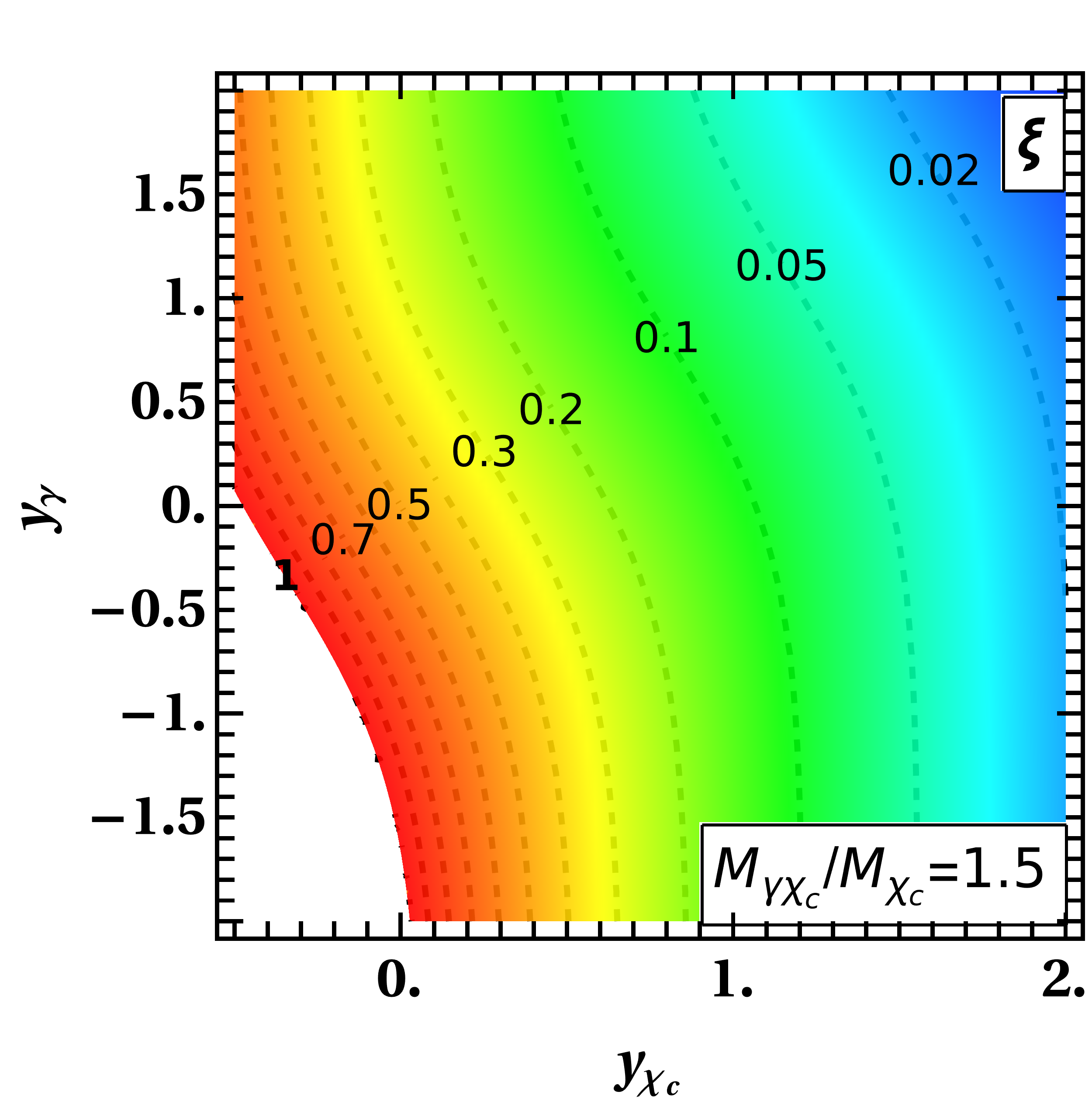}\includegraphics[totalheight=6cm]{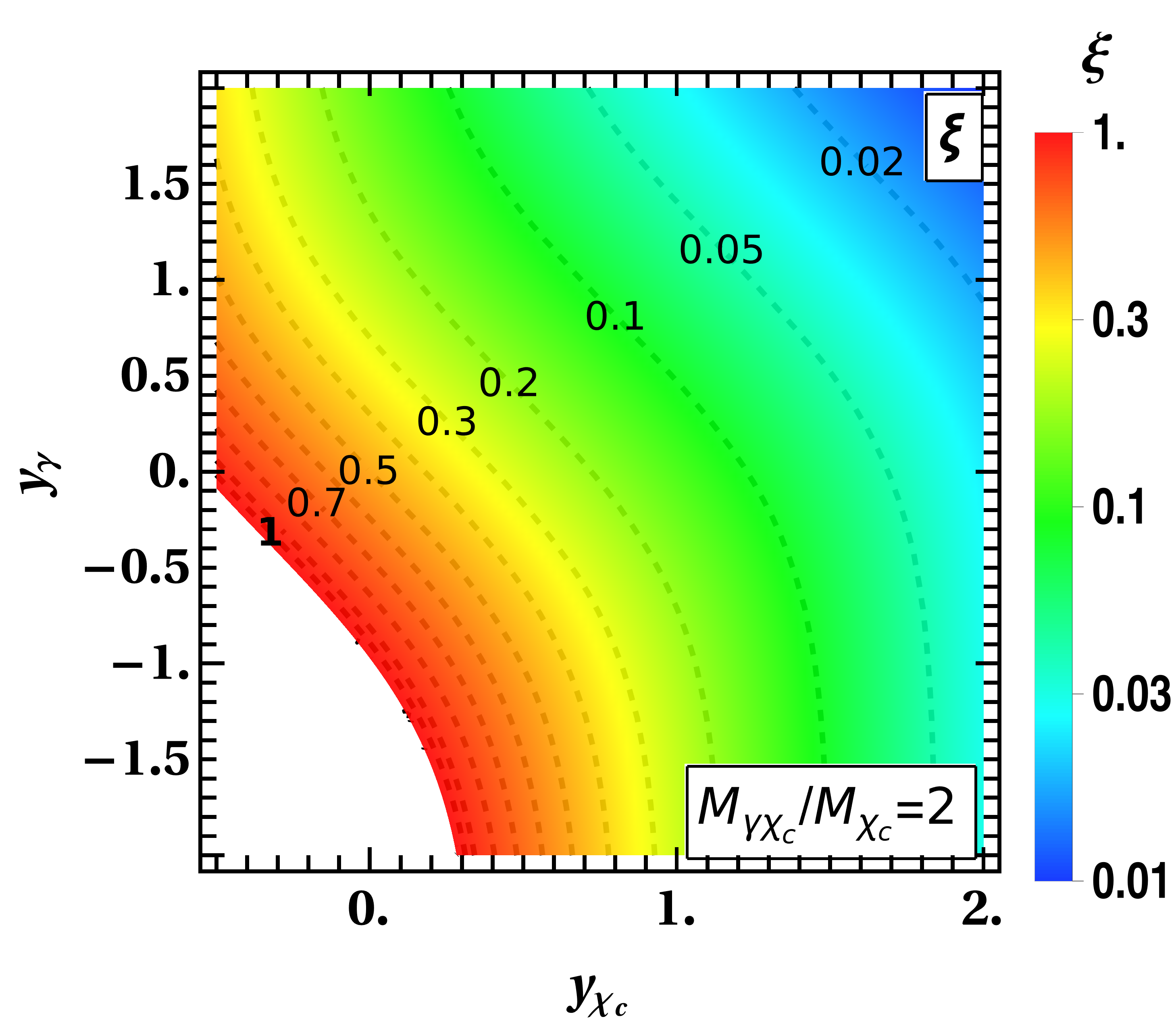}

\includegraphics[totalheight=6cm]{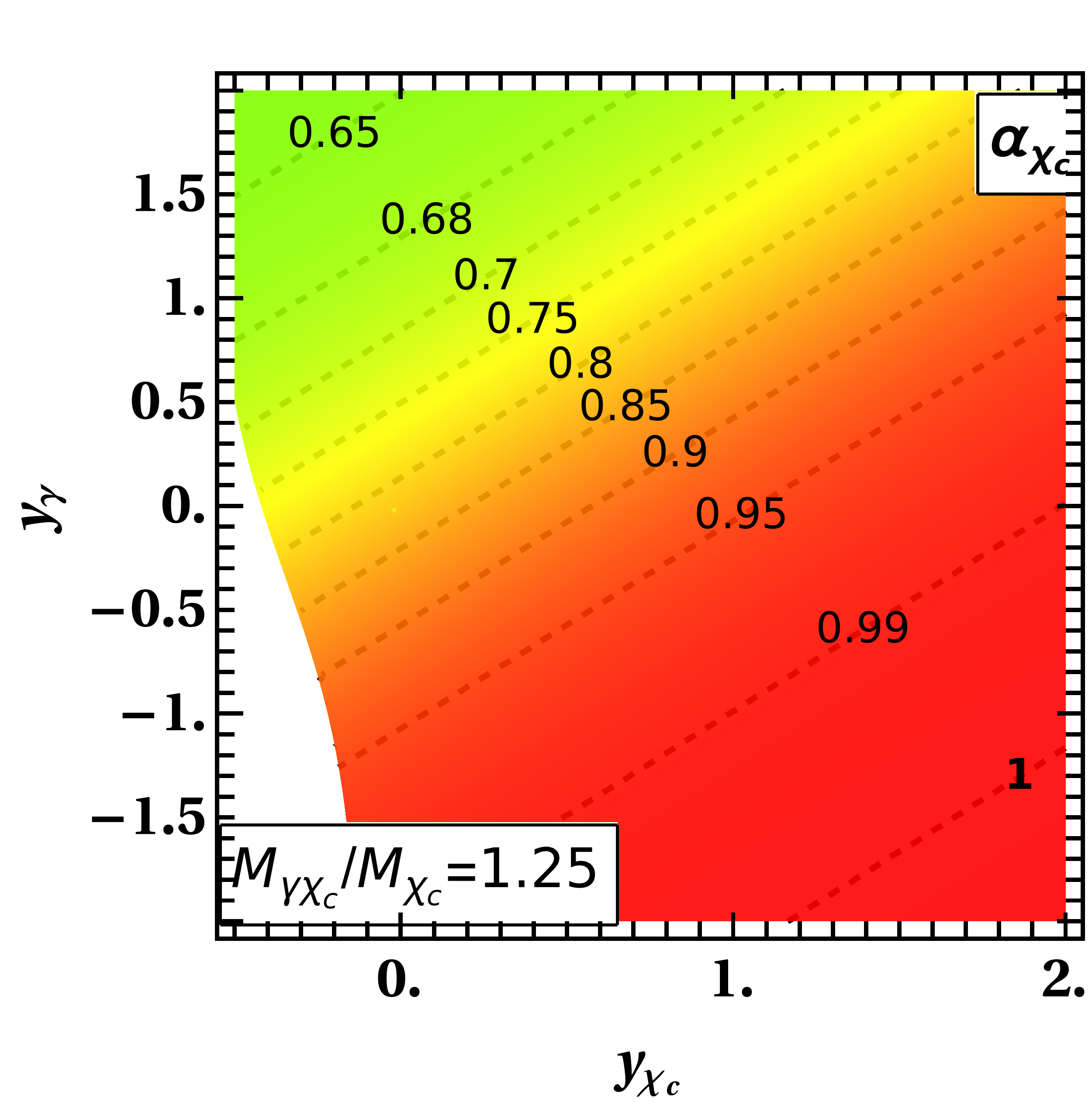}\includegraphics[totalheight=6cm]{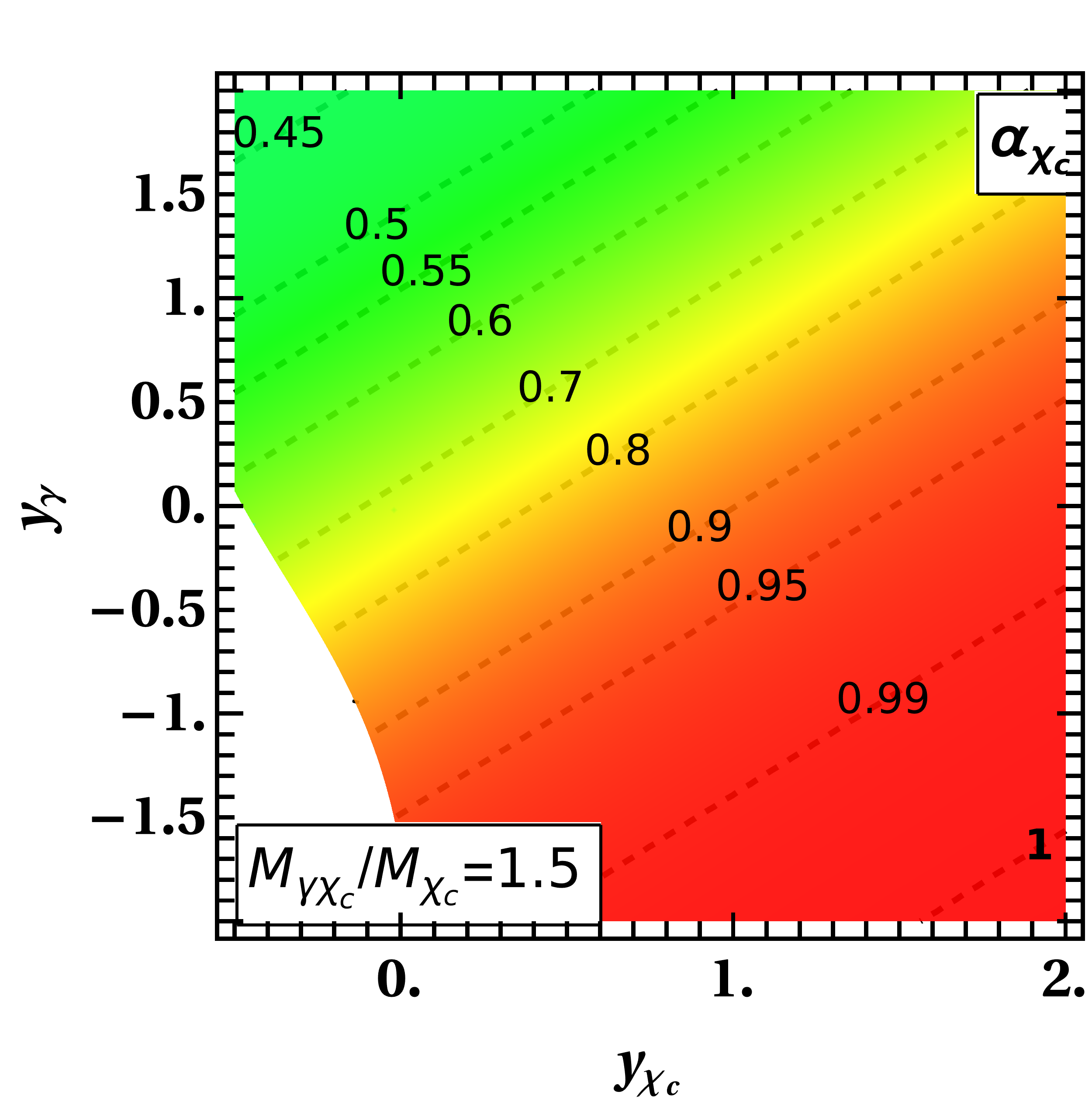}\includegraphics[totalheight=6cm]{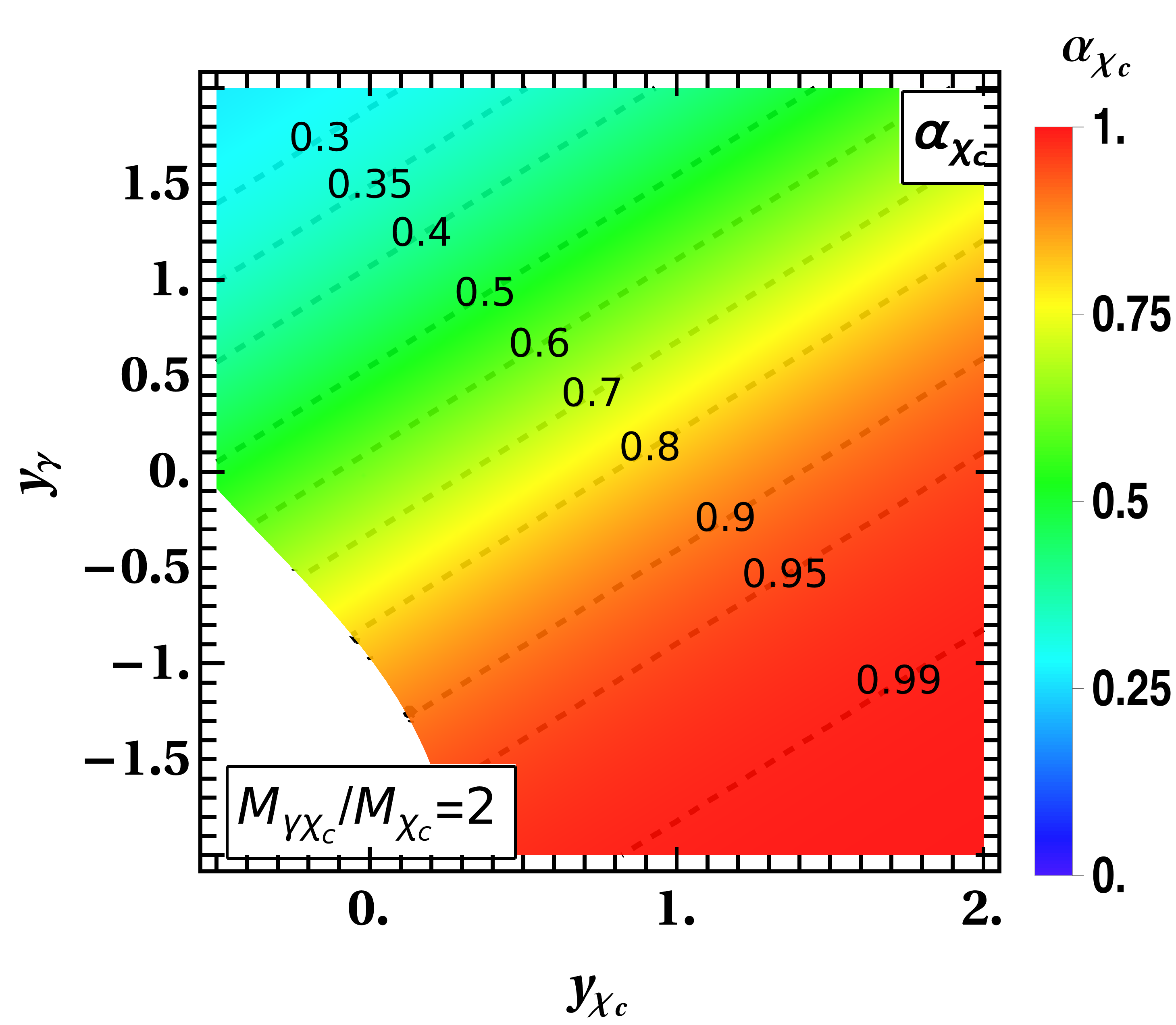}\caption{\protect\label{fig:xiRapidity}(Color online) The contour plot shows
the dependence of the skewedness $\xi$ (upper row) and the variable
$\alpha_{\chi_{c}}$ (lower row) on the rapidities of the final state
particles $y_{\chi_{c}},\,y_{\gamma}$ at different fixed values of
the invariant mass $M_{\gamma\chi_{c}}$. The labels near the contour
lines (and corresponding color shading) show the values of $\xi,\,\alpha_{\chi_{c}}$.
For both particles positive sign of rapidity corresponds to motion
in the direction of the incoming photon. For the sake of simplicity
we consider the kinematics with zero transverse momentum $\Delta_{\perp}$,
which gives the dominant contribution in the total cross-section.
}
\end{figure}

The parametrization~~(\ref{eq:q}-\ref{eq:k}) satisfies various
nontrivial constraints on momenta of the produced particles, such
as onshellness, conservation of the energy-momentum, etc. In order
to clarify the physical meaning of the introduced variables and the
role of the aforementioned constraints, in the Figures~\ref{fig:Domain}
and \ref{fig:xiRapidity} we have shown the values of transverse momenta
$\boldsymbol{p}_{\chi_{c}}^{\perp}=-\boldsymbol{p}_{\perp}-\boldsymbol{\Delta}_{\perp},\,\boldsymbol{p}_{\gamma}^{\perp}=\boldsymbol{p}_{\perp}-\boldsymbol{\Delta}_{\perp}/2$
which may be achieved at fixed rapidities of the final-state particles
in the kinematically allowed domains (colored bands). The color of
each pixel in the Figure~~(\ref{fig:Domain}) in the left and right
panels illustrates the value of the azimuthal angle $\phi$ between
the transverse momenta of $\chi_{c},\gamma$ and the invariant mass
$M_{\gamma\chi_{c}}$, respectively. In the Figure~\ref{fig:xiRapidity}
we also have shown the relation of the variables $\xi,\,\alpha_{\chi_{c}}$
to rapidities $y_{\chi_{c}},y_{\gamma}$ .

\begin{figure}
\includegraphics[width=10cm]{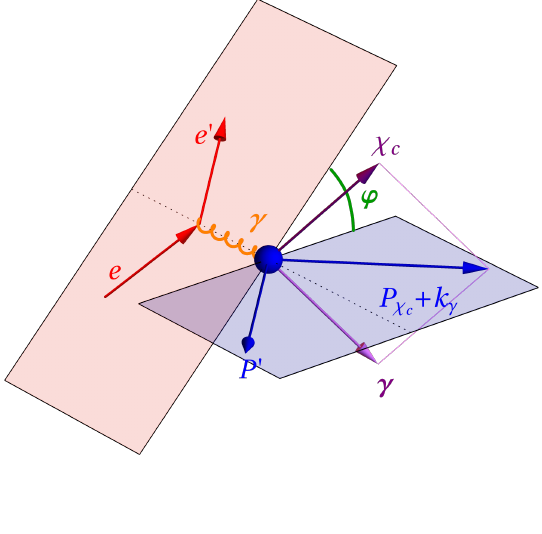}

\caption{(Color online) Definition of the angle $\varphi$ between the lepton
and hadron scattering planes for electroproduction of $\chi_{c}\gamma$
pairs. The target is shown as a colored sphere, which rests in its
reference frame or moves along a dotted line towards the collision
point with photon $\gamma^{*}$. The lepton plane is formed by the
momenta of the electron before and after the scattering (marked with
labels $e,e'$) and also include the momentum of the photon $\gamma^{*}$
due to momentum conservation. . The hadron plane by definition is
formed by the momenta of incoming photon $\gamma^{*}$ and the recoiled
proton $P_{{\rm out}}$. The momenta of quarkonium and emitted photon
in general do not collinear to the hadronic plane, however their sum
$p_{\chi_{c}}+k_{\gamma}$ in view of the momentum conservation also
belongs to the hadron scattering plane.}
\label{fig:Planes} \label{fig:CGCBasic-2}
\end{figure}

In what follows we will focus on the kinematics where the variables
$m_{N},\,\left|\Delta_{\perp}\right|,t$ are negligibly small, whereas
all the other variables are parametrically large, $\sim M_{\chi_{c}}$.
In this kinematics it is possible to simplify the light-cone decomposition~(\ref{eq:q}-\ref{eq:k})
and obtain approximate relations of the above-mentioned variables
with $\alpha_{\chi_{c}},\boldsymbol{p}_{\perp}$ , viz: 
\begin{align}
-t' & \approx\alpha_{\chi_{c}}M_{\gamma\chi_{c}}^{2}-M_{\chi_{c}}^{2}{\color{red}},\qquad-u'\approx\left(1-\alpha_{\chi_{c}}\right)M_{\gamma\chi_{c}}^{2},\label{eq:KinApprox}\\
\boldsymbol{p}_{\perp}^{2} & \approx\left(1-\alpha_{\chi_{c}}\right)\left(\alpha_{\chi_{c}}M_{\gamma\chi_{c}}^{2}-M_{\chi_{c}}^{2}\right)=-t'\left(1-\alpha_{\chi_{c}}\right),\qquad M_{\gamma\chi_{c}}^{2}\approx2s\xi\label{eq:KinApprox-2}
\end{align}
The physical constraint for the momenta of real (on shell) photons
\begin{equation}
t'=\left(q-k\right)^{2}=-2q\cdot k=-2\,\left|\boldsymbol{q}\right|\left|\boldsymbol{k}\right|\left(1-\cos\theta_{\boldsymbol{q},\,\boldsymbol{k}}\right)\le0,\label{eq:tPrimeConstraint}
\end{equation}
together with constraint $\boldsymbol{p}_{\perp}^{2}\ge0$ implies
that the variable $\alpha_{\chi_{c}}$ is bound by 
\begin{equation}
M_{\chi_{c}}^{2}/M_{\gamma\chi_{c}}^{2}\le\alpha_{\chi_{c}}\le1,\label{eq:alphaConstraint}
\end{equation}
so for the Mandelstam variable $u'$ we get 
\begin{equation}
0\le-u'\le\left(-u'\right)_{{\rm max}}=M_{\gamma\chi_{c}}^{2}-M_{\chi_{c}}^{2}-t.
\end{equation}
The produced $\chi_{c}$ and emitted photon are kinematically well-separated
from the recoil proton, if the corresponding invariant masses satisfy
\begin{align}
\left(P_{{\rm out}}+p_{\chi_{c}}\right)^{2} & \approx M_{\chi_{c}}^{2}+s\alpha_{\chi_{c}}\left(1-\xi\right)\gg\left(m_{N}+M_{\chi_{c}}\right)^{2},\qquad\left(P_{{\rm out}}+k\right)^{2}\approx s\left(1-\alpha_{\chi_{c}}\right)\left(1-\xi\right)\gg m_{N}^{2}
\end{align}
These conditions are satisfied everywhere except a border region $\alpha_{\chi_{c}}\approx1$,
or equivalently $u'\approx0$. The invariant cross-section of the
process $\gamma p\to\chi_{c}\gamma p$ can be rewritten in terms of
these variables as
\begin{equation}
\frac{d\sigma_{\gamma p\to\chi_{c}\gamma p}^{(\lambda,\sigma,H)}}{dt\,dt'\,dM_{\gamma\chi_{c}}}\approx\frac{\left|\mathcal{A}_{\gamma p\to\chi_{c}\gamma p}^{(\lambda,\sigma,H)}\right|^{2}}{128\pi^{3}W^{4}M_{\gamma\chi_{c}}}\label{eq:Photo}
\end{equation}
where $\mathcal{A}_{\gamma p\to\chi_{c}\gamma p}^{(\lambda,\sigma,H)}$
is the invariant amplitude , $\lambda,\sigma$ are the helicities
of the incoming and outgoing photons, and $H$ is the helicity of
$\chi_{c}$ meson. Similarly, the \textit{electro}production cross-section
is given by a convolution of the leptonic tensor $L_{\mu\nu}$ with
the corresponding amplitudes, which in the heliclity basis can be
rewritten as
\begin{equation}
\frac{d\sigma_{ep\to e\chi_{c}\gamma p}^{(\sigma,H)}}{d\Omega}\approx\frac{1}{Q^{4}}\sum_{\lambda,\bar{\lambda}}L_{\lambda,\bar{\lambda}}\frac{\left(\mathcal{A}_{\gamma p\to\chi_{c}\gamma p}^{(\lambda,\sigma,H)}\right)^{*}\mathcal{A}_{\gamma p\to\chi_{c}\gamma p}^{(\bar{\lambda},\sigma,H)}}{128\pi^{3}W^{4}M_{\gamma\chi_{c}}},\label{eq:Photo-1}
\end{equation}
the shorthand notation $d\Omega$ stands for the phase volume $d\Omega\equiv d\ln W^{2}dQ^{2}\,dt\,dt'\,dM_{\gamma\chi_{c}}d\varphi$,
the angle $\varphi$ is the angle between the leptonic and hadronic
scattering planes (see Figure~\ref{fig:Planes} for more accurate
definition), and $\lambda,\bar{\lambda}$ are the helicities of the
incoming photon in the amplitude and its conjugate. As was discussed
in~~\cite{Goloskokov:2009ia,Mantysaari:2020lhf}, the leptonic
tensor $L_{\lambda,\bar{\lambda}}$ has nondiagonal components in
helicity basis. Due to these interference contribution, the cross-section
of the electroproduction may have a nontrivial dependence on the angle
$\varphi$, namely 
\begin{align}
\frac{d\sigma_{ep\to e\gamma\chi_{c}p}}{d\Omega} & =\epsilon\frac{d\sigma^{(L)}}{d\Omega}+\frac{d\sigma^{(T)}}{d\Omega}+\sqrt{\epsilon(1+\epsilon)}\cos\varphi\frac{d\sigma^{(LT)}}{d\Omega}+\sqrt{\epsilon(1+\epsilon)}\sin\varphi\frac{d\sigma^{(L'T)}}{d\Omega}+\label{eq:sigma_def}\\
 & +\epsilon\cos2\varphi\frac{d\sigma^{(TT)}}{d\Omega}+\epsilon\sin2\varphi\frac{d\sigma^{(T'T)}}{d\Omega},\nonumber 
\end{align}
where we introduced the superscript letters $L,T,T'$ in the right-hand
side of~(\ref{eq:sigma_def}) to show the contributions of the longitudinal
and transverse polarizations $\lambda,\bar{\lambda}$ (as well as
their possible interference). The shorthand notation $\epsilon$ in~(\ref{eq:sigma_def})
is the usual ratio of the longitudinal and transverse photon fluxes,
\begin{equation}
\epsilon\approx\frac{1-y}{1-y+y^{2}/2},\label{eq:RatioFluxes}
\end{equation}
and $y$ is the fraction of the electron energy which passes to the
virtual photon (the so-called inelasticity). This variable may be
related to the invariant energy $\sqrt{s_{ep}}$ of the electron-proton
collision as 
\begin{equation}
y=\frac{W^{2}+Q^{2}-m_{N}^{2}}{s_{ep}-m_{N}^{2}}.
\end{equation}
Since we are mostly interested in the small-$Q$ kinematics, which
gives the dominant contribution, we may disregard the contributions
of the longitudinal polarization and rewrite the cross-section~(\ref{eq:sigma_def})
as
\begin{align}
\frac{d\sigma_{ep\to e \gamma \chi_c  p}}{d\Omega} & \approx\frac{d\sigma^{(T)}}{d\Omega}+\epsilon\cos2\varphi\frac{d\sigma^{(TT)}}{d\Omega}+\epsilon\sin2\varphi\frac{d\sigma^{(T'T)}}{d\Omega}\approx\frac{d\sigma^{(T)}}{d\Omega}\left(1+c_{2}\cos2\varphi+s_{2}\sin2\varphi\right)\label{eq:sigma_def-1}
\end{align}
where the leading (angular independent) contribution may be related
to the photoproduction cross-section~(\ref{eq:Photo}) in the equivalent
photon approximation~\cite{Weizsacker:1934,Williams:1935,Budnev:1975poe}
as
\begin{equation}
\frac{d\sigma^{(T)}}{d\ln W^{2}dQ^{2}\,dt\,dt'\,dM_{\gamma\chi_{c}}d\varphi}\approx\frac{\alpha_{{\rm em}}}{2\pi^{2}\,Q^{2}}\,\left(1-y+\frac{y^{2}}{2}-(1-y)\frac{Q_{{\rm min}}^{2}}{Q^{2}}\right)\frac{d\sigma_{\gamma p\to \gamma\chi_c  p}}{dt\,dt'\,dM_{\gamma\chi_{c}}},\label{eq:LTSep}
\end{equation}
the variable $Q_{{\rm min}}^{2}=m_{e}^{2}y^{2}/\left(1-y\right)$,
$m_{e}$ is the mass of the electron. The coefficients (angular harmonics)
$c_{2},\,s_{2}$ in~(\ref{eq:sigma_def-1}) may be related to different
components of the amplitude $\mathcal{A}_{\gamma p\to\chi_{c}\gamma p}^{(\lambda,\sigma,H)}$
as 
\begin{equation}
c_{2}=\frac{2\epsilon\,\,{\rm Re}\left(\mathcal{A}_{\gamma p\to\chi_{c}\gamma p}^{(-,+,H)*}\mathcal{A}_{\gamma p\to\chi_{c}\gamma p}^{(+,+,H)}\right)}{\left|\mathcal{A}_{\gamma p\to\chi_{c}\gamma p}^{(+,+,H)}\right|^{2}+\left|\mathcal{A}_{\gamma p\to\chi_{c}\gamma p}^{(-,+,H)}\right|^{2}},\quad s_{2}=\frac{2\epsilon\,\,{\rm Im}\left(\mathcal{A}_{\gamma p\to\chi_{c}\gamma p}^{(-,+,H)*}\mathcal{A}_{\gamma p\to\chi_{c}\gamma p}^{(+,+,H)}\right)}{\left|\mathcal{A}_{\gamma p\to\chi_{c}\gamma p}^{(+,+,H)}\right|^{2}+\left|\mathcal{A}_{\gamma p\to\chi_{c}\gamma p}^{(-,+,H)}\right|^{2}},\label{eq:c2s2}
\end{equation}
and for derivation of~(\ref{eq:c2s2}) we used explicit expressions
for the components of the leptonic tensor $L_{\lambda,\bar{\lambda}}$
in helicity basis~\cite{Mantysaari:2020lhf}
\begin{equation}
L_{\pm,\pm}=e^{2}\frac{Q^{2}\left(2-2y+y^{2}\right)}{y^{2}},\qquad L_{\pm,\mp}=e^{2}e^{\pm2i\varphi}\frac{2Q^{2}\left(1-y\right)}{y^{2}}.
\end{equation}
The ratio of $c_{2}$ and $s_{2}$ gives us access to he relative
phase of $\mathcal{A}_{\gamma p\to\chi_{c}\gamma p}^{(+,+,H)}$ and
$\mathcal{A}_{\gamma p\to\chi_{c}\gamma p}^{(-,+,H)}$ amplitudes,
namely
\begin{equation}
\frac{s_{2}}{c_{2}}={\rm tan}\left[{\rm arg}\left(\mathcal{A}_{\gamma p\to\chi_{c}\gamma p}^{(+,+,H)}\right)-{\rm arg}\left(\mathcal{A}_{\gamma p\to\chi_{c}\gamma p}^{(-,+,H)}\right)\right]
\end{equation}
which presents a new observable.

\subsection{The amplitude of the $\chi_{c}\gamma$ pair production}

\label{subsec:Amplitudes}
\begin{figure}
\includegraphics[scale=0.4]{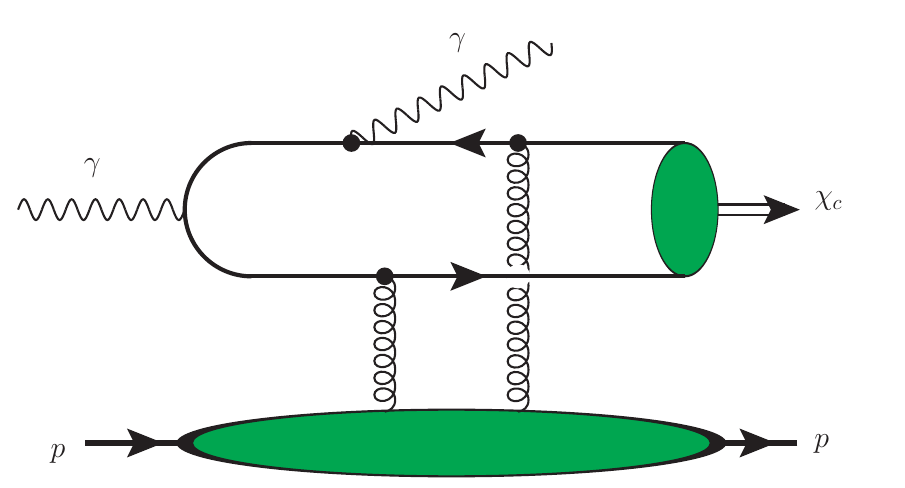}\includegraphics[scale=0.4]{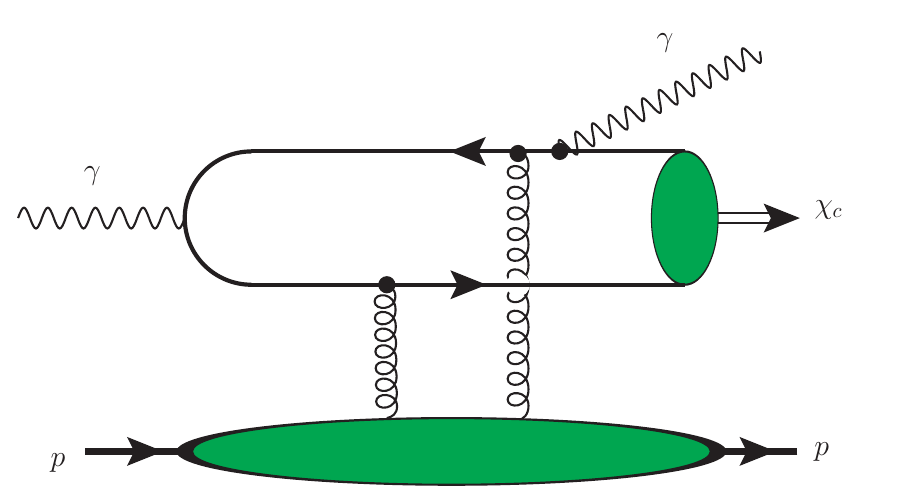}\includegraphics[scale=0.4]{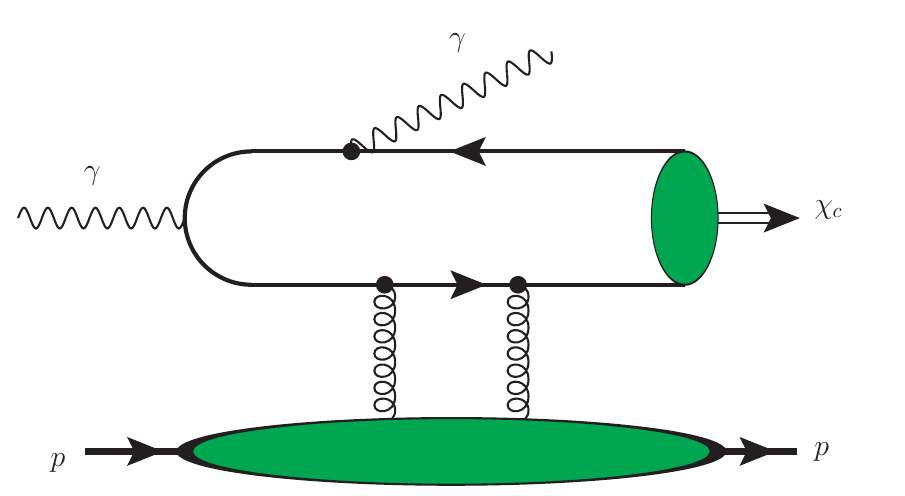}

\includegraphics[scale=0.4]{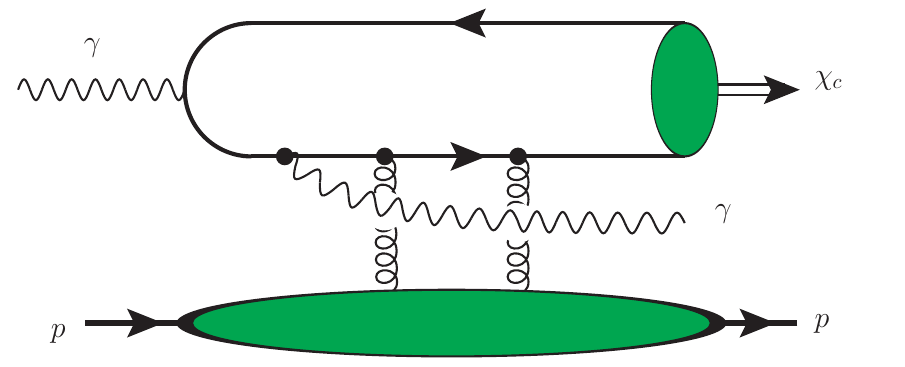}\includegraphics[scale=0.4]{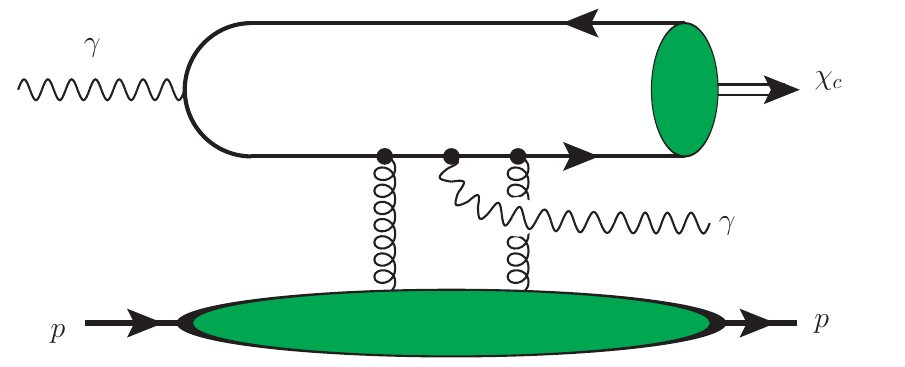}\includegraphics[scale=0.4]{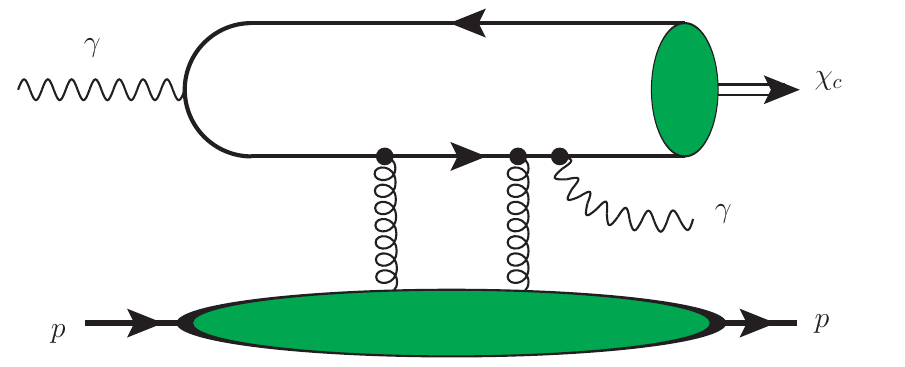}

\caption{\protect\label{fig:Photoproduction-A}The diagrams which contribute
to $\chi_{c}\gamma$ photoproduction. It is implied that each diagram
should be supplemented with a charge conjugate diagram (with inverted
heavy quark lines) and additional contribution with permuted gluon
vertices in $t$-channel (24 diagrams in total, see Appendix~\ref{sec:CoefFunction}
for more details). We assume that the electromagnetic field is given
in axial gauge $\epsilon\cdot p_{\chi_{c}}=0$ and thus do not take
into account contributions with coupling of the electromagnetic field
to the Wilson line inside $\chi_{c}$.}
\end{figure}

The evaluation of the amplitude $\mathcal{A}_{\gamma p\to\chi_{c}\gamma p}$
follows the standard rules of the collinear factorization framework,
and allows to express it in terms of the GPDs of the target~\cite{Diehl:2000xz,Diehl:2003ny,Guidal:2013rya,Boer:2011fh,Burkert:2022hjz}
convoluted with process-dependent partonic amplitudes (coefficient
functions). As discussed earlier, in the heavy quark mass limit we
may disregard the proton mass of the proton $m_{N}$ and all components
of the momentum transfer $\Delta^{\mu}$. We will consider that the
invariant mass $M_{\gamma\chi_{c}}$ is large enough to get rid of
the feed-down contributions from radiative decays of higher state
charmonia. In our evaluations we may disregard the fact that the
frame~(\ref{eq:q}-\ref{eq:k}) differs from the symmetric frame
where the GPDs conventionally are introduced, because the transverse
Lorentz boost which relates the two frames is given by 
\begin{align}
\ell^{+} & \to\ell^{+},\qquad\ell^{-}\to\ell^{-}+\ell^{+}\beta_{\perp}^{2}+\boldsymbol{\ell}_{\perp}\cdot\boldsymbol{\beta}_{\perp},\qquad\boldsymbol{\ell}_{\perp}\to\boldsymbol{\ell}_{\perp}+\ell^{+}\boldsymbol{\beta}_{\perp},\quad{\rm where}\quad\boldsymbol{\beta_{\perp}}=-\boldsymbol{\Delta_{\perp}}/\left(2\bar{P}_{{\rm SRF}}^{+}\right),
\end{align}
and the momentum $\boldsymbol{\Delta}_{\perp}$ is eventually disregarded
in coefficient function. For this reason, the momenta of the active
parton (gluon) before and after interaction can be approximated as
\begin{align}
k_{i} & =\left((x+\xi)\bar{P}^{+},\,0,\,0\right),\quad k_{f}=\left((x-\xi)\bar{P}^{+},\,0,\,\boldsymbol{\Delta}_{\perp}\right),
\end{align}
where $\bar{P}^{+}=\left(P_{{\rm in}}^{+}+P_{{\rm out}}^{+}\right)/2$
and the variable $x$ is the light-cone fraction of the average parton
momentum defined as $x=\left(k_{i}^{+}+k_{f}^{+}\right)/\left(2\bar{P}^{+}\right)$.
For unpolarized proton the straightforward spinor algebra yields for
the square of the amplitude~\cite{Belitsky:2001ns}
\begin{align}
\sum_{{\rm spins}}\left|\mathcal{A}_{\gamma p\to M_{1}M_{2}p}^{(\lambda,\,\sigma,H)}\right|^{2} & =\left[4\left(1-\xi^{2}\right)\left(\left|\mathcal{H}_{\gamma\chi_{c}}^{(\lambda,\,\sigma,H)}\right|^{2}+\left|\tilde{\mathcal{H}}_{\gamma\chi_{c}}^{(\lambda,\,\sigma,H)}\right|^{2}\right)-2\xi^{2}{\rm Re}\left(\mathcal{H}_{\gamma\chi_{c}}^{(\lambda,\,\sigma,H)}\mathcal{E}_{\gamma\chi_{c}}^{(\lambda,\,\sigma,H)*}+\tilde{\mathcal{H}}_{\gamma\chi_{c}}^{(\lambda,\,\sigma,H)}\tilde{\mathcal{E}}_{\gamma\chi_{c}}^{(\lambda,\,\sigma,H)*}\right)\right.\label{eq:AmpSq}\\
 & \qquad\left.-\left(\xi^{2}+\frac{t}{4m_{N}^{2}}\right)\left|\mathcal{E}_{\gamma\chi_{c}}^{(\lambda,\,\sigma,H)}\right|^{2}-\xi^{2}\frac{t}{4m_{N}^{2}}\left|\tilde{\mathcal{E}}_{\gamma\chi_{c}}^{(\lambda,\,\sigma,H)}\right|^{2}\right],\nonumber 
\end{align}
where, inspired by previous studies of Compton scattering and single-meson
production~\cite{Belitsky:2001ns,Belitsky:2005qn}, we introduced
shorthand notations for the convolutions of the partonic amplitudes
with GPDs

\begin{align}
\mathcal{H}_{\gamma\chi_{c}}^{(\lambda,\,\sigma,H)}\left(\xi,\,t\right) & =\int dx\,C_{\gamma\chi_{c}}^{(\lambda,\,\sigma,H)}\left(x,\,\xi\right)H_{g}\left(x,\xi,t\right),\qquad\mathcal{E}_{\gamma\chi_{c}}^{(\lambda,\,\sigma,H)}\left(y_{1},y_{2},t\right)=\int dx\,C_{\gamma\chi_{c}}^{(\lambda,\,\sigma,H)}\left(x,\,\xi\right)E_{g}\left(x,\xi,t\right),\label{eq:Ha}\\
\tilde{\mathcal{H}}_{\gamma\chi_{c}}^{(\lambda,\,\sigma,H)}\left(\xi,\,t\right) & =\int dx\,\tilde{C}_{\gamma\chi_{c}}^{(\lambda,\,\sigma,H)}\left(x,\,\xi\right)\tilde{H}_{g}\left(x,\xi,t\right),\qquad\tilde{\mathcal{E}}_{\gamma\chi_{c}}^{(\lambda,\,\sigma,H)}\left(y_{1},y_{2},t\right)=\int dx\,\tilde{C}_{\gamma\chi_{c}}^{(\lambda,\,\sigma,H)}\left(x,\,\xi\right)\tilde{E}_{g}\left(x,\xi,t\right).\label{eq:ETildeA}
\end{align}
The partonic amplitudes $C_{\gamma\chi_{c}}^{(\lambda,\,\sigma,H)},\,\tilde{C}_{\gamma\chi_{c}}^{(\lambda,\,\sigma,H)}$
can be evaluated perturbatively, taking into account the diagrams
shown in the Figure~\ref{fig:Photoproduction-A}. For evaluation
of these amplitudes we assume that the quarkonium is described by
the wave function $\Phi_{\chi_{c}}\left(z,\,\boldsymbol{k}_{\perp}\right)$,
and use explicit expressions for different helicity components of
the latter from~\cite{Benic:2024}. Since the quarkonium moves with
nonzero transverse momentum in chosen frame, the relative momentum
between the quark and antiquark can be expressed in terms of $z,\,\boldsymbol{k}_{\perp}$
as
\begin{equation}
k_{Q}^{\mu}-k_{\bar{Q}}^{\mu}=k^{\mu}=\left(\frac{\left(1-2z\right)\left(M_{\chi_{c}}^{2}+\boldsymbol{p}_{\perp\chi_{c}}^{2}\right)-2\,\boldsymbol{k}_{\perp}\cdot\boldsymbol{p}_{\perp\chi_{c}}}{\sqrt{2}p_{\chi_{c}}^{-}},\quad\sqrt{2}p_{\chi_{c}}^{-}\left(2z-1\right),\quad\boldsymbol{k}_{\perp}\right)\label{eq:VecK}
\end{equation}
and is orthogonal to the vector $p_{\chi_{c}}^{\mu}$. In the heavy
quark mass limit the internal motion of the quarks becomes slow, so
the the wave function $\Phi_{\chi_{c}}\left(z,\,\boldsymbol{k}_{\perp}\right)$
becomes very narrow in the momentum space. For this reason, we can
make expansion of the partonic amplitudes over the small components
$\boldsymbol{k}_{\perp},\,\left(1-2z\right)$ up to the first nonvanishing
order. Due to $C$-parity all the odd powers of $\left(1-2z\right),\,\boldsymbol{k}_{\perp}$
eventually cancel and the expansion starts from the second order terms
in $\boldsymbol{k}_{\perp}^{2}$, $\left(2z-1\right)^{2}$. Matching
of this procedure with NRQCD framework in the heavy quark mass limit
is discussed in Appendix~\ref{subsec:WF} (we remind that for $P$-wave
the NRQCD projector is more complicated than for the $S$-wave and
includes differential operators). Using the relations~(\ref{eq:Rel},~\ref{eq:LDME})
for light-cone amplitudes and explicit parametrization~(\ref{eq:Explicit})
provided in Appendix, we may express the averages
\begin{equation}
\left\langle \left(z-\frac{1}{2}\right)^{2}\right\rangle =\int\frac{dz\,d^{2}\boldsymbol{k}}{z(1-z)8\pi^{2}}\phi\left(z,\,\boldsymbol{k}_{\perp}\right)\left(z-\frac{1}{2}\right)^{2},\,\left\langle \boldsymbol{k}_{\perp}^{2}\right\rangle =\int\frac{dz\,d^{2}\boldsymbol{k}}{z(1-z)8\pi^{2}}\phi\left(z,\,\boldsymbol{k}_{\perp}\right)\boldsymbol{k}_{\perp}^{2}
\end{equation}
in terms of the NRQCD LDME $\left\langle \mathcal{\mathcal{O}}_{\chi_{c}}^{[1]}\left(^{3}P_{J}^{[1]}\right)\right\rangle $.
The remaining steps of evaluation of these amplitudes are straightforward
and are discussed in more detail in Appendix~\ref{sec:CoefFunction}.
The final expressions for the coefficient functions conveniently can
be represented as

\begin{align}
C_{\gamma\chi_{c}}^{(\lambda,\,\sigma,H)} & =\frac{m_{c}^{2}}{M_{\chi_{c}}^{2}}c_{\gamma\chi_{c}}^{(\lambda,\,\sigma,\,H)}+d_{\gamma\chi_{c}}^{(\lambda,\,\sigma,\,H)}\approx\frac{1}{4}c_{\gamma\chi_{c}}^{(\lambda,\,\sigma,\,H)}+d_{\gamma\chi_{c}}^{(\lambda,\,\sigma,\,H)}\label{eq:Monome}\\
\tilde{C}_{\gamma\chi_{c}}^{(\lambda,\,\sigma,H)} & =\frac{m_{c}^{2}}{M_{\chi_{c}}^{2}}\tilde{c}_{\gamma\chi_{c}}^{(\lambda,\,\sigma,\,H)}+\tilde{d}_{\gamma\chi_{c}}^{(\lambda,\,\sigma,\,H)}\approx\frac{1}{4}\tilde{c}_{\gamma\chi_{c}}^{(\lambda,\,\sigma,\,H)}+\tilde{d}_{\gamma\chi_{c}}^{(\lambda,\,\sigma,\,H)}\label{eq:Monome-1}
\end{align}
where the terms $c_{\gamma\chi_{c}}^{(\lambda,\,\sigma,\,H)},\,d_{\gamma\chi_{c}}^{(\lambda,\,\sigma,\,H)}$
correspond to contributions which stem from expansion over $\left(z-\frac{1}{2}\right)$
and $\boldsymbol{k}_{\perp}$, respectively. The explicit expressions
for the $c_{\gamma\chi_{c}}^{(\lambda,\,\sigma,\,H)},\,d_{\gamma\chi_{c}}^{(\lambda,\,\sigma,\,H)},$
$\tilde{c}_{\gamma\chi_{c}}^{(\lambda,\,\sigma,\,H)}$ and $\tilde{d}_{\gamma\chi_{c}}^{(\lambda,\,\sigma,\,H)}$
are given in Appendix~\ref{sec:CoefFunction} in view of their extensive
size. In general the coefficient functions are rational functions
of the variable $\zeta=x/\xi$  with isolated poles at $\zeta=\pm1\mp i0$
and $\zeta=\pm\kappa\mp i0$ ($x=\pm\xi\mp i0$ and $x=\pm\kappa\xi\mp i0$
in conventional notations), where
\begin{equation}
\kappa=1-\frac{1-1/r^{2}}{1-\alpha/2}=\frac{1}{r^{2}}\frac{2-\alpha r^{2}}{2-\alpha},\qquad r=M_{\gamma\chi_{c}}/M_{\chi_{c}}=\frac{W}{M_{\chi_{c}}}\sqrt{\frac{2\xi}{1+\xi}\left(1-\frac{m_{N}^{2}}{W^{2}}\right)}\approx\frac{W}{M_{\chi_{c}}}\sqrt{\frac{2\xi}{1+\xi}}.\label{eq:kappa}
\end{equation}
The values of $\kappa$ are limited by $|\kappa|<1$ in the physically
relevant kinematics ($r>1,\,\alpha\in(1/r^{2},1)$). In contrast to
our previous results for $\eta_{c}\gamma$ production~\cite{Siddikov:2024blb},
the poles are of the second order. This happens because for $S$-wave
quarkonia we just disregarded the relative motion of the quarks in
the heavy quark mass limit, whereas for the $P$-wave quarkonia we
expand individual diagrams over the relative momentum (velocity) and
pick up higher order terms, thus increasing the order of the poles
when such expansion comes from denominator. Fortunately, the standard
$m^{2}\to m^{2}-i0$ prescription in Feynman propagators provides
contour deformation prescription near these poles, and allows to avoid
the potential ambiguities in the integration. Afterwards the integration
in the vicinity of these poles formally can be carried out using a
set of well-known identities
\begin{align}
\int_{-1}^{1}dx\frac{F\left(x,\xi,t\right)}{\left(x-x_{0}\pm i0\right)^{2}} & =-\int_{-1}^{1}dx\,F\left(x,\xi,t\right)\frac{d}{dx}\frac{1}{\left(x-x_{0}\pm i0\right)}=-\left.\frac{F\left(x,\xi,t\right)}{\left(x-x_{0}\pm i0\right)}\right|_{x=-1}^{x=+1}+\int_{-1}^{1}dx\,\frac{\partial_{x}F\left(x,\xi,t\right)}{\left(x-x_{0}\pm i0\right)}.\label{eq:DoublePole}\\
 & \int_{-1}^{1}dx\,\frac{H\left(x,\xi,t\right)}{\left(x-x_{0}\pm i0\right)}={\rm P.V.}\int_{-1}^{1}dx\,\frac{H\left(x,\xi,t\right)}{\left(x-x_{0}\right)}\mp i\pi\,H\left(x_{0},\xi,t\right)
\end{align}
where $x_{0}$ is some constant which may depend parametrically on
$\alpha,\,r,\,\,\xi$. At special point $\alpha=2/r^{2}$ the value
of $\kappa$ vanishes and the poles $x=\pm\kappa\xi\mp i0$ apparently
start pinching the integration contour. However, this does not lead
to a physical singularity in the amplitude because each individual
term which contributes additively to $C_{\gamma\chi_{c}}^{(\lambda,\sigma,H)}$
has an isolated pole either at $x=\kappa\xi-i0$ or $x=-\kappa\xi+i0$. 

\begin{figure}

\includegraphics[width=6cm]{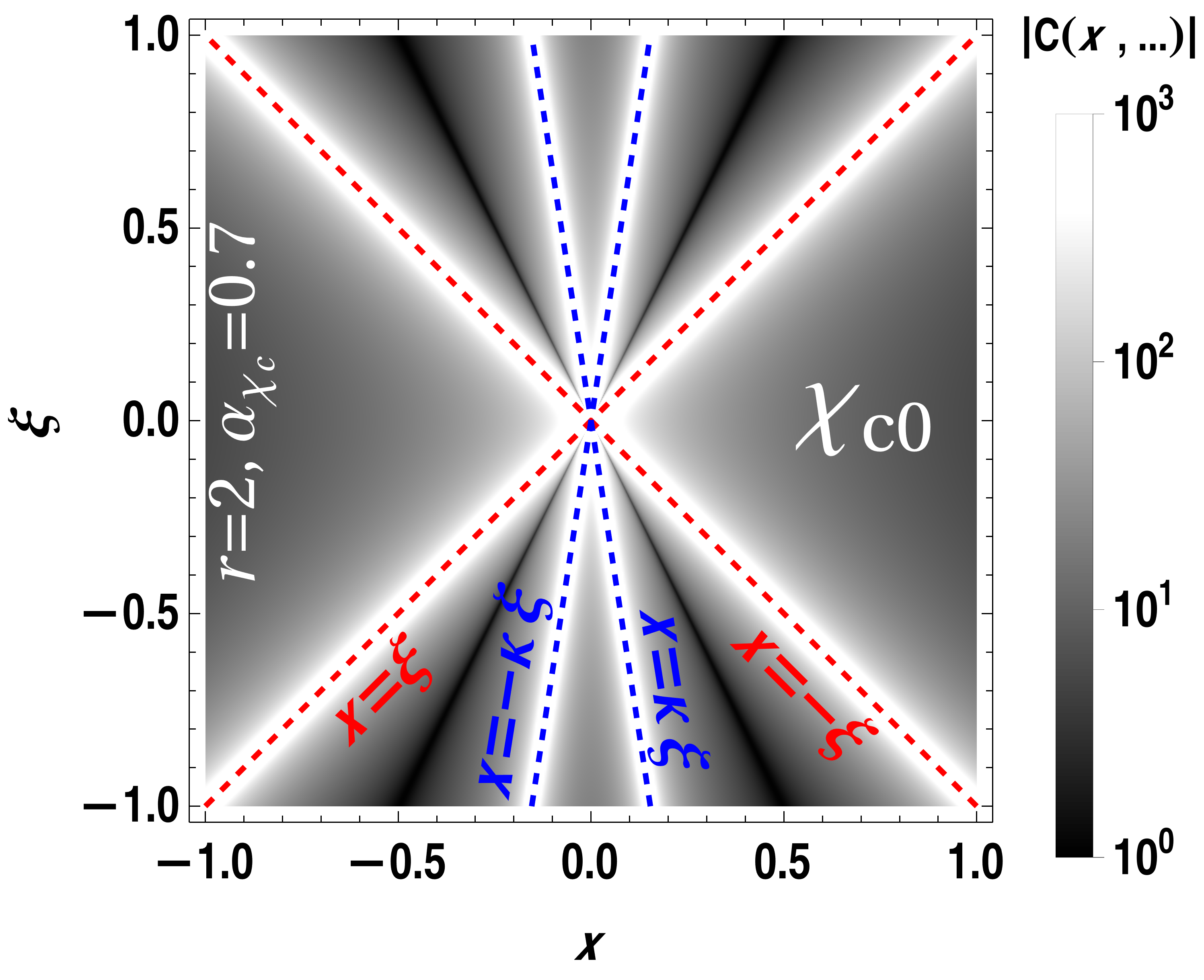}\includegraphics[width=6cm]{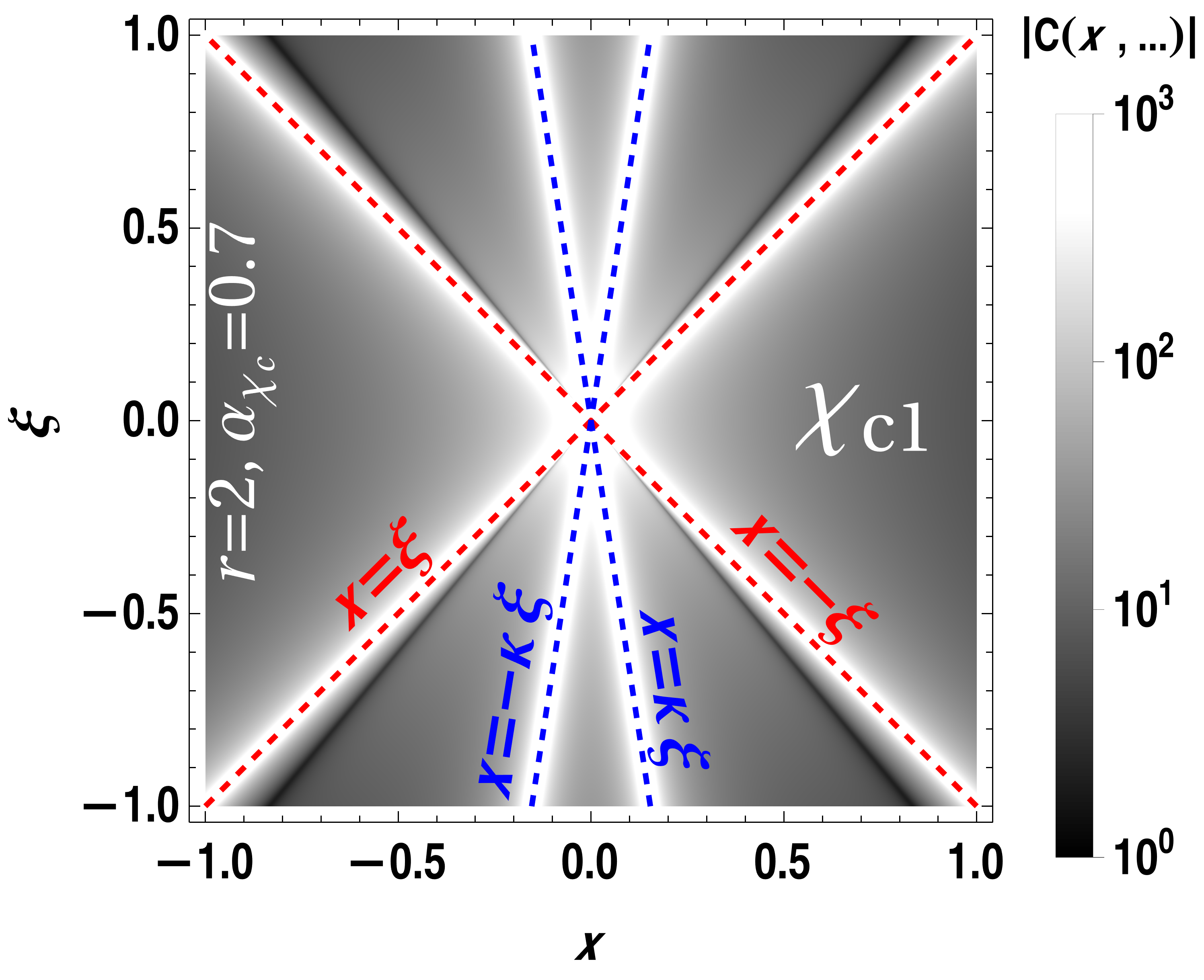}\includegraphics[width=6cm]{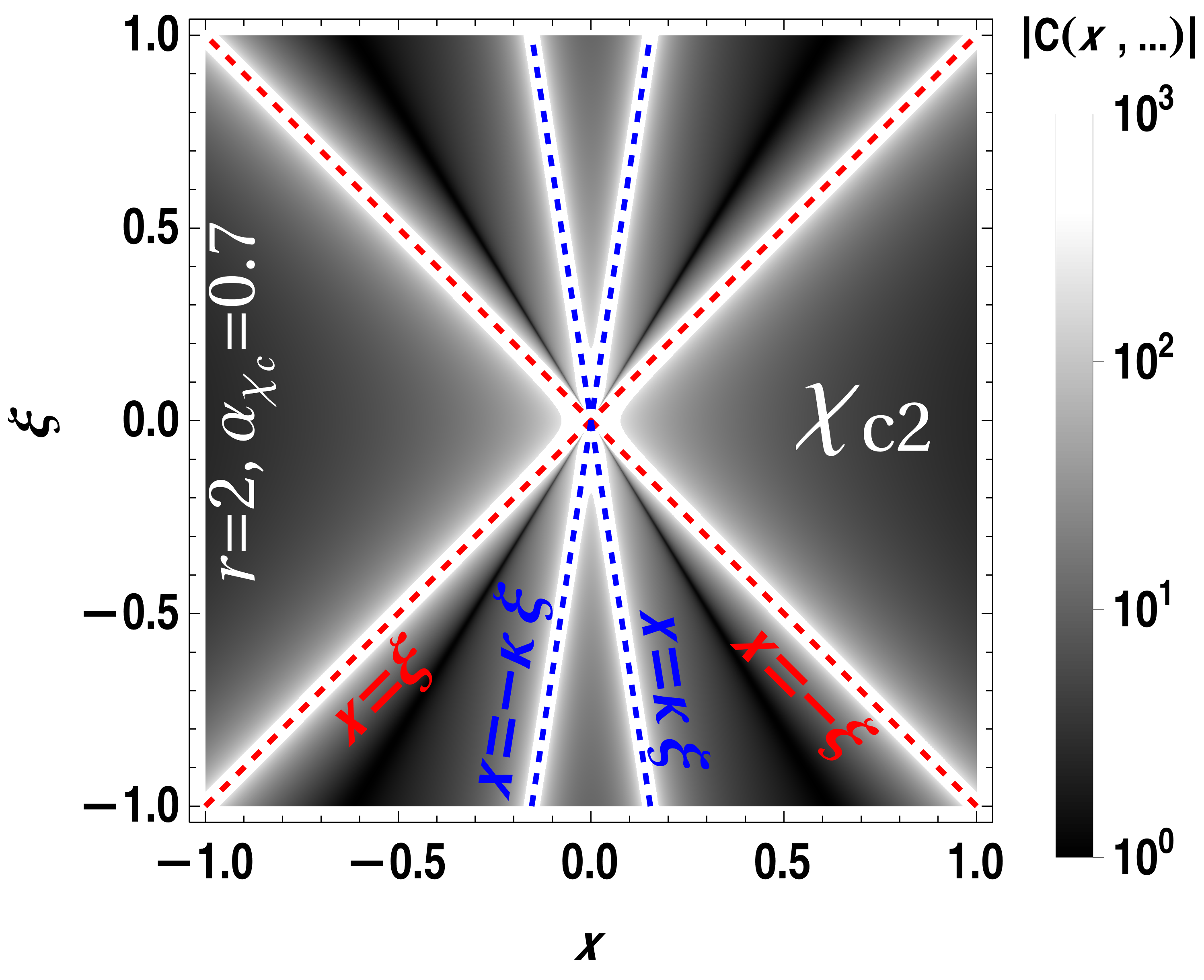}

\includegraphics[width=6cm]{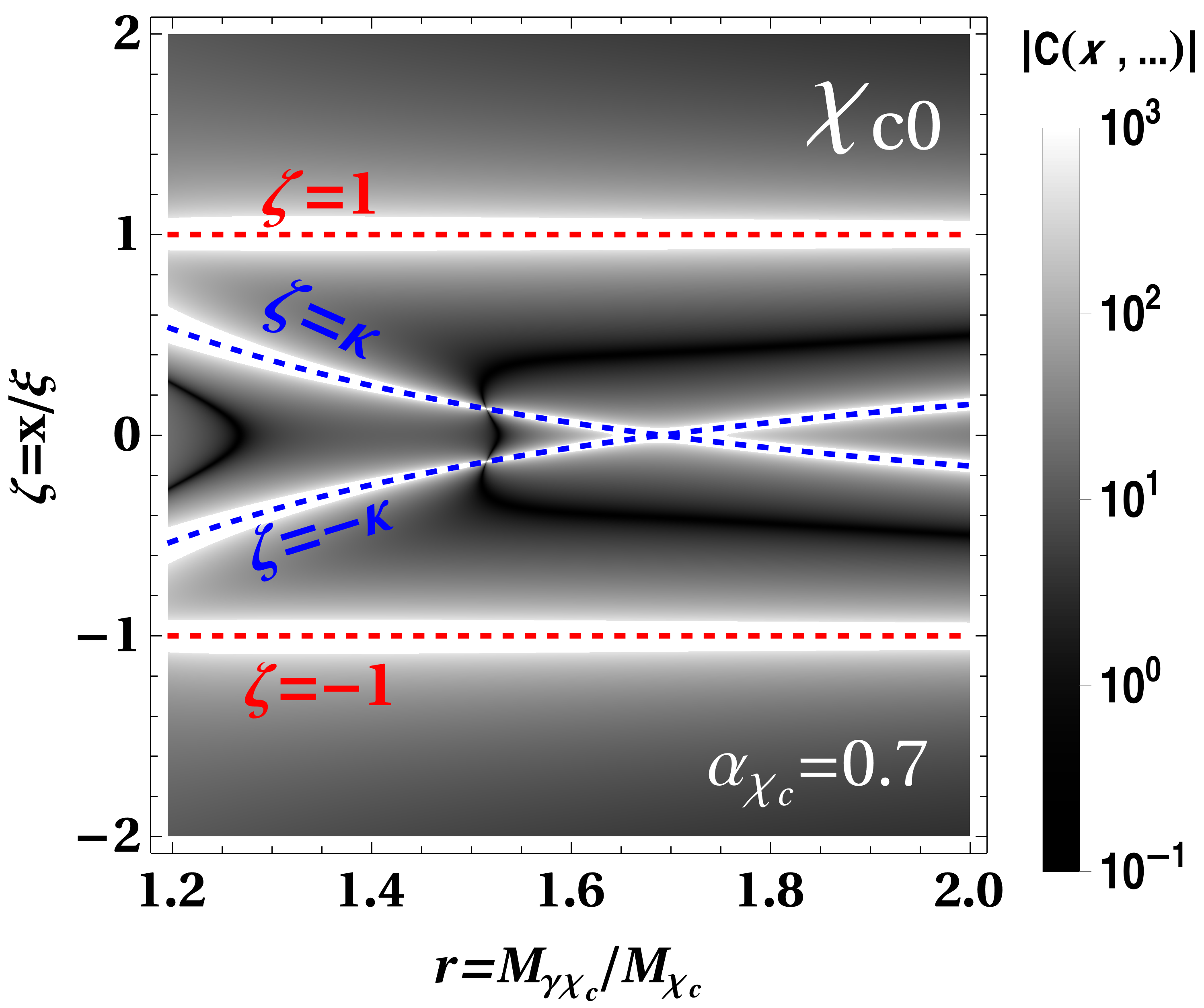}\includegraphics[width=6cm]{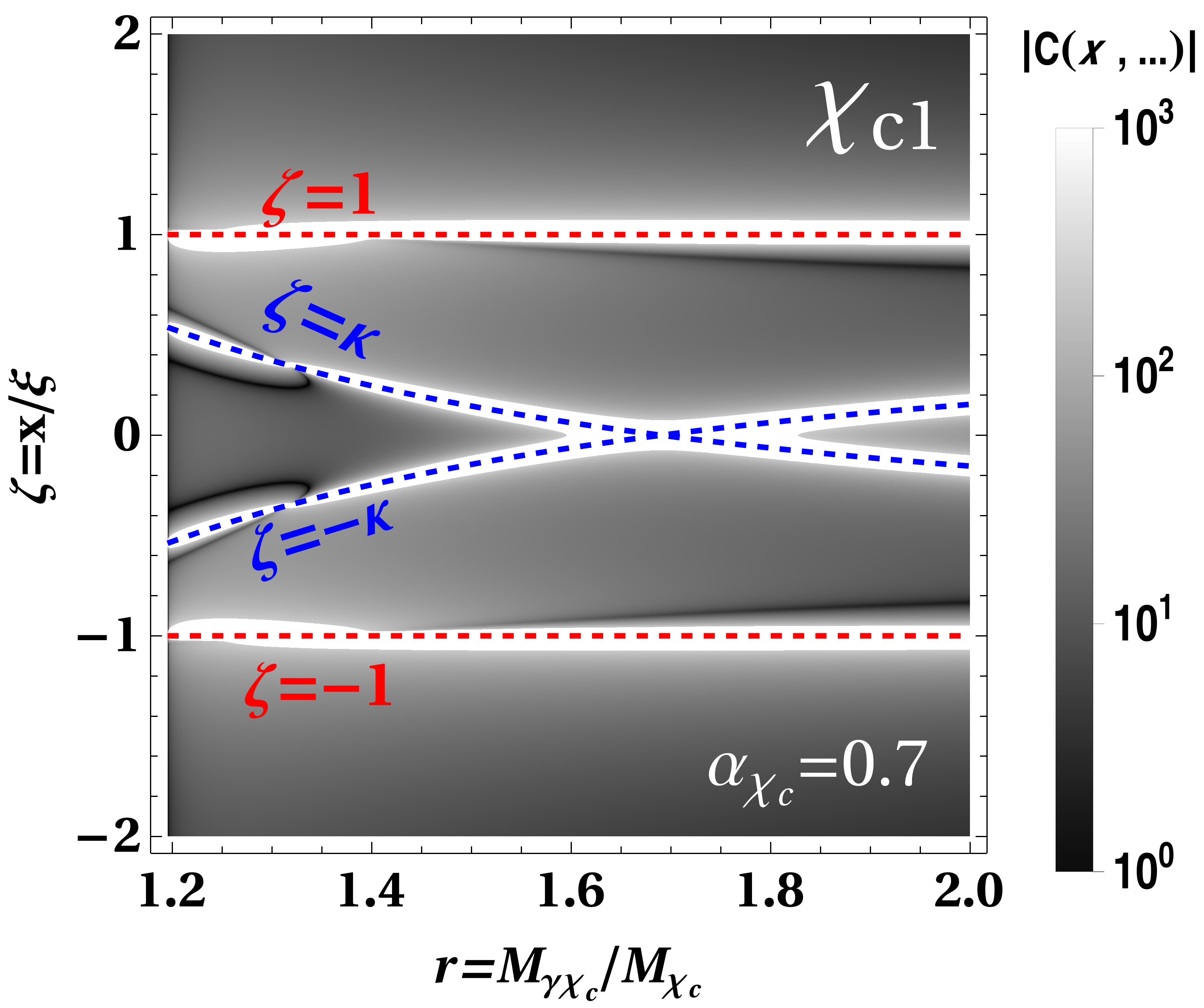}\includegraphics[width=6cm]{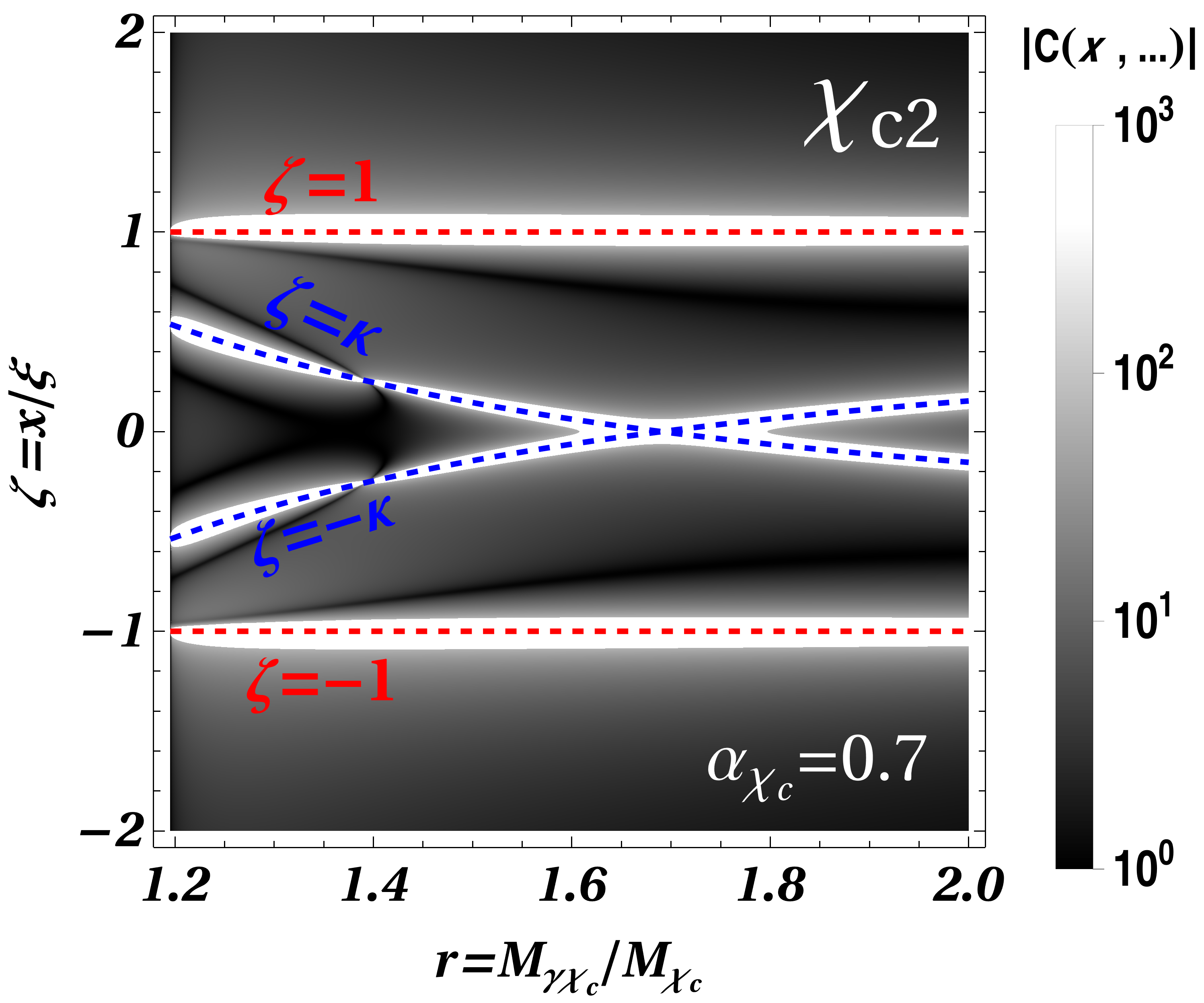}\caption{\protect\label{fig:CoefFunction}The density plot which illustrates
the coefficient function $C_{\gamma\chi_{c}}^{(++)}$ (in relative
units). Upper row: dependence on variables $x,\,\xi$ at fixed $r=M_{\gamma\chi_{c}}/M_{\chi_{c}}$,
$\alpha_{\chi_{c}}=(-u')/M_{\gamma\chi_{c}}$. Lower row: Dependence
on the ratio $\zeta=x/\xi$ and $r$ at fixed $\alpha_{\chi_{c}}$.
In both rows for $\chi_{c1}$ and $\chi_{c2}$ mesons we have shown
the coefficient function of the dominant component with helicity $H_{\chi_{c}}=+1$
(same as for the incoming photon). White strips with colored dashed
lines effectively demonstrate the position of the double poles $x=\pm\xi,\,x=\pm\kappa\,\xi$
($\zeta=\pm1,\,\pm\kappa$) in the coefficient function~(\ref{eq:Monome}).
The parameter $\kappa$ depends on the kinematics of the process and
is given in~(\ref{eq:kappa}); it is always bound by the constraint
$|\kappa|<1$ in the physically relevant kinematics.}
\end{figure}

In the Figure~\ref{fig:CoefFunction} we have shown dependence of
$C_{\gamma\chi_{c}}^{(+,+,H)}$ on its arguments for different mesons.
Due to space limitations, for $\chi_{c1}$ and $\chi_{c2}$ mesons
we have shown only the results for the component with helicity $H_{\chi_{c}}=+1$,
which are expected to give the largest contribution. While in general
the coefficient functions depend significantly on the spin and helicity
state of the produced quarkonia, the position of the poles (white
strips with dashed lines inside) remains largely the same because
it stems from denominators of the Feynman diagrams. We can see that
the functions $C_{\gamma\chi_{c}}^{(\lambda,\sigma,H)}$ have pronounced
peaks in the vicinity of their poles, and decrease rapidly when we
move away from them. This implies that position of the poles determines
the region which gives the dominant contribution in convolution integrals~(\ref{eq:Ha}-\ref{eq:ETildeA}).
Varying the invariant mass $M_{\gamma\chi_{c}}$ and the value of
$\alpha_{\chi_{c}}$, it is possible to change the parameter $\kappa$
in the region $|\kappa|<1$ and in this way probe the behavior of
the gluon GPDs $H_{g}$ in the whole ERBL domain.

\section{Numerical estimates}

\label{sec:Numer}

\subsection{Differential cross-sections}

\label{subsec:diff}
\begin{figure}
\includegraphics[width=6cm]{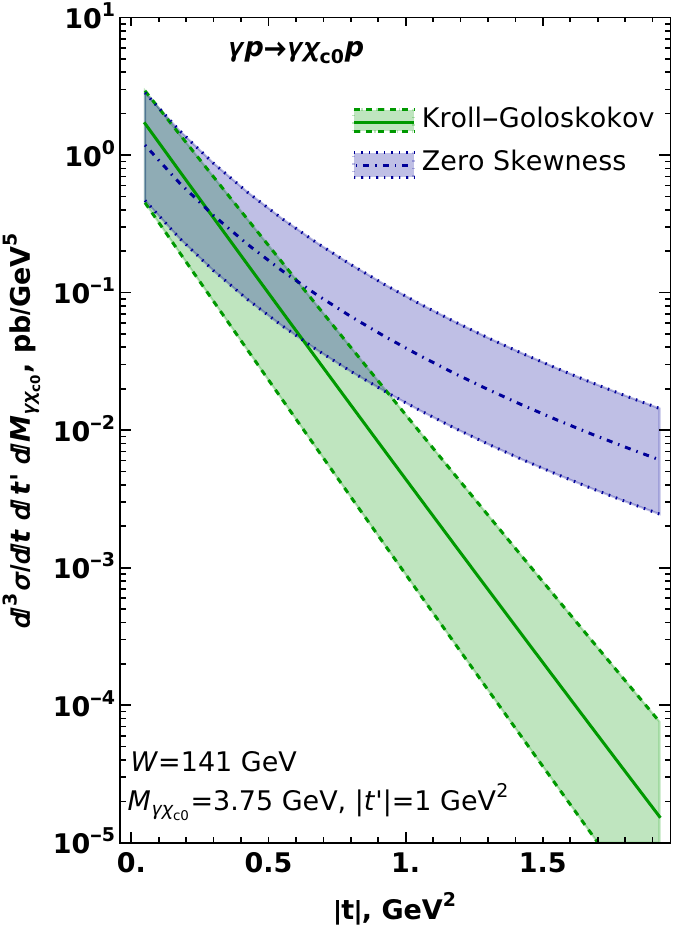}\includegraphics[width=6cm]{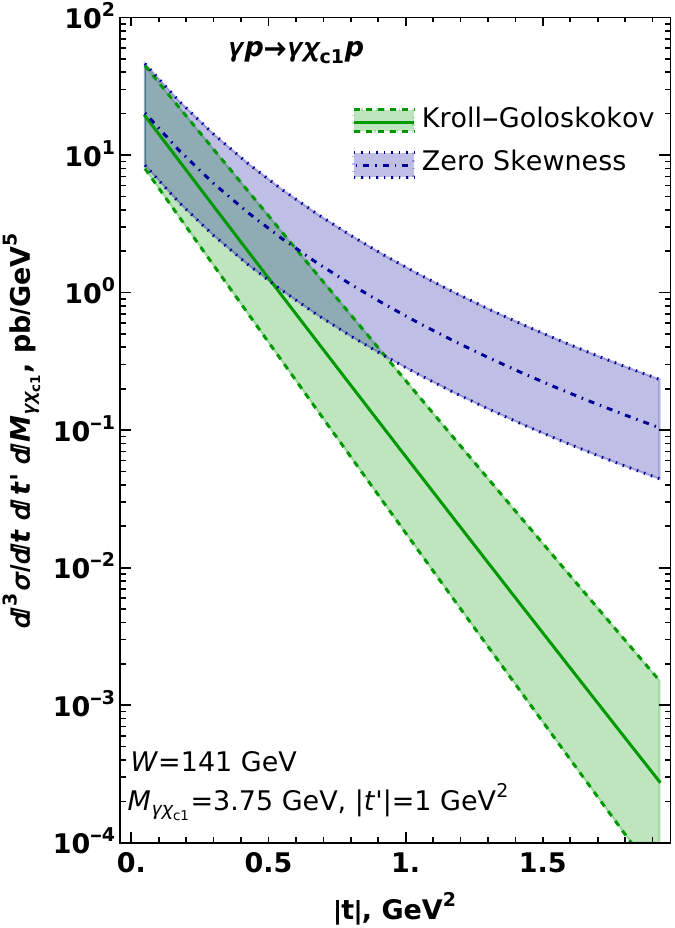}\includegraphics[width=6cm]{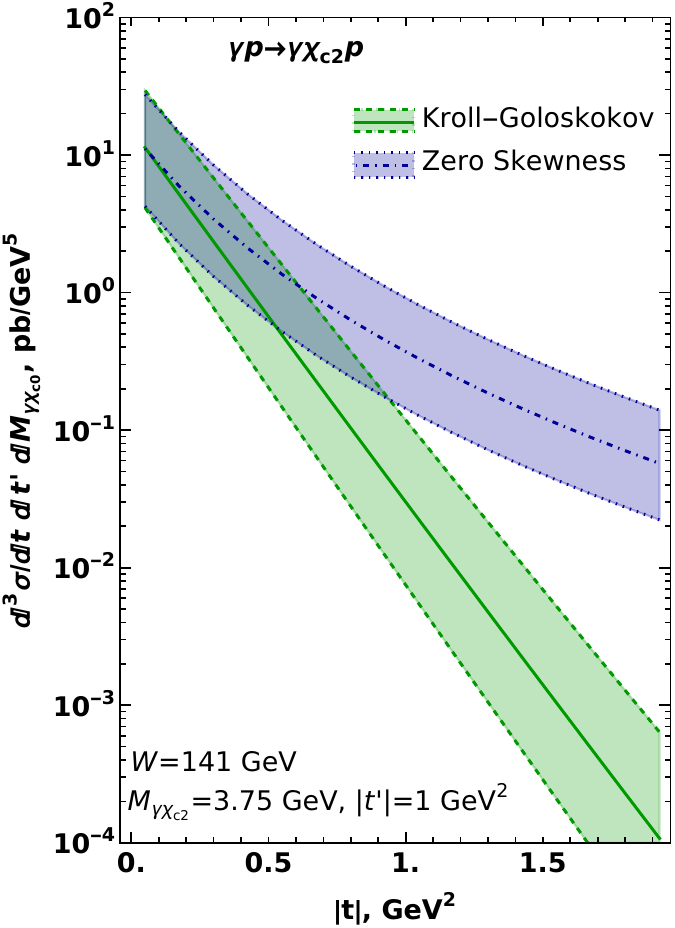}

\caption{\protect\label{fig:tDep-1}Dependence of the unpolarized photoproduction
cross-section~(\ref{eq:Photo}) on the invariant momentum transfer
$t$ to the target at fixed invariant mass $M_{\gamma\chi_{c}}.$
For $\chi_{c_{1}}$ and $\chi_{c_{2}}$ summation over all helicities
is implied. The upper and lower colored bands correspond to Zero Skewness
and Kroll-Goloskokov paramertizations of the gluon GPD. The width
of each band reflects uncertainty due to the omitted higher order
corrections and was obtained varying the factorization scale $\mu$
in the range $\mu\in\left(0.5\,...\,2\right)M_{\gamma\chi_{c}}$.}
\end{figure}

\begin{figure}
\includegraphics[width=9cm]{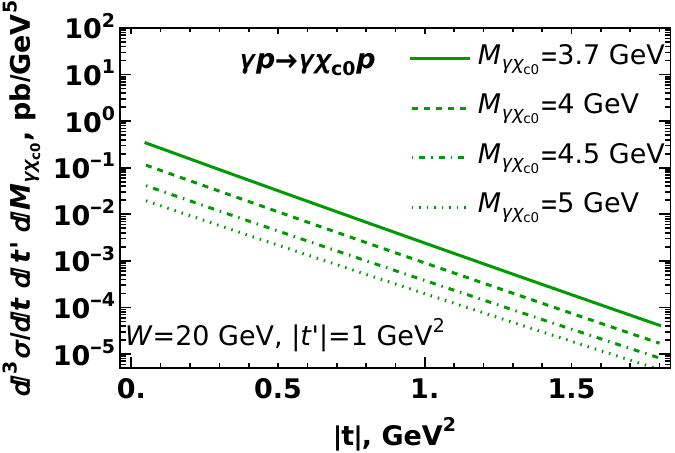}\includegraphics[width=9cm]{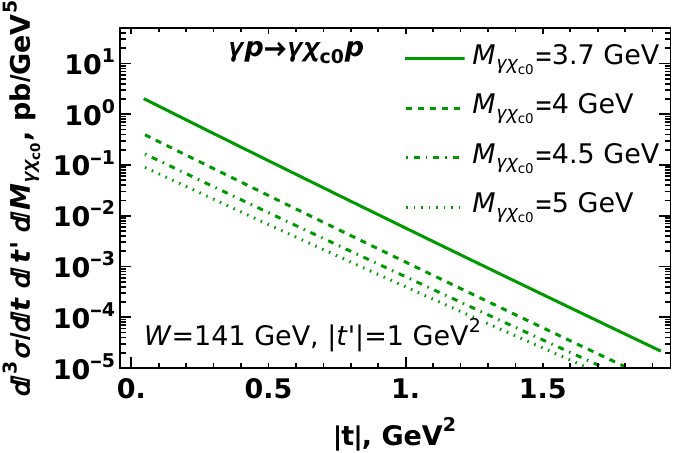}

\includegraphics[width=9cm]{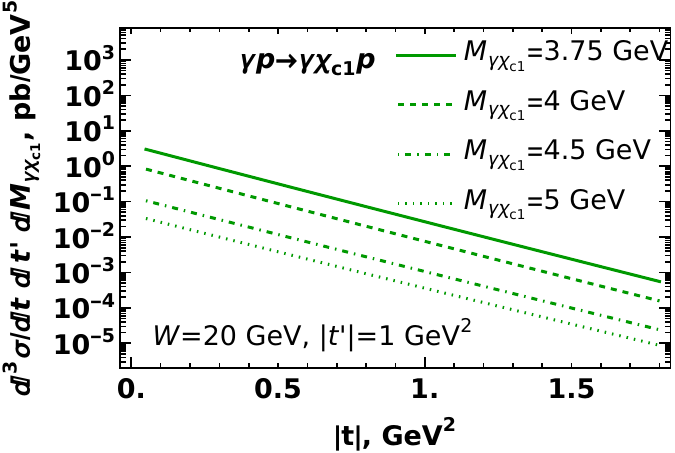}\includegraphics[width=9cm]{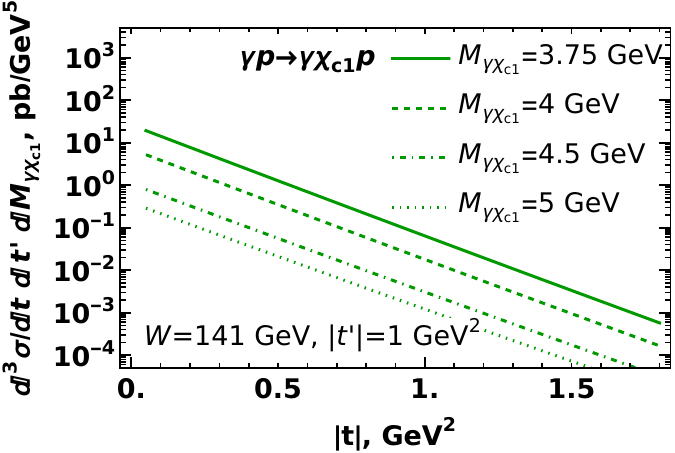}

\includegraphics[width=9cm]{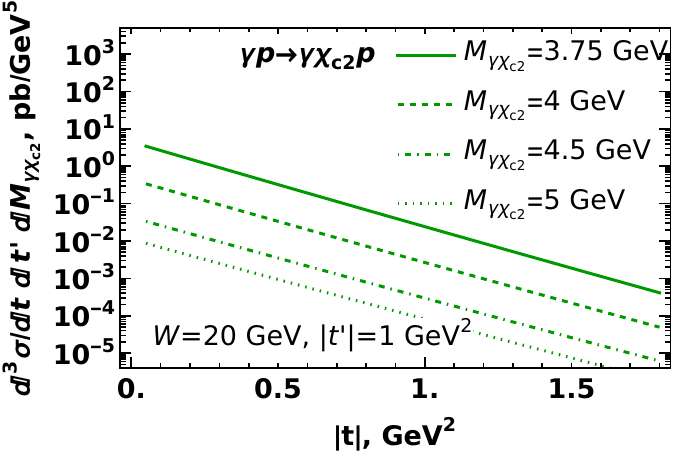}\includegraphics[width=9cm]{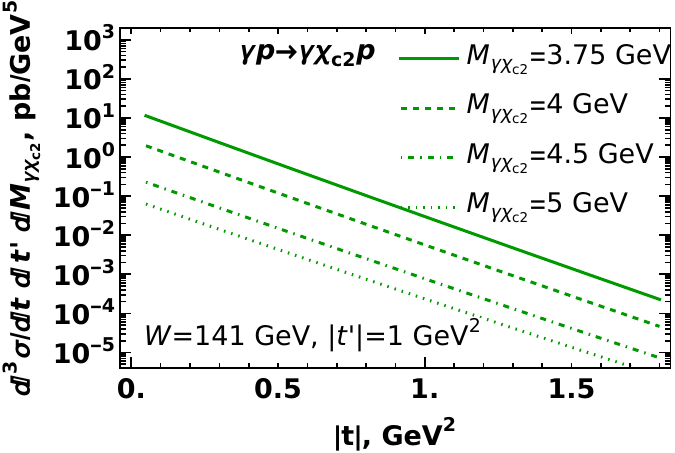}

\caption{\protect\label{fig:tDep}The differential cross-section~(\ref{eq:Photo})
as a function of the invariant momentum transfer $t$ for different
fixed invariant mass $M_{\gamma\chi_{c}}$. The upper, central and
lower rows correspond to $\chi_{c_{0}},\,\chi_{c_{1}}$ and $\chi_{c_{2}}$
mesons. The left and right columns differ by the value of collision
energy $W$. We also checked that the shape of the $t$-dependence
has negligible dependence on the energy $W$. }
\end{figure}

At present, the largest uncertainty for our predictions
stems from the choice of the Generalized Parton Distributions. In
order to illustrate the magnitude of this choice, in the Figure~\ref{fig:tDep-1}
we have shown the differential cross-section~(\ref{eq:Photo}) evaluated
with different parametrizations of GPD and plotted as a function of
the invariant momentum transfer $t$ to the target. The dependence
on this variable is the simplest for interpretation because the latter
is disregarded in coefficient function in the collinear kinematics,
and thus the observed dependence on $|t|$ is entirely due to the
implemented parametrization of the GPD. For comparison, we considered
the Kroll-Goloskokov parametrization from~\cite{Goloskokov:2006hr,Goloskokov:2007nt,Goloskokov:2008ib,Goloskokov:2009ia,Goloskokov:2011rd}
and the so-called Zero Skewness parametrization, which disregards
the dependence on the variable $\xi$ altogether and models the GPD
of each flavor as a product of the nucleon form factor and forward
Parton Distribution Function. We can see that the difference between
the predictions of the two parametrizations grows as a function $t$,
when the off-forward effects become more pronounced. Though the cross-section~(\ref{eq:Photo},\ref{eq:AmpSq})
obtains contributions of GPDs with different helicity states, for
unpolarized target the contribution of the GPD $H^{g}\left(x,\xi,t\right)$
is clearly the dominant. The width of the band in these plots illustrates
the uncertainty due to omitted higher order corrections and was found
varying the factorization scale $\mu$ in the range $\mu\in(0.5-2)\,M_{\gamma\chi_{c}}$.
In the Figure~\ref{fig:tDep} we show that the shape of the $t$-dependence
almost does not depend on the total spin state of $\chi_{c}$, the
energy of collision and invariant mass $M_{\gamma\chi_{c}}$. We also
found that the slope of the $t$-dependence does not depend on the
choice of the variable $t'$, and on helicity of $\chi_{c1},\,\chi_{c2}$.
Such sharply decreasing $t$-dependence is common to many exclusive
processes, and, as was explained in~\cite{Lepage:1980fj}, stems
from the fact that the (large) transverse momentum in exclusive process
should be shared equally between all partons inside the target in
order to avoid its destruction, and each such exchange is suppressed
by the (perturbative) propagators at large $t$. The $t$-dependence
implemented in phenomenological parametrizations of the GPDs is fitted
to the DVCS and DVMP data, which confirm such pronounced decrease.
Technically, this implies that the $\gamma\chi_{c}$ pairs predominantly
are produced in the kinematics where the variables $\boldsymbol{\Delta}_{\perp}^{2},\,t$
are negligibly small, so the transverse momenta $\boldsymbol{p}_{\perp}^{\chi_{c}},\,\boldsymbol{k}_{\perp}^{\gamma}$
are oppositely directed and have comparable magnitudes. Due to simplicity
of $t$-dependence, in our further presentation we will tacitly assume
that $|t|=|t_{{\rm min}}|$, or consider the observables in which
the dependence on $t$ is integrated out.

\begin{figure}
\includegraphics[width=9cm]{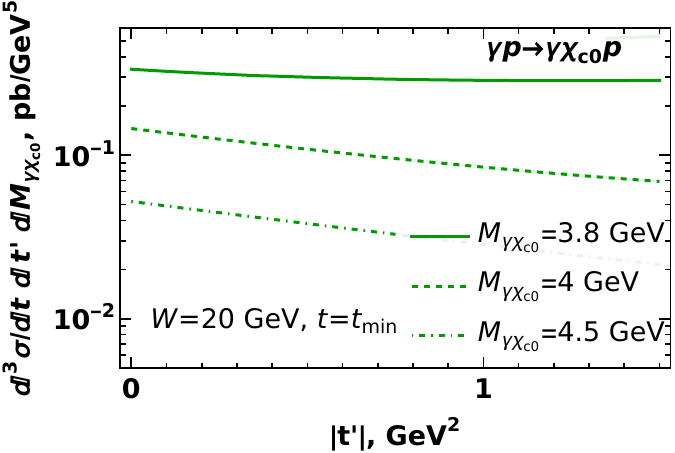}\includegraphics[width=9cm]{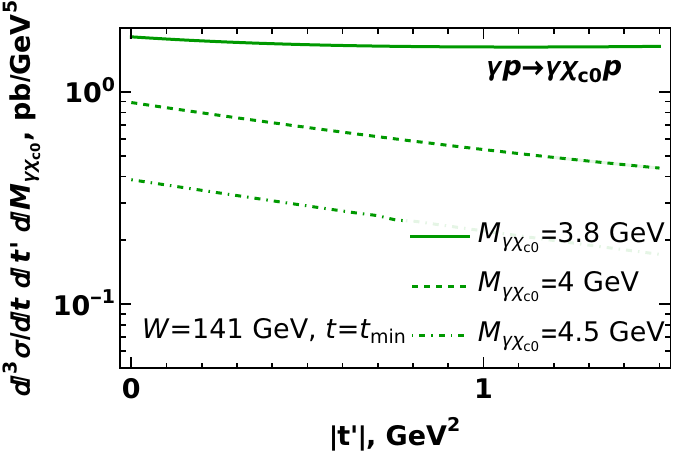}

\includegraphics[width=9cm]{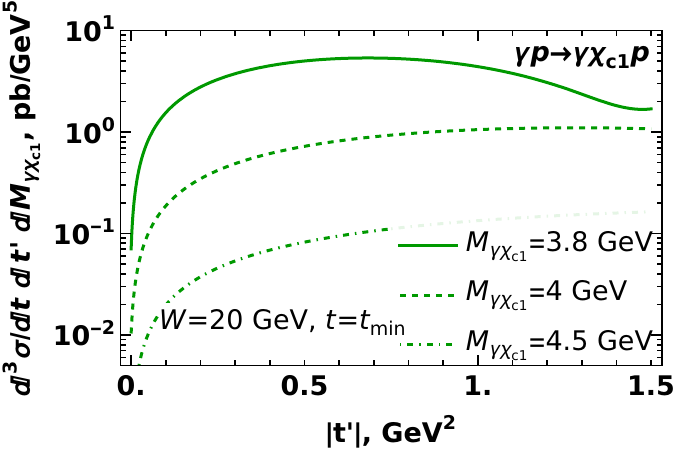}\includegraphics[width=9cm]{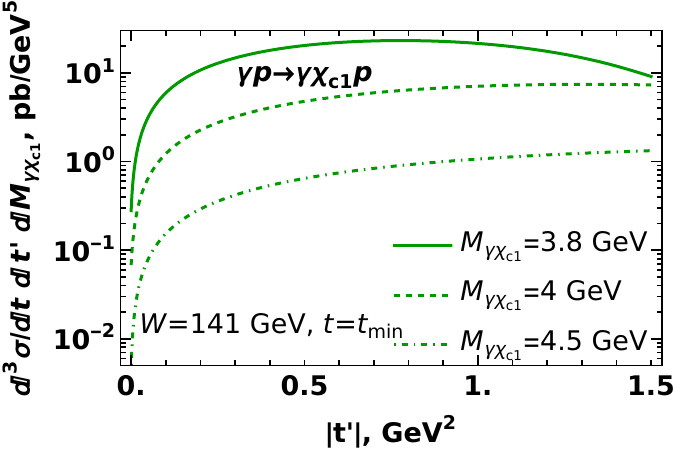}

\includegraphics[width=9cm]{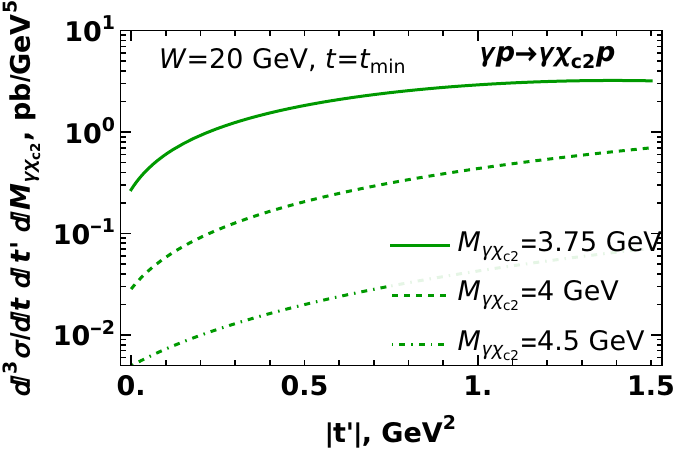}\includegraphics[width=9cm]{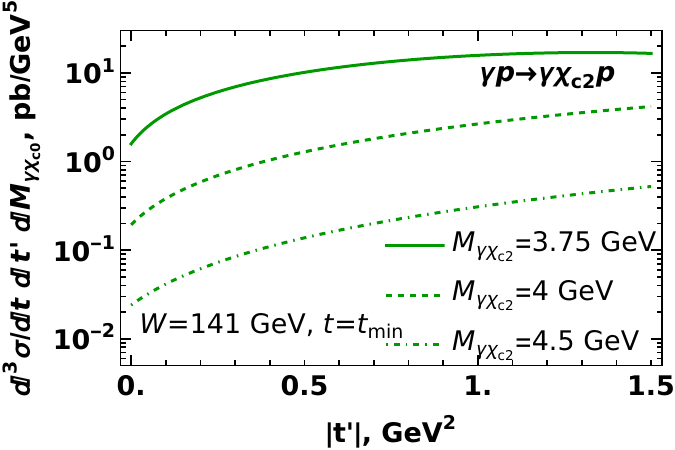}

\caption{\protect\label{fig:tPrimeDep}The unpolarized photoproduction cross-section~(\ref{eq:Photo})
as a function of the the variable $|t'|$ defined in~~(\ref{eq:Photo}).
The upper, central and lower rows correspond to $\chi_{c_{0}},\,\chi_{c_{1}}$
and $\chi_{c_{2}}$mesons; the left and right columns correspond to
different values of the collision energy $W$. See the next Figure~\ref{fig:tPrimeDep-1}
for $t'$-dependence of the individual helicity components in $\chi_{c1},\chi_{c2}$.}
\end{figure}

\begin{figure}
\includegraphics[width=9cm]{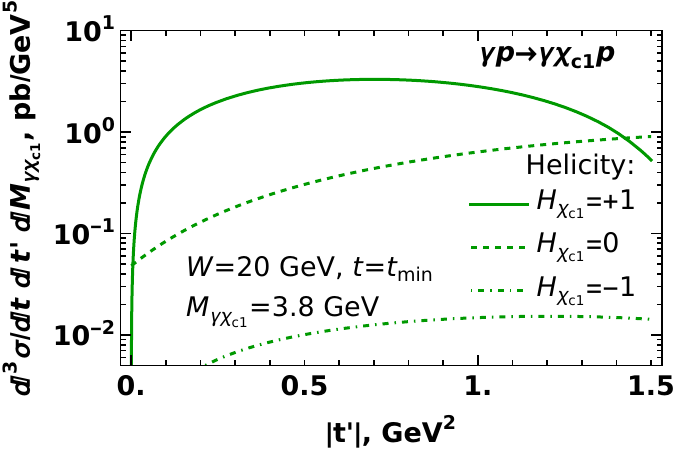}\includegraphics[width=9cm]{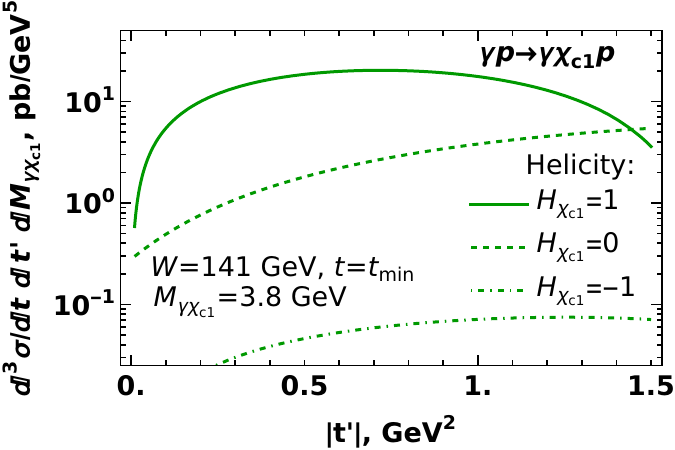}

\includegraphics[width=9cm]{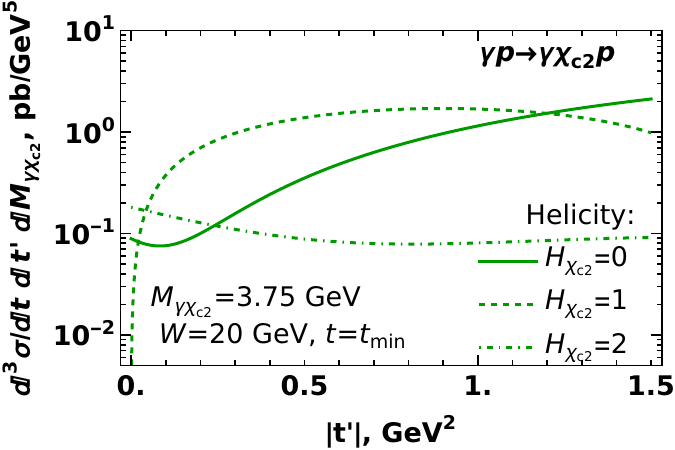}\includegraphics[width=9cm]{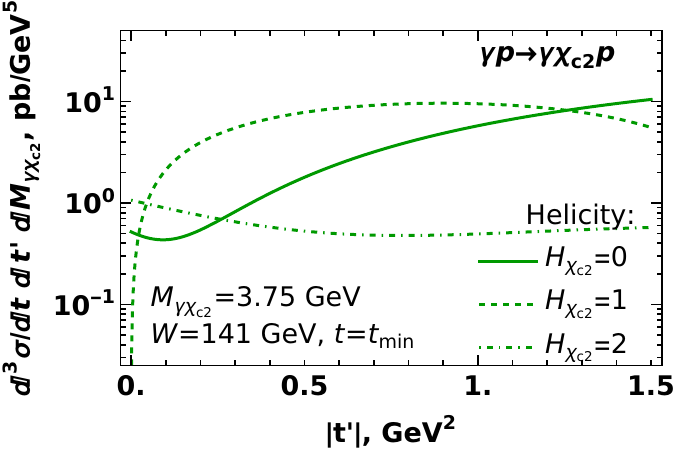}

\caption{\protect\label{fig:tPrimeDep-1}The photoproduction cross-sections
of $\chi_{c1}$ and $\chi_{c2}$ with different helicity projections
$H_{\chi_{c}}$. We assume that the incoming photon has helicity +1
(for helicity -1 should invert the sign of all helicities). For $\chi_{c2}$,
the cross-sections of states with helicities $H_{\chi_{c2}}=-1$ and
$H_{\chi_{c2}}=-2$ are negligible, and were omitted for the sake
of legibility. The left and right columns correspond to different
collision energy $W$. See the text for discussion of the observed
$t'$-dependence.}
\end{figure}

In the Figure~\ref{fig:tPrimeDep} we show the dependence of the
cross-section~(\ref{eq:Photo}) on the variable $\left|t'\right|$
for unpolarized case, and in the Figure~\ref{fig:tPrimeDep-1} we
also show the $t'$ dependence for the cross-sections with definite
(fixed) helicity $H_{\chi_{c}}$ at fixed helicity $+1$ of the incoming
photon. The observed behavior could be partially understood from conservation
of angular momentum. In general $\chi_{cJ}$ meson and the photon
move with nonzero (and parametrically large) transverse momenta $\sim\left|\boldsymbol{p}_{\perp}\right|$.
However, we may observe from (\ref{eq:KinApprox},~$\ref{eq:KinApprox-2}$)
that $\left|\boldsymbol{p}_{\perp}\right|$ may vanish at points $\alpha_{\chi_{c}}\approx\alpha_{{\rm min}}=1/r^{2}$,
and $\alpha_{\chi_{c}}\approx\alpha_{{\rm max}}=1$ which correspond
to $t'\approx0$ and $t'\approx\left|t'\right|_{{\rm max}}=M_{\gamma\chi_{c}}^{2}-M_{\chi_{c}}^{2}$,
respectively. At the point $\alpha_{\chi_{c}}\approx\alpha_{{\rm min}}$
the final-state $\chi_{c}$ and $\gamma$ move along the collision
axis in nearly the same direction, and for this reason a conservation
of total angular momentum (its projection on collision axis) requires
a conservation of a sum of helicities. Similarly, at the point $\alpha_{\chi_{c}}\approx\alpha_{{\rm max}}$
the $\chi_{c}$ and $\gamma$ move along the collision axis in opposite
directions, so the conservation of angular momentum leads to conservation
of difference of helicities. This consideration allows to understand
the observed suppression of the dominant components with $H_{\chi_{c}}=+1$
at small-$t'$, and suppression at $|t'|\gtrsim1.5\,{\rm GeV^{2}}$
for the curve $M_{\gamma\chi_{c}}=3.8\,{\rm GeV}$. Such suppression
also occurs for larger values of $M_{\gamma\chi_{c}}$, though starts
later and is not seen in the plot. Technically, this suppression happens
due to explicit factors $\sim\left|\boldsymbol{p}_{\perp}\right|$
and $\left(1-\alpha_{\chi_{c}}r^{2}\right)$ in the coefficient function.

\begin{figure}
\includegraphics[width=9cm]{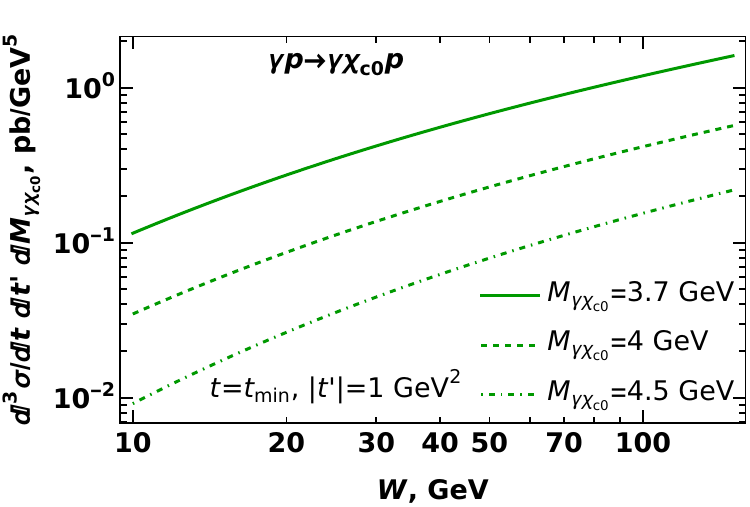}\includegraphics[width=9cm]{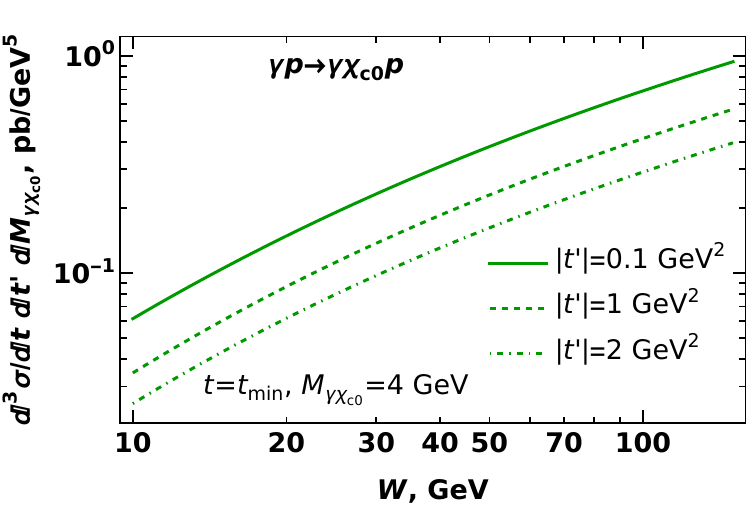}

\includegraphics[width=9cm]{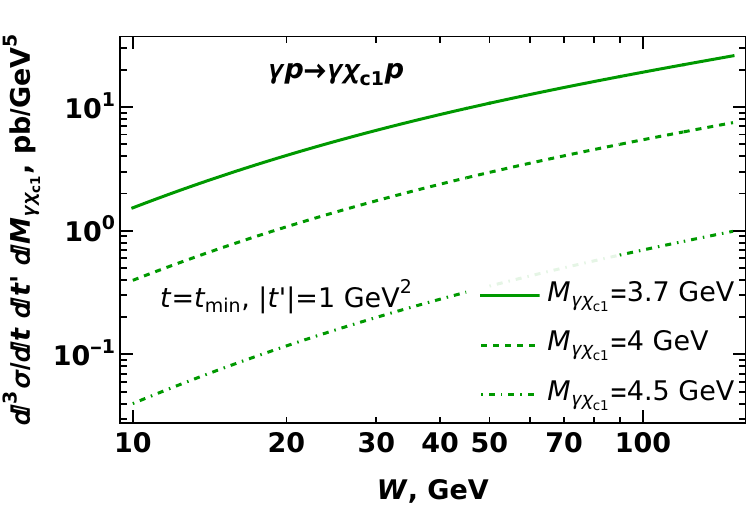}\includegraphics[width=9cm]{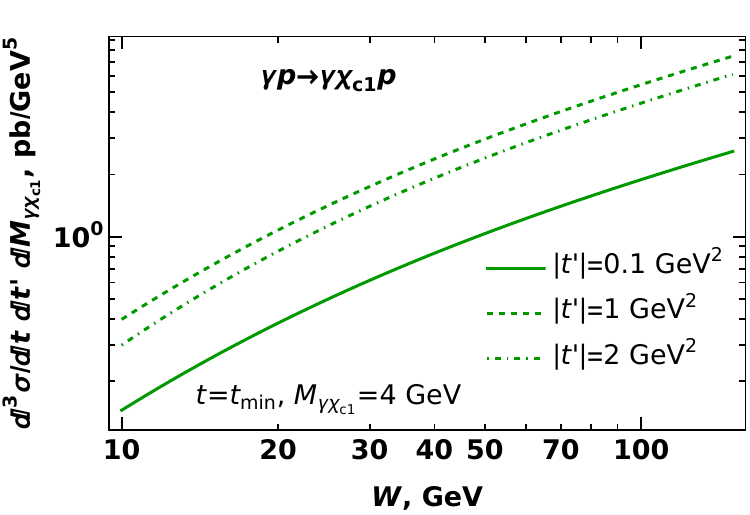}

\includegraphics[width=9cm]{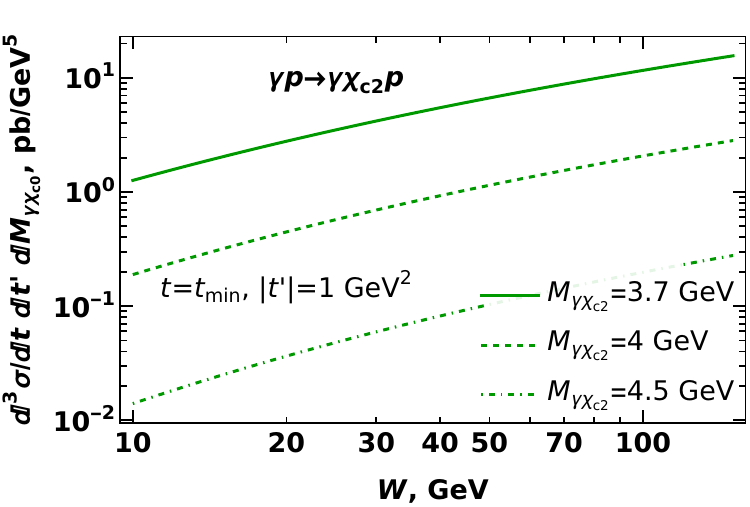}\includegraphics[width=9cm]{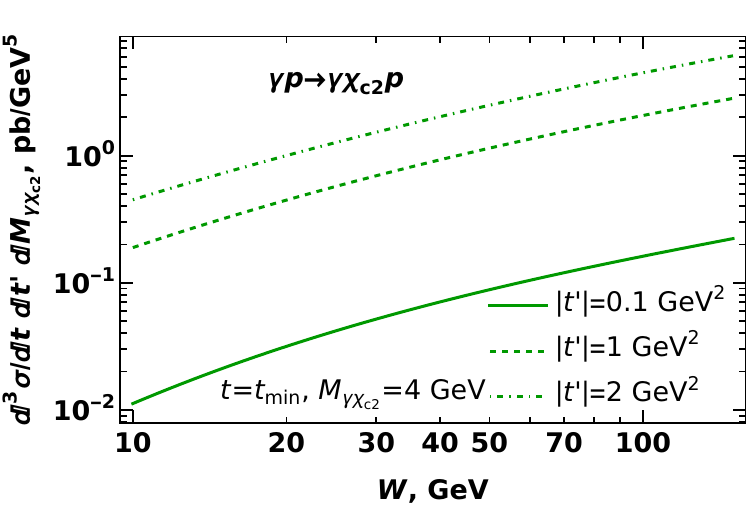}\caption{\protect\label{fig:WDep} The cross-section of $\chi_{cJ}$ production
as a function of the invariant collision energy $W$ at different
values of the invariant mass $M_{\gamma\chi_{c}}$ (left row) and
variable $|t'|$ (right row). The almost linear dependence (in double
logarithmic coordinates) with approximately the same slope indicates
that the dependence of the cross-section on $W$ largely factorizes
and can be approximated as $d\sigma(W)/dt\,dt'\,dM_{\gamma\chi_{c}}\sim W^{\lambda}$,
where the slope parameter $\lambda\approx0.8-1.1$ depends mildly
on other variables.}
\end{figure}

In the Figure~\ref{fig:WDep} we have shown the dependence of the
cross-section on the invariant energy $W$. The nearly linear growth
in double logarithmic coordinates suggests that the cross-section
has a power law dependence on energy, $d\sigma(W)\sim W^{\lambda}$,
where the parameter $\lambda$ has a mild dependence on other kinematic
variables. This behavior may be understood if we take into account
that the amplitude gets the dominant contribution from the gluon GPD
near $|x|\sim\xi\sim1/W^{2}$, and the fact that the gluon GPDs grows
as a function of the light-cone fraction $x$. In the implemented
phenomenological parametrization, the small-$x$ behavior of the GPD
roughly may be approximated as $H_{g}(x,\xi,t)\sim x^{-\delta_{g}}$~\cite{Goloskokov:2006hr}.
After convolution with coefficient functions~(\ref{eq:Monome}-\ref{eq:Monome-1}),
this translates into a power law behavior for the cross-section, 
\begin{equation}
\frac{d\sigma(W)}{dt\,dt'\,dM_{\gamma\chi_{c}}}\sim\xi^{-2\delta_{g}}\sim W^{4\delta_{g}},\qquad\lambda=4\delta_{g},\label{eq:diffW}
\end{equation}
in agreement with results of numerical evaluation. The typical values
of the parameter $\lambda$ are $\lambda\approx0.8-1.1$.

\begin{figure}
\includegraphics[width=6cm]{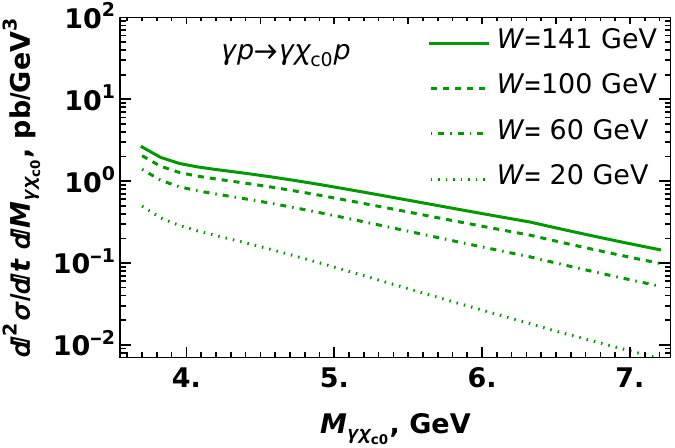}\includegraphics[width=6cm]{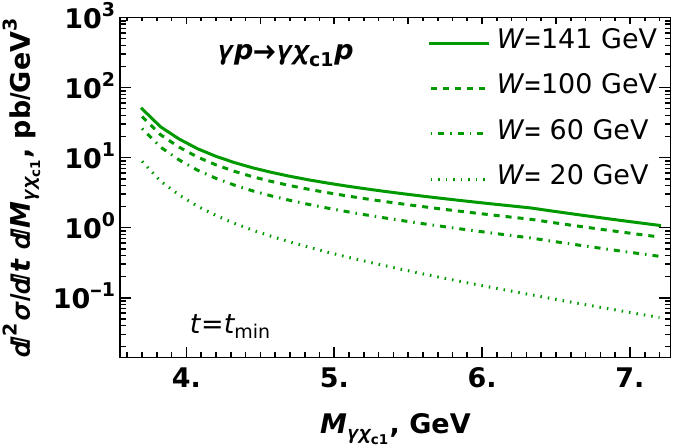}\includegraphics[width=6cm]{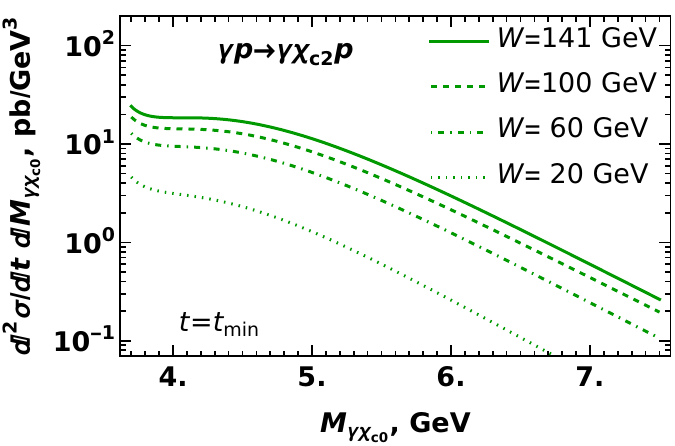}

\caption{\protect\label{fig:M12Dep} Dependence of the cross-section on the
invariant masses of produced $\gamma\chi_{c}$ pair, for several proton
energies $W$ .}
\end{figure}

\begin{figure}
\includegraphics[width=9cm]{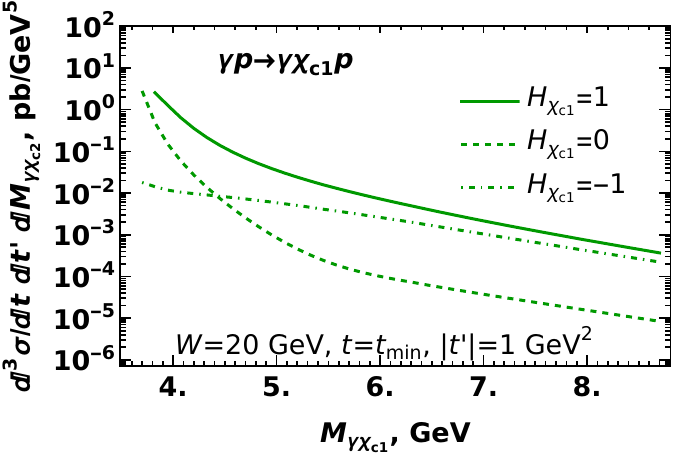}\includegraphics[width=9cm]{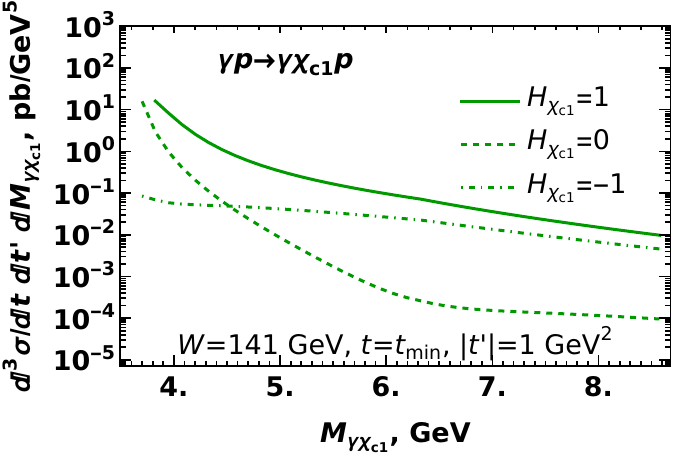}

\includegraphics[width=9cm]{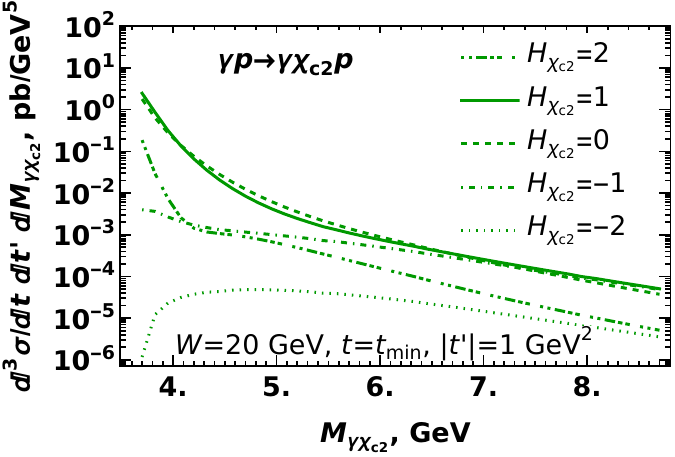}\includegraphics[width=9cm]{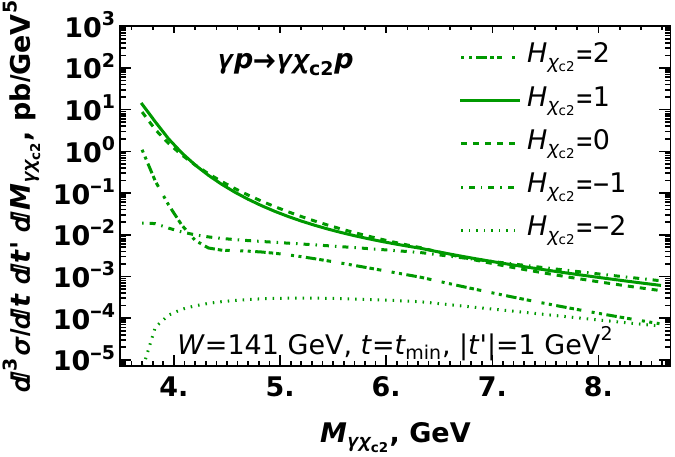}

\caption{\protect\label{fig:M12Dep-1} The dependence of different helicity
components in the cross-sections of $\chi_{c1}$ and $\chi_{c2}$
production on the invariant masses of produced $\gamma\chi_{c}$ pair.
Left and right columns differ by value of the collision energy $W$.
The sign of helicity of the hadron is positive, if it coincides with
helicity of the incoming photon.}
\end{figure}

In the Figure~\ref{fig:M12Dep} we show the dependence of the cross-section
on the invariant mass $M_{\gamma\chi_{c}}$. This dependence largely
stems from the prefactor $\sim1/\left(r^{2}-1\right)=M_{\chi_{c}}^{2}/\left(M_{\gamma\chi_{c}}^{2}-M_{\chi_{c}}^{2}\right)$
in the coefficient functions and additional factor $\sim1/M_{\gamma\chi_{c}}$
in the cross-section~(\ref{eq:Photo}), which together lead to a
strong suppression of the cross-section. On the other hand, the dependence
on $M_{\gamma\chi_{c}}$ in the kinematical constraint~(\ref{eq:alphaConstraint})
implies that the phase volume increases as a function of $M_{\gamma\chi_{c}}$
and partially attenuates the pronounced suppression of the cross-section
discussed earlier, however it is not sufficient to change the observed
decrease of the cross-section. The observed strong suppression of
the cross-section as a function of the invariant mass agrees with
findings of~~\cite{GPD2x3:9,GPD2x3:7} for other photon-meson channels.
In the Figure~\ref{fig:M12Dep-1} we have shown the contributions
of different helicity components for $\chi_{c1}$ and $\chi_{c2}$mesons.
The increase of $M_{\gamma\chi_{c}}$ at fixed values of other kinematic
variables decreases the variable $\alpha_{\chi_{c}}=\left(M_{\chi_{c}}^{2}+|t'|\right)/M_{\gamma\chi_{c}}^{2}$
and increases the transverse momentum $\left|\boldsymbol{p}_{\perp}\right|$,
as could be seen from~(\ref{eq:KinApprox}). At very small values
of $|\boldsymbol{p}_{\perp}|$, the final-sate meson and photon move
almost in the same direction as incoming photon. A production of $\chi_{c}$
with negative helicity in this kinematics is suppressed, since it
would require to change the helicities of both quarks, leading to
a strong suppression. At larger masses $M_{\gamma\chi_{c}}$, the
angle between the oncoming photon and the produced quarkonia increases
and may achieve large values (up to $\pi/2$), to the direction of
the incoming photon, for this reason the cross-sections with opposite
helicities start approaching each other. 

\begin{figure}
\includegraphics[width=9cm]{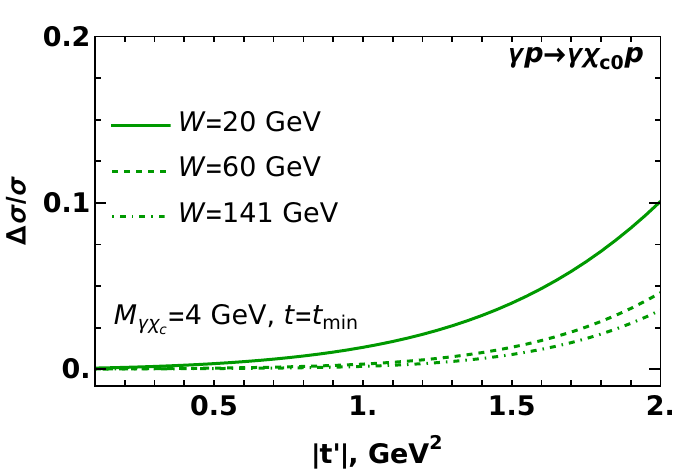}\includegraphics[width=9cm]{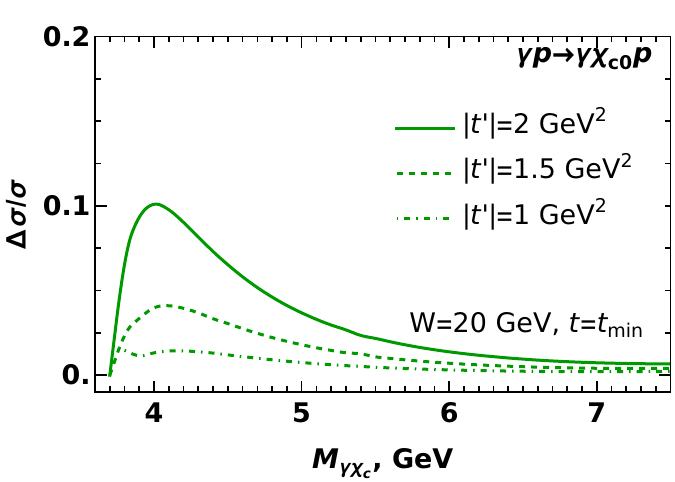}

\includegraphics[width=9cm]{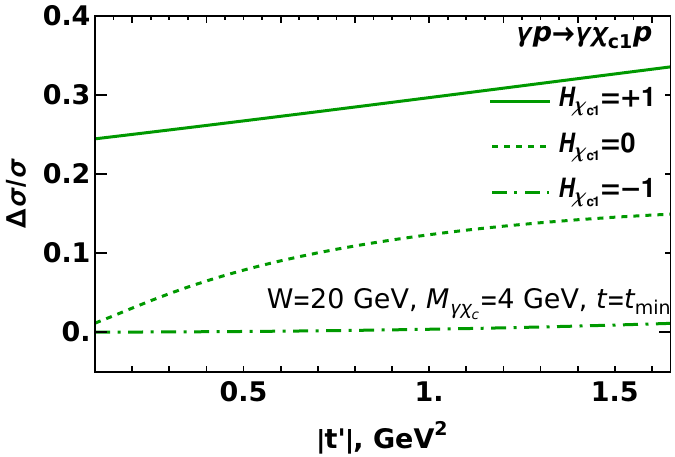}\includegraphics[width=9cm]{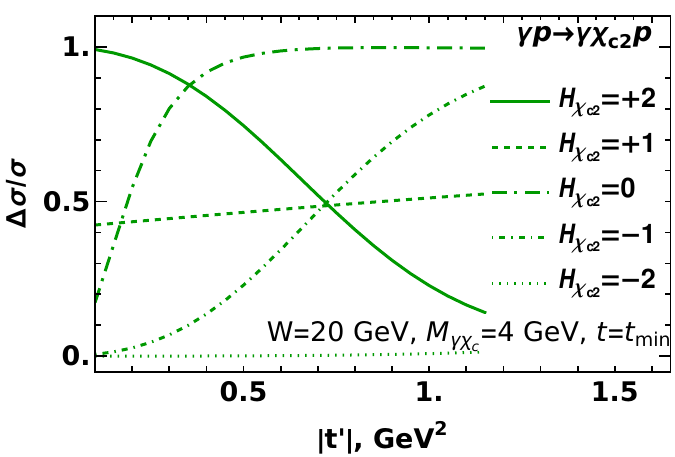}

\caption{\protect\label{fig:Polarized}The fraction of the contribution with
photon helicity flip $\Delta\sigma$ in the total cross-section. The
upper row shows this fraction for $\chi_{c0}$. In the left plot we
show dependence on the variable $t'$, which controls the angles between
the final state $\chi_{c},\gamma$ and the direction of the incoming
photon at fixed energies. Similarly, the helicity flip components
are suppressed at higher energies for $\chi_{c0}$ due to decreasing
scattering (production) angles; similar dependence was also observed
for $\chi_{c1},\chi_{c2}$. The plots in the bottom row show the dependence
of the above-mentioned fraction on $t'$ for $\chi_{c1},\chi_{c2}$
(in different harmonics). The harmonics with helicities $H_{\chi_{c}}<0$
are numerically negligible. The ratio remains small (below 10\%) for
$\chi_{c0}$, in the whole range analyzed in this paper, yet may be
sizable for $\chi_{c1},\chi_{c2}$ production with certain helicity
projections}
\end{figure}

Finally, we would like to discuss the relative contribution of different
polarizations of the final state photon to the total cross-section.
In the Figure~\ref{fig:Polarized} we plotted the fraction of the
contribution with photon helicity flip in the total cross-section
\begin{equation}
\frac{\Delta\sigma}{\sigma}\equiv\frac{d\sigma_{\gamma p\to\chi_{c}\gamma p}^{(+,\,-,H)}}{d\sigma_{\gamma p\to\chi_{c}\gamma p}^{(+,\,-,H)}+d\sigma_{\gamma p\to\chi_{c}\gamma p}^{(+,\,+,H)}}=\frac{\left|\mathcal{A}_{\gamma p\to\chi_{c}\gamma p}^{(+,\,-,H)}\right|^{2}}{\left|\mathcal{A}_{\gamma p\to\chi_{c}\gamma p}^{(+,\,-,H)}\right|^{2}+\left|\mathcal{A}_{\gamma p\to\chi_{c}\gamma p}^{(+,\,+,H)}\right|^{2}}.\label{eq:fraction}
\end{equation}
As we explained earlier, in general the produced $\chi_{c}$ meson
and photon have a relatively large transverse momentum $\pm\boldsymbol{p}_{\perp}$,
as could be seen from~(\ref{eq:q}-\ref{eq:k}), and thus move with
nonzero angle w.r.t. direction of the incoming photon. However, for
special points where $\boldsymbol{p}_{\perp}=0$, the final state
particle move collinear (or anticollinear) w.r.t. incoming photon,
and for this reason the conservation of angular momentum requires
suppression of certain modes in which a sum (or a difference) of helicities
does not conserve. Since the angle which forms the particle with collision
axis decreases with energy $W$, we expect that at very high energies
$W$ or small $\boldsymbol{p}_{\perp}$ (small $t'$) the contributions
which break the above-mentioned selection rule may be suppressed.
In agreement with our expectations, the ratio~(\ref{eq:fraction})
decreases as a function of energy $W$ for all $\chi_{c}$ mesons.
This selection rule also allows to understand the observed $t'$-dependence
of different harmonics in $\chi_{c2}$ meson: the helicity conservation
requires that state with helicity $H_{\chi_{c2}}=+2$ at small $t'$
(small angles) should proceed only via the photon helicity flip, whereas
for $H_{\chi_{c2}}=0$ the helicity conservation requires that the
photon's helicity is conserved. However, as we increase $|t'|$, the
contributions of the ``forbidden'' (suppressed) modes become more
and more important. For the mode with $H_{\chi_{c1}}=+1$, both helicity
flip and non-flip contributions are suppressed at small $t'$, however
this suppression is exactly the same for both modes, and the ratio~(\ref{eq:fraction})
for this mode has a relatively mild dependence on $t'$, with comparable
(within factor of 2) contributions of helicity flip and non-flip parts.
The harmonics with helicities $H_{\chi_{c}}<0$ are numerically negligible
and thus do not present much interest.

\begin{figure}
\includegraphics[width=8.5cm]{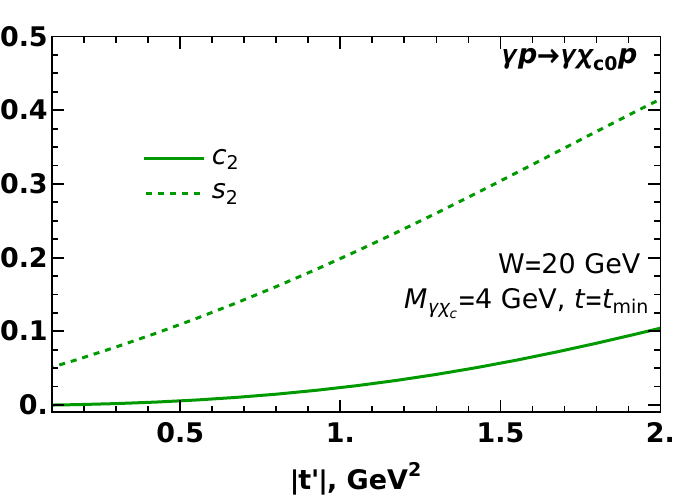}\includegraphics[width=8.5cm]{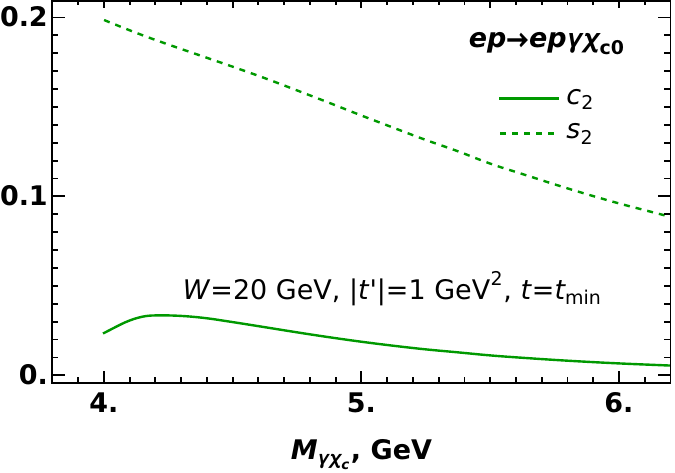}

\includegraphics[width=8.5cm]{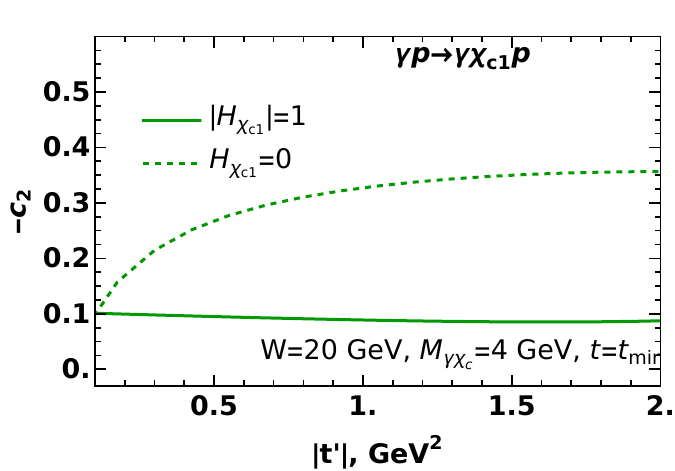}\includegraphics[width=8.5cm]{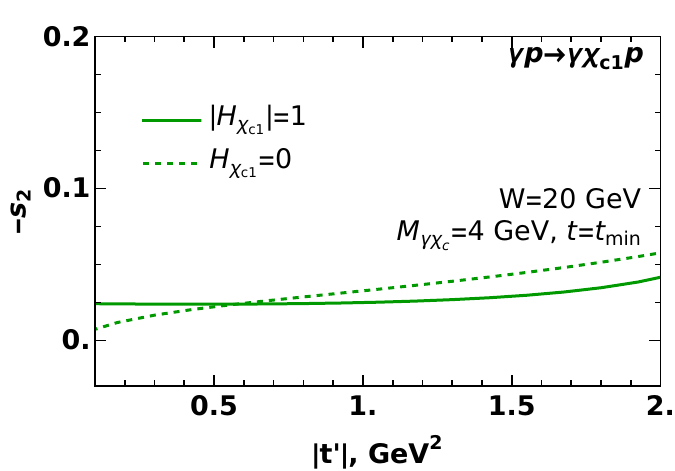}

\includegraphics[width=8.5cm]{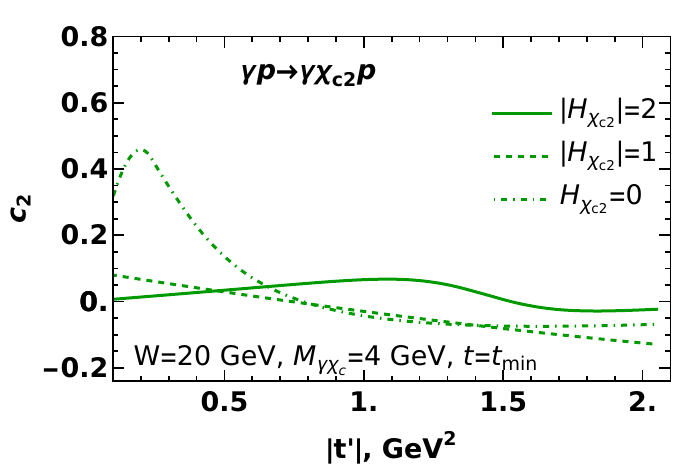}\includegraphics[width=8.5cm]{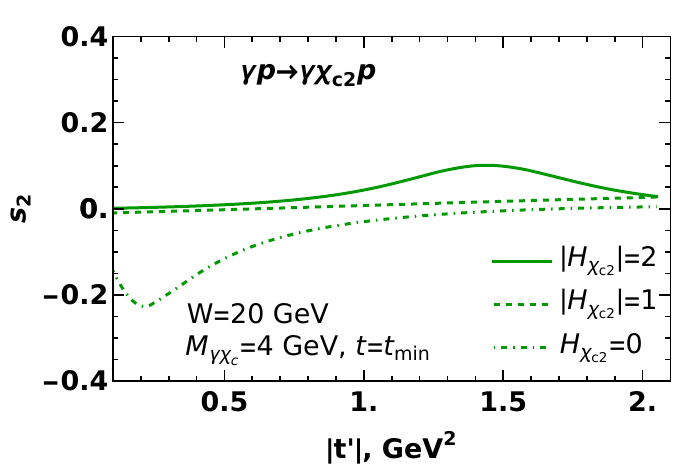}

\caption{\protect\label{fig:Polarized-1}The angular harmonics $c_{2},\,s_{2}$
defined in~(\ref{eq:sigma_def}-\ref{eq:c2s2}). In this Figure we
assume that the detector sums different polarizations of the final-state
photons. For $\chi_{c1},\chi_{c2}$ in view of $P$-parity the cross-sections
with different signs of helicity $H_{\chi_{C}}$ are equal. Top, central
and bottom rows correspond to $\chi_{c0},\chi_{c1}$ and $\chi_{c2}$
respectively. For the sake of definiteness we assumed that the parameter
$\epsilon$ defined in~(\ref{eq:RatioFluxes}) equals $\epsilon=0.5$.}
\end{figure}

\begin{figure}
\includegraphics[width=9cm]{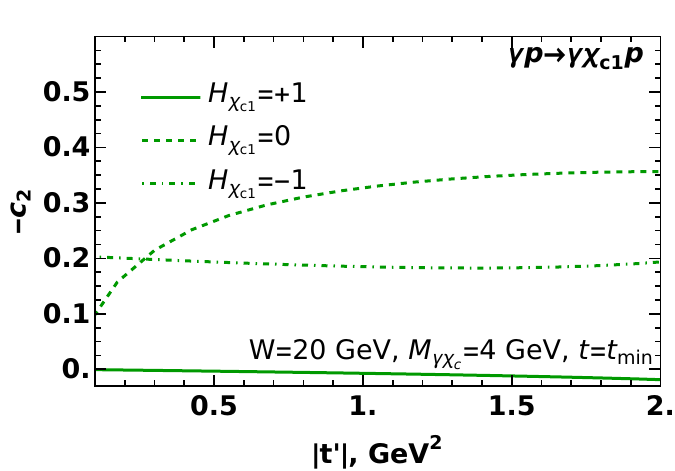}\includegraphics[width=9cm]{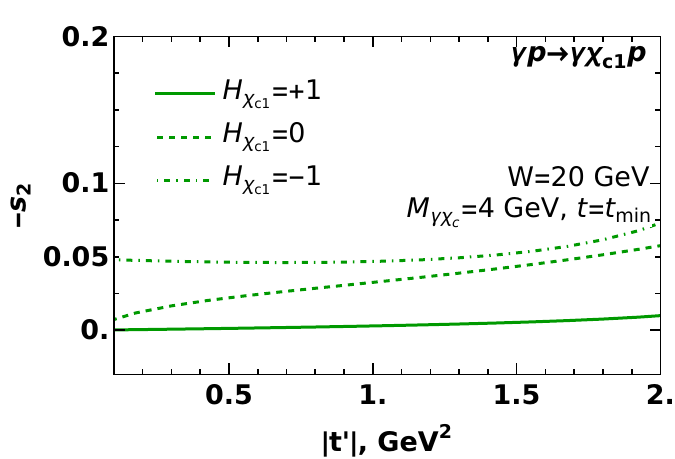}

\includegraphics[width=9cm]{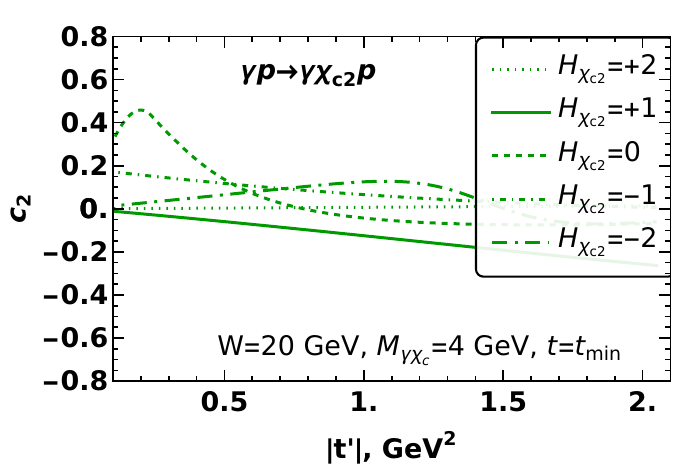}\includegraphics[width=9cm]{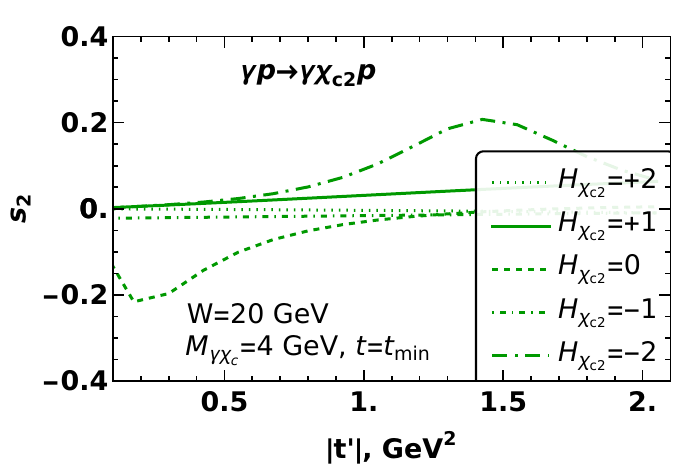}

\caption{\protect\label{fig:Polarized-2}Sensitivity of the angular harmonics
$c_{2},s_{2}$ defined in~(\ref{eq:sigma_def}-\ref{eq:c2s2}) to
the polarization of the final-state photon (for the sake of definiteness
we consider polarization \textquotedblleft +1\textquotedblright ).
The notation $H_{\chi_{c}}$ is the helicity of the produced $\chi_{c}$
meson. Top and bottom rows correspond to $\chi_{c1}$ and $\chi_{c2}$
respectively (for spinless $\chi_{c0}$, the cross-section is not
sensitive to the sign of the final-state helicity). For the sake of
definiteness we assumed that the parameter $\epsilon$ defined in~(\ref{eq:RatioFluxes})
equals $\epsilon=0.5$.}
\end{figure}

As we discussed earlier, in \textit{electro}production experiments
it is impossible to fix the sign of the virtual (incoming) photon,
for this reason the ratio~(\ref{eq:fraction}) reveals itself via
angular asymmetries~$c_{2},\,s_{2}$ defined in~(\ref{eq:c2s2}).
In the Figure~\ref{fig:Polarized-1} we have shown these harmonics
for the case when the detector does not distinguish ($\approx$sums
over) the polarizations of the final state photon. In this setup,
the cross-sections of $\chi_{c}$ production do not distinguish the
sign of helicity $H_{\chi_{c}}$. In the Figure~\ref{fig:Polarized-2}
we have shown the results for $\chi_{c1}$ and $\chi_{c2}$ when the
detector picks up only events with definite polarization of the final
state photon (+1 for definiteness). As we can infer from~~(\ref{eq:c2s2}),
a sizable asymmetry is possible only if the amplitudes $\left|\mathcal{A}_{++,\,H}\right|$
and $\left|\mathcal{A}_{-+,\,H}\right|=\left|\mathcal{A}_{+-,\,-H}\right|$
are comparable by magnitude. Apparently this condition is satisfied
only for $\chi_{c2}$ meson with $H_{\chi_{c2}}=0$ in the kinematics
of small $t'$.

\subsection{Integrated cross-sections \& counting rates}

\label{subsec:integr}
\begin{figure}
\includegraphics[width=6cm]{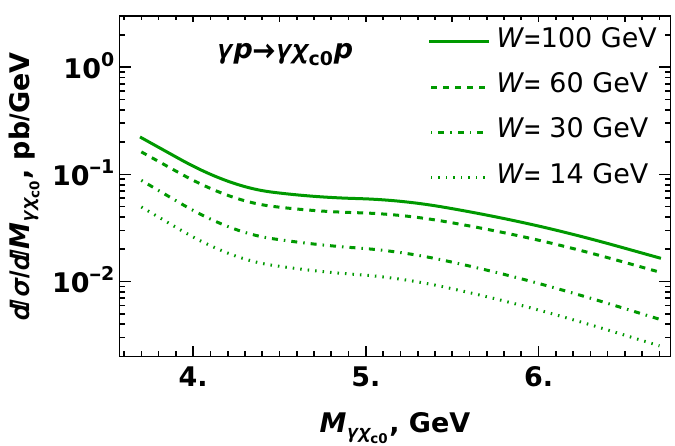}\includegraphics[width=6cm]{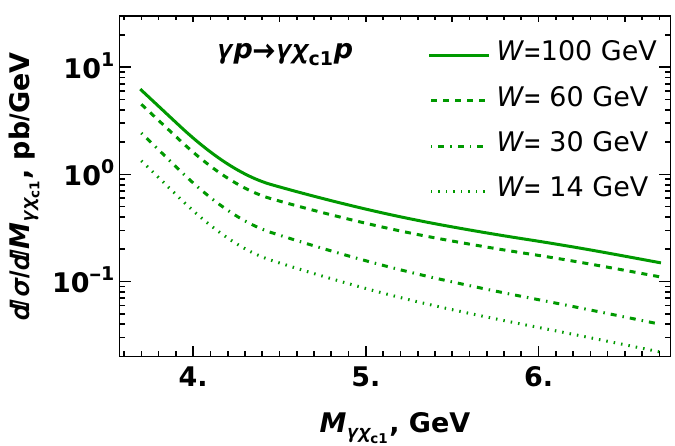}\includegraphics[width=6cm]{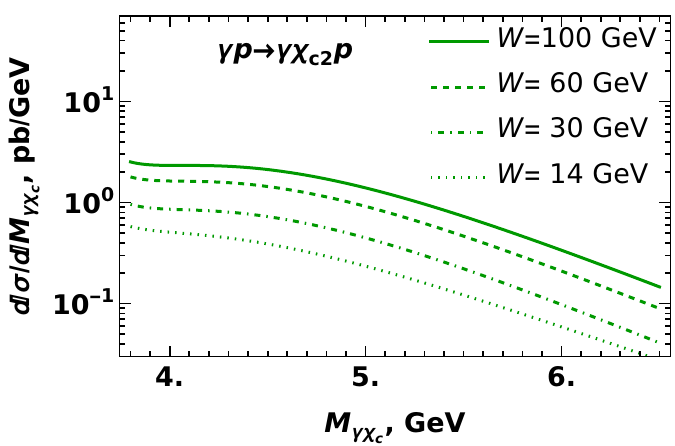}

\caption{\protect\label{fig:M12}The single-differential cross-section $d\sigma/dM_{\gamma\chi_{c}}$
as a function of the invariant mass of $\chi_{c}\gamma$ pair at different
collision energies.}
\end{figure}

So far we considered the differential cross-sections,
which are best suited for theoretical studies. Unfortunately, it is
difficult to measure such small cross-sections because of insufficient
statistics, and for this reason now we will provide predictions for
the yields integrated over some or all kinematic variables. In the
left panel of the Figure~\ref{fig:M12} we have shown the cross-section
$d\sigma/dM_{\gamma\chi_{c}}$ for different energies. The observed
dependence on $M_{\gamma\chi_{c}}$ largely repeats the $M_{\gamma\chi_{c}}$-dependence
of the differential cross-section shown in the Figure~\ref{fig:M12}.
This is a consequence of our earlier observations that dependence
on different variables in the differential cross-section largely factorize,
i.e. the \textit{shape} of dependence on any of the variables $t,t',M_{\gamma\chi_{c}}$
depends very weakly on the other variables. 

\begin{figure}
\includegraphics[width=6cm]{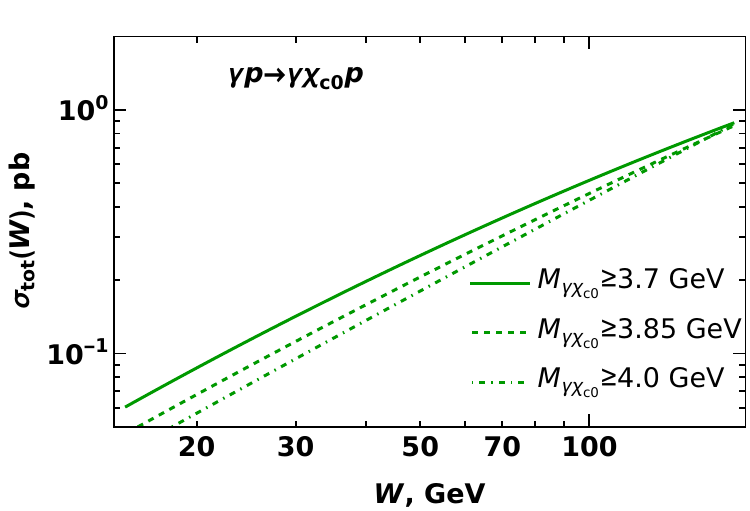}\includegraphics[width=6cm]{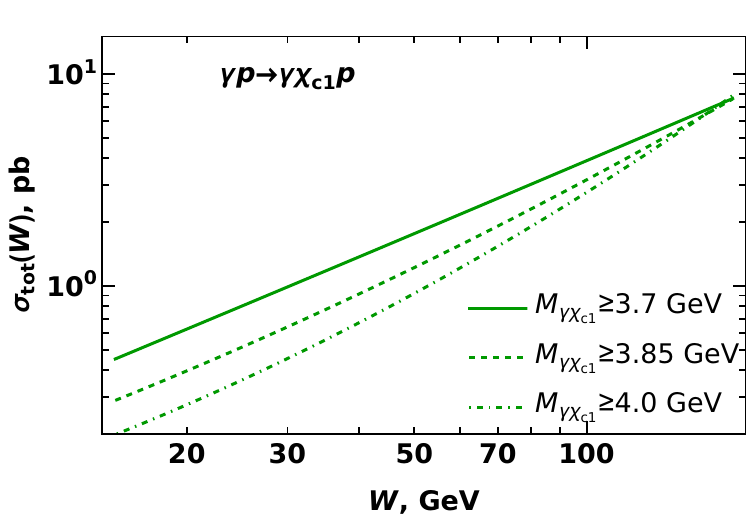}\includegraphics[width=6cm]{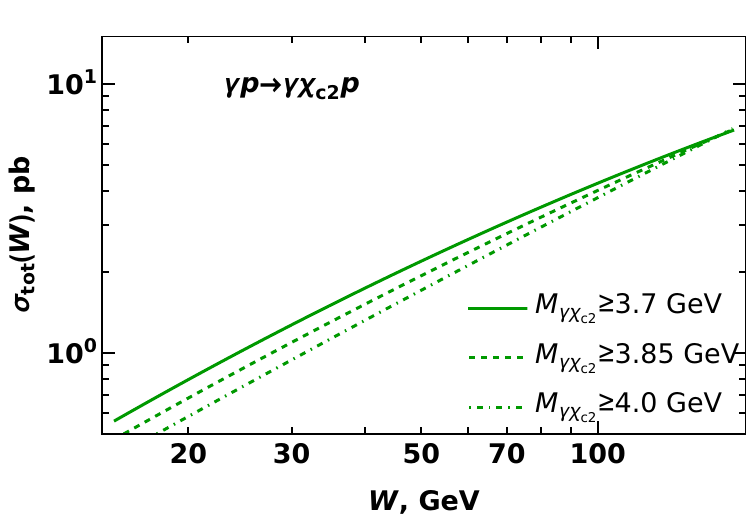}

\includegraphics[width=6cm]{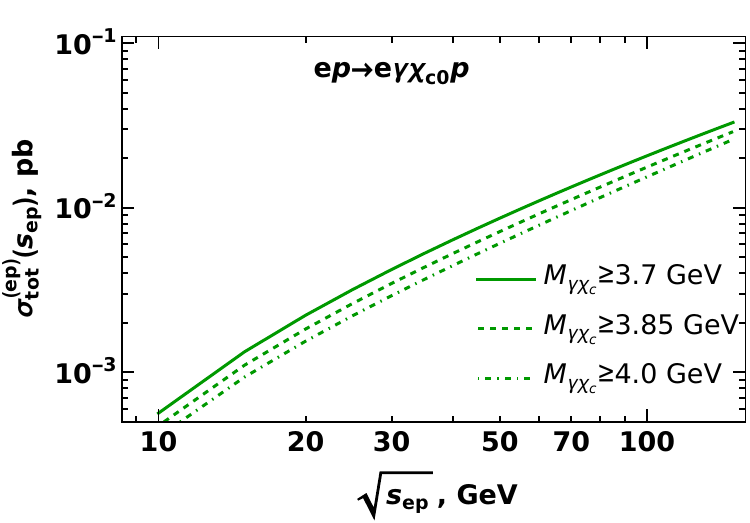}\includegraphics[width=6cm]{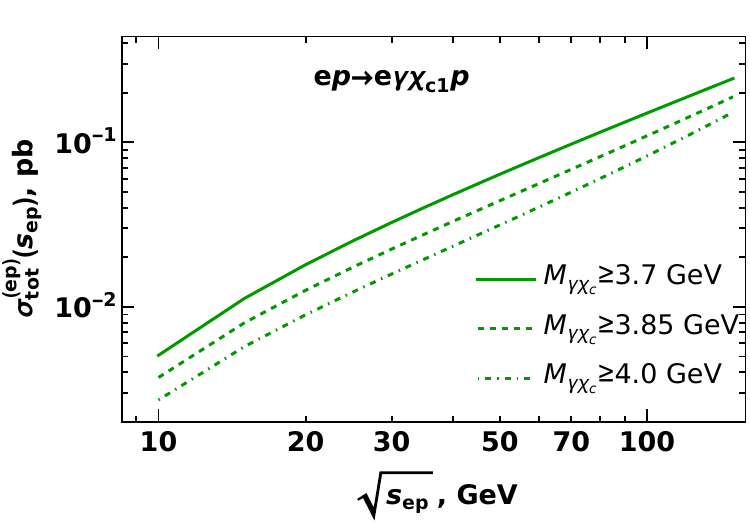}\includegraphics[width=6cm]{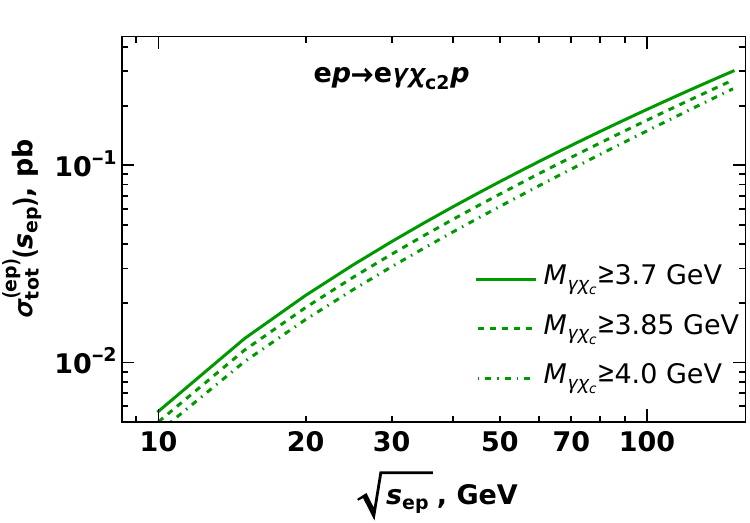}

\caption{\protect\label{fig:Total}Predictions for the total (\textquotedblleft fiducial\textquotedblright )
cross-section of photoproduction (upper row) and electroproduction
(lower row) as a function of energy $W$ for different cutoffs on
the invariant mass of $\chi_{c}\gamma$ pairs.}
\end{figure}

In the left panel of the Figure~\ref{fig:Total} we show the energy
dependence of the total (fully integrated) cross-section $\sigma_{{\rm tot}}(W)$
for the photoproduction of $\chi_{c}\gamma$. As we discussed earlier,
the suggested approach cannot be applied when the emitted photon is
soft, for this reason we consider only the kinematics where the invariant
mass $M_{\gamma\chi_{c}}$ of the photon-meson pair is sufficiently
large. The choice of the minimal value $\left(M_{\gamma\chi_{c}}\right)_{{\rm min}}$
is somewhat arbitrary, for this reason we have shown the result for
several possible cutoffs. This dependence is stronger at small energies
due to phase space contraction for production of $\chi_{c}\gamma$
pair with large invariant mass. At large $W$, for all cutoffs the
$W$-dependence may be approximated as 
\begin{align}
\sigma_{{\rm tot}}\left(W,\,M_{\gamma\chi_{c0}}\ge3.7\,{\rm GeV}\right) & \approx0.48\,{\rm pb}\,\left(\frac{W}{100\,{\rm GeV}}\right)^{0.97},\\
\sigma_{{\rm tot}}\left(W,\,M_{\gamma\chi_{c1}}\ge3.7\,{\rm GeV}\right) & \approx3.7\,{\rm pb}\,\left(\frac{W}{100\,{\rm GeV}}\right)^{1.14},\\
\sigma_{{\rm tot}}\left(W,\,M_{\gamma\chi_{c2}}\ge3.7\,{\rm GeV}\right) & \approx4.2\,{\rm pb}\,\left(\frac{W}{100\,{\rm GeV}}\right)^{0.9},
\end{align}
in agreement with our earlier findings~(\ref{eq:diffW}) for the
energy dependence of the differential cross-sections. In the right
panel of the same Figure~\ref{fig:Total} we have shown our estimates
for the fully integrated cross-section~\footnote{The evaluation of the total electroproduction cross-section requires to integrate over all possible virtualities $Q$ of the incoming photons. In our evaluations we disregarded this $Q$-dependence for simplicity, since the flux of equivalent photons falls off rapidly as a function of $Q$, and the $Q$-dependence of the  \textit{photo}production part is controlled by the scale $M_{\chi_{c}}$ (namely is very mild for $Q<M_{\chi_{c}}$ and drops sharply for $Q>M_{\chi_{c}}$).  For our numerical estimate we assumed that the $Q$-dependence of the \textit{photo}production cross-section is given by $d\sigma(Q)=\Theta\left(M_{\chi_{c}}-Q\right)d\sigma(Q=0)$,
where $\Theta$ is a Heaviside step function.} of the electroproduction $ep\to e\gamma\chi_{c}p$ as a function
of the invariant electron-proton collision energy $\sqrt{s_{ep}}$.

\begin{table}
\begin{tabular}{|c|c|c|c|c|c|c|c|}
\hline 
 & $\sigma_{{\rm tot}}^{({\rm ep})}$ & $dN/dt$ & $N$ & ${\rm Br}\left(\chi_{c}(1P)\to J/\psi\gamma\right)$ & ${\rm Br}\left(J/\psi\to\mu^{+}\mu^{-}\right)$ & $dN_{d}/dt$ & $N_{d}$\tabularnewline
\hline 
$\chi_{c0}$ & $31\,{\rm fb}.$ & 27/day & 3.1$\times10^{3}$ & 1.4~\% &  & 0.65/month & 2.5\tabularnewline
\cline{1-5}\cline{7-8}
$\chi_{c1}$ & $230\,{\rm fb}$ & 199/day & 2.3$\times10^{4}$ & 34.3~\% & 5.9 \% & 120/month & 460\tabularnewline
\cline{1-5}\cline{7-8}
$\chi_{c2}$ & $250\,{\rm fb}$ & 216/day & 2.5$\times10^{4}$ & 19~\% &  & 73/month & 280\tabularnewline
\hline 
\end{tabular}

\caption{\protect\label{tab:Estimates}Estimates of the cross-section, production
and counting rates in the kinematics of the future EIC collider. Values
of the cross-section $\sigma_{{\rm tot}}^{({\rm ep})}$ are evaluated
at $\sqrt{s_{ep}}=141\,{\rm GeV}$ with cutoff on the invariant mass
$M_{\gamma\chi_{c0}}\ge3.7\,{\rm GeV}$. For estimates of the production
rate $dN/dt$ we used the instantaneous luminosity $\mathcal{L}=10^{34}\,{\rm cm^{-2}s^{-1}}=10^{-5}{\rm fb}^{-1}s^{-1}$
, and the total number of produced events $N$ was estimated using
integrated luminosity $\protect\int dt\,\mathcal{L}=100{\rm fb^{-1}}$~\cite{Accardi:2012qut,AbdulKhalek:2021gbh,Navas:2024X}.
The detection rate $N_{d}$ and the total number of events was obtained
multiplying these numbers by a product of branchings ${\rm Br}\left(\chi_{c}(1P)\to J/\psi\gamma\right){\rm Br}\left(J/\psi\to\mu^{+}\mu^{-}\right)$,
whose values were taken from~\cite{BESIII:2019eyx,Navas:2024X}.}
\end{table}

For the sake of reference in the Table~\ref{tab:Estimates} we provided
the estimates of the cross-section for the upper EIC energy $\sqrt{s_{ep}}\approx141\,$GeV,
together with corresponding production and detection rates. The small
values of the electroproduction cross-sections $\sigma_{{\rm tot}}^{({\rm ep})}$
are a consequence of $\sim\alpha_{{\rm em}}$ in the leptonic prefactor
(see~(\ref{eq:LTSep})) and a very rapid decrease of the $t$-dependence
in the exclusive process. For estimates of the detection rates we
assumed that the $\chi_{c}$ meson is reconstructed from its radiative
decays into $J/\psi$, and the latter is reconstructed from its decays
into $\mu^{+}\mu^{-}$ or $e^{+}e^{-}$ . For $\chi_{c0}$ mesons
due to small value of ${\rm Br}\left(\chi_{c0}(1P)\to J/\psi\gamma\right)$
we got very small values of the counting rates. However, $\chi_{c0}$
has a number of other 2-, 3- and 4-body decay channels with larger
branchings, e.g. ${\rm Br}(\chi_{c}(1P)\to\pi^{+}\pi^{-}\pi^{0}\pi^{0})=3.3\%$,
${\rm Br}(\chi_{c}(1P)\to\rho^{+}\pi^{-}\pi^{0})=2.9\%$, ${\rm Br}(\chi_{c}(1P)\to\pi^{+}\pi^{-}K^{+}K^{-})=1.8\%$,
${\rm Br}(\chi_{c}(1P)\to\rho^{+}K^{-}K^{0})=1.21\%$ . If these channels
will be used for reconstruction of $\chi_{c0}$, the expected counting
rates will be at least a factor of 20 larger than what is shown in
the Table~\ref{tab:Estimates}.

A significantly larger counting rates could be achieved in the ultraperpheral
kinematics at LHC, due to enhancement of the gluonic densities in
small-$x$ kinematics. However, the collinear factorization approach
is not reliable in that kinematics due to onset of saturation effects,
for this reason we do not make predictions for LHC energies.

\subsection{Comparison with $\chi_{c}$ photoproduction}

\label{subsec:odderon}The exclusive photoproduction of $C$-even
quarkonia, such as $\gamma p\to\eta_{c}p$ and $\gamma p\to\chi_{c}p$
are considered in the literature as possible tools for analysis of
the 3-gluon exchanges in $t$-channel (the so-called odderons)~\cite{Odd5,Odd6,Odd7}.
While the odderons have been predicted many years ago~\cite{Odd1,Odd2,Odd3},
for a long time the magnitude of the odderon-mediated processes remained
largely unknown because it is controlled by a completely new nonperturbative
amplitude (see~\cite{Odd7,Benic:2023} for a short overview). A
recent comparison of $pp$ and $p\bar{p}$ elastic cross-sections
measured at LHC and Tevatron~\cite{OddTotem1,OddTotem2} apparently
corroborates the existence of odderons, though includes sizable uncertainties
introduced by the extrapolation procedure, and does not allow to study
them in detail using perturbative methods due to lack of a hard scale.
For this reason, at present the searches of odderons mostly focus
on production channels which proceed via the $C$-odd exchanges in
$t$-channels . The photoproduction $\gamma p\to\chi_{c}p$ is a rather
clean channel for study of odderons which has been suggested in~\cite{Benic:2024}
as alternative to $\eta_{c}$ photoproduction. The cross-section of
the odderon-mediated $\chi_{c}$ photoproduction is on par with that
of $\eta_{c}$, yet it is preferable for experimental studies due
to smaller contamination by the photon-photon fusion (the so-called
Primakoff mechanism).

The $\chi_{c}\gamma$ photoproduction proposed in this manuscript
deserves a lot of interest as a potential background to $\chi_{c}$
photoproduction. Indeed, since the acceptance for photons at modern
detectors is below unity, there is a sizable probability that the
final photon is not observed, and so the $\chi_{c}\gamma$ photoproduction
could be misinterpreted as exclusive $\chi_{c}$ production. Formally
the cross-section of $\chi_{c}\gamma$ is suppressed by factor $\mathcal{O}\left(\alpha_{{\rm em}}\right)$,
however this suppression can be partially compensated by lack of the
small $C$-odd exhanges in $t$-channel, thus leading to a comparable
sizable contribution. In general the accurate estimate of this background
requires detailed knowledge of the detector's geometry and acceptance.
In what follows we will assume for simplicity that \emph{all} photons
remain undetected, and will integrate over the momentum of the produced
photon. Such approach gives an upper estimate for the $\chi_{c}\gamma$
background.

\begin{figure}
\includegraphics[width=18cm]{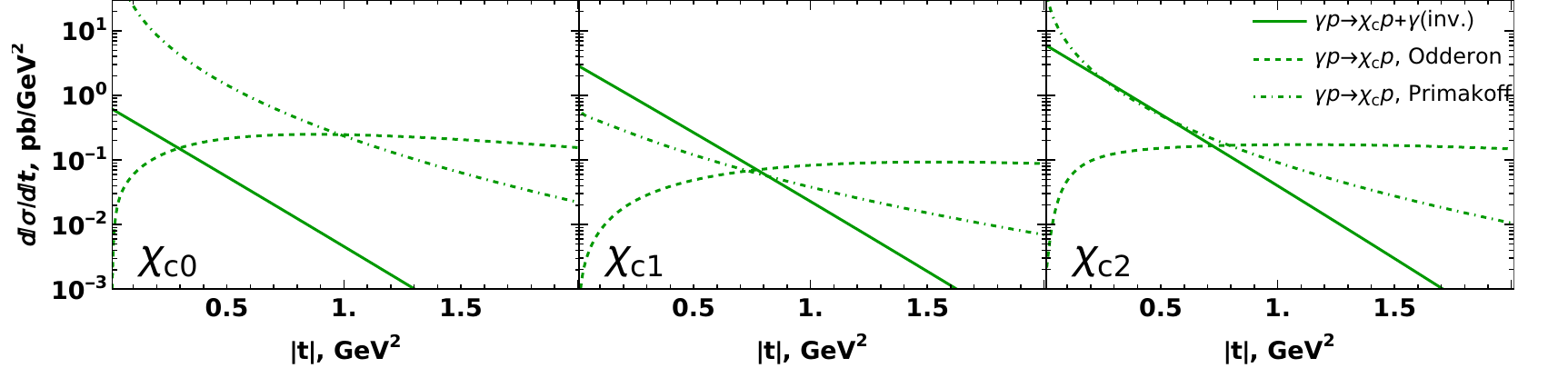}\caption{\protect\label{fig:Odd} The cross-section of the process $\gamma p\to\chi_{c}\gamma p$
with invisible photon is compared with the cross-sections of the $\gamma p\to\chi_{c}p$
production via odderon exchange and Primakoff mechanism. The cross-sections
of $\gamma p\to\chi_{c}p$ with different spins are taken from~\cite{Benic:2024}
(Fig 3 for $x=10^{-2}$ ); the cross-section of $\gamma p\to\chi_{c}\gamma p$
is evaluated at the same invariant energy $W=\sqrt{M_{\chi_{c}}^{2}/x}\approx35\,{\rm GeV}$.}
\end{figure}

In the Figure~(\ref{fig:Odd}) we have shown the cross-sections of
the $\chi_{c}$ and $\chi_{c}\gamma$ with unobserved photon in the
final state. For $\chi_{c1}$ and $\chi_{c2}$ the cross-section of
the process $\chi_{c}\gamma$ exceeds the cross-section of the $\chi_{c}$
photoproduction in the small-$t$ kinematics ($|t|\lesssim1\,{\rm GeV}^{2}$)
, though is exponentially suppressed at large $t$ due to implemented
GPD model~\cite{Goloskokov:2006hr,Goloskokov:2007nt,Goloskokov:2008ib,Goloskokov:2009ia,Goloskokov:2011rd}.
The fact that odderon contribution in the small-$t$ kinematics can
be smaller than contributions of other mechanisms is known from the
literature, and the Primakoff mechanism has been considered as a major
background to odderon-mediated production~\cite{Benic:2023,Benic:2024}.
Our findings indicate that for $\chi_{c1}$ and $\chi_{c2}$, the
production via $\gamma p\to\chi_{c}\gamma p$ mechanism gives a comparable
background. Furthermore, in contrast to Primakoff contribution, this
mechanism gives the same contribution for proton and neutron targets,
and thus cannot be eliminated using neutron-rich target.

\section{Conclusions}

\label{sec:Conclusions}In this paper we analyzed the exclusive $\chi_{c}\gamma$
photoproduction in the collinear factorization approach and calculated
separately contributions of different helicity components. We provide
the analytic expressions for the coefficient function of all helicity
components, which are the rational functions of their arguments. They
have a pair of ``classical'' poles at $x=\pm\xi$, and a pair of
supplementary poles at $x=\pm\kappa\,\xi$, where the parameter $\kappa$
is a function of kinematics and is restricted by $|\kappa|<1$ in
the physically allowed region . The complex structure of the coefficient
function does not permit direct deconvolution (inversion of the integrals
in~\ref{eq:Ha}-\ref{eq:ETildeA} for extraction of GPDs). Nonetheless,
it has been observed that the coefficient functions are mostly sensitive
to the behavior of the GPDs in the vicinity of its poles, and by varying
the kinematics and thereby altering the position of the poles, the
proposed process can be utilized to constrain existing phenomenological
parametrizations of GPDs across the entire ERBL region. We also demonstrated
that in electroproduction for some helicity components there are sizable
angular asymmetries which can be used as complementary observables
for experimental study.

Numerically, the cross-sections for all spins on unpolarized target
get the dominant contribution from the unpolarized gluon GPD $H_{g}$.
We found that the $\chi_{c1}$ and $\chi_{c2}$ mesons have sufficiently
large cross-sections for experimental study, and that predominantly
these mesons are produced with the same sign of helicity as the incoming
photon. The typical cross-sections for these mesons are of order a
few picobarns in the kinematics of low-energy EIC experiments, comparable
by magnitude to cross-sections of $\gamma^{*}p\to\gamma Mp$, $M=\eta_{c},\,\pi,\,\rho$
processes evaluated for the same (large) values of invariant masses
of meson-photon pair. We believe that this process is within the reach
of experimental studies: for instance, for $ep$ collisions at EIC
we expect the production rate of a few thousand $\chi_{c}\gamma$
pairs per each 100 ${\rm fb}^{-1}$ of integrated luminosity. For
$\chi_{c0}$ meson, the production cross-section and the expected
detection rates are significantly smaller and apparently insufficient
for reliable experimental study.

 We believe that the suggested process can be studied experimentally
in the ultraperipheral kinematics at LHC, in electron-proton collisions
at the future Electron Ion Collider (EIC)~\cite{Accardi:2012qut,AbdulKhalek:2021gbh,Burkert:2022hjz}
and possibly at JLAB after future 22 GeV upgrade~\cite{Accardi:2023chb}.

\section*{Acknowldgements}

We thank our colleagues at UTFSM university for encouraging discussions.
This research was partially supported by ANID PIA/APOYO AFB230003
(Chile) and Fondecyt (Chile) grants 1220242 and 1251975. I. Z. also
acknowledges support of the fellowship ``ANID Beca de Doctorado
Nacional'' , and expresses his gratitude to  the Institute of Physics
of PUCV. \textquotedbl Powered@NLHPC: This research was partially
supported by the supercomputing infrastructure of the NLHPC (ECM-02)\textquotedbl .

\appendix

\section{Wave function of the $\chi_{c}$ meson}

\label{subsec:WF}In the literature there are several complementary
descriptions of heavy quarkonia dynamics, in terms of the light cone
wave functions (LCWFs), and in terms of the NRQCD matrix elements,
which may be obtained from the LCWFs in the heavy quark mass limit~\cite{Ma:2006hc,Wang:2013ywc,Wang:2017bgv}.
We will start our discussion with the former in view of it simplicity
and will discuss its matching with NRQCD at the end of this section.

\begin{figure}
\includegraphics[width=4cm]{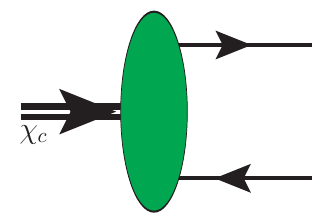}\includegraphics[width=4cm]{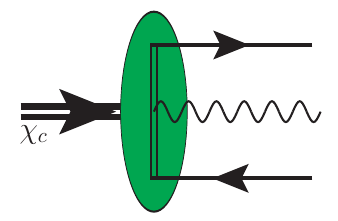}\includegraphics[width=4cm]{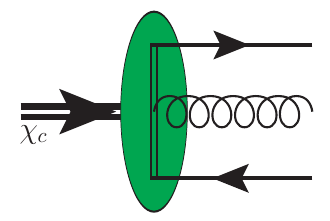}

\label{ChicWF}\caption{\protect\label{fig:ChicWF}The diagrammatic representation of the
leading twist Fock components of the $\chi_{c}$ meson, in notations
of~\cite{Collins:2011zzd}. The last two diagrams explicitly show
contributions of the photons and gluons connected to Wilson line (a
double line connecting two quarks). In NRQCD picture these diagrams
schematically illustrate the contributions of operators which include
gauge field in the covariant derivatives $D_{\mu}$. }
\end{figure}

Following the historical tradition, in this section we will temporarily
choose the frame in which the $\chi_{c}$ meson moves with infinitely
large momentum in ``-'' direction and zero transverse momentum,
\begin{align}
p_{\chi_{c}}^{\mu} & =\left(\frac{M_{\chi_{c}}^{2}}{2p_{\chi_{c}}^{-}},\,p_{\chi_{c}}^{-},\,\boldsymbol{0}_{\perp}\right).
\end{align}
This frame is related to any other frame in which the quarkonium has
nonzero transverse momentum $\boldsymbol{p}_{\perp}$ by a transverse
boost which keeps the $"-"$ component of the arbitrary vector $v$,
\begin{equation}
v^{-}\to v^{-},\,\,v^{+}\to v^{+}+\boldsymbol{\beta}_{\perp}\cdot\boldsymbol{v}_{\perp}+v^{-}\beta_{\perp}^{2}/2,\quad\boldsymbol{v}_{\perp}\to\boldsymbol{v}_{\perp}+v^{-}\boldsymbol{\beta}_{\perp}.\label{eq:TransverseBoost}
\end{equation}
with $\boldsymbol{\beta}_{\perp}=-\boldsymbol{p}_{\perp}/p^{-}$.
The leading twist quarkonia wave functions may be defined in momentum
space
\begin{align}
\Phi_{\chi_{cJ}}\left(z,\,\boldsymbol{k}_{\perp}\right) & =\int d^{2}\boldsymbol{r}_{\perp}e^{-i\boldsymbol{k}_{\perp},\cdot\boldsymbol{r}_{\perp}}\int\frac{d\lambda}{2\pi}e^{izp_{\chi_{c}}^{+}\lambda}\times\label{eq:DAdefEta}\\
 & \times\left\langle 0\left|\bar{\Psi}\left(-\frac{\lambda}{2}n_{+}-\frac{\boldsymbol{r}_{\perp}}{2}\right)\Gamma_{\chi_{cJ}}\mathcal{L}\left(-\frac{\lambda}{2}n_{+}-\frac{\boldsymbol{r}_{\perp}}{2},\,\frac{\lambda}{2}n_{+}+\frac{\boldsymbol{r}_{\perp}}{2}\right)\Psi\left(\frac{\lambda}{2}n_{+}+\frac{\boldsymbol{r}_{\perp}}{2}\right)\right|\chi_{c}(p)\right\rangle ,\nonumber \\
 & \mathcal{L}\left(x\,y\right)\equiv\mathcal{P}{\rm exp}\left(ig\int_{y}^{x}d\zeta\,A_{a}^{-}\left(\zeta\right)t^{a}+ie\int_{y}^{x}d\zeta\,A^{-}\left(\zeta\right)\right).\label{eq:Wilson}
\end{align}
where $z$ is the fraction of its momentum carried by the $c$-quark,
$\boldsymbol{k}_{\perp}=\boldsymbol{k}_{Q}^{\perp}-\boldsymbol{k}_{\bar{Q}}^{\perp}$
is the transverse component of the relative momentum, and $\mathcal{L}$
is the usual Wilson link introduced to guarantee the gauge invariance
of the result. In presence of the electromagnetic field, in order
to guarantee the local gauge invariance of the defined wave function,
the Wilson link must also include the second term $ie\int_{y}^{x}d\zeta\,A_{({\rm em})}^{-}\left(\zeta\right)$
in the argument of the exponent. In diagrammatic language, as explained
in~\cite{Collins:2011zzd}, this Wilson link implies additional
contributions with coupling of the gauge boson, as shown in the Figure~\ref{fig:ChicWF}.
After integration over $\boldsymbol{k}_{\perp}$ (which corresponds
to the limit $\boldsymbol{r}_{\perp}=0$ in the configuration space),
these objects reduce to the light-cone distribution amplitudes (LCDAs)
discussed in detail in~\cite{LightQuarkDA:1,LightQuarkDA:2,LightQuarkDA:3,LightQuarkDA:4,LightQuarkDA:5}.
The matrix $\Gamma_{\chi_{c}}$ which appears in~(\ref{eq:DAdefEta})
depends on the spin $J$ of the $\chi_{c}$ meson and its helicity
$H$, and is given explicitly by 
\begin{equation}
\Gamma_{\chi_{cJ}}=\left\{ \begin{array}{cc}
1, & \qquad J=0\\
i\gamma_{5}\gamma_{\mu}E^{\mu}, & \qquad J=1\\
\frac{1}{2}\left(\gamma_{\mu}k_{\nu}+\gamma_{\nu}k_{\mu}\right)E^{\mu\nu}, & \qquad J=2
\end{array}\right.\label{eq:Gamma}
\end{equation}
where the vector $E^{\mu}$ and tensor $E^{\mu\nu}$ are defined in
arbitrary frame as
\begin{align}
E_{H=\pm1}^{\mu} & =\left(\frac{\boldsymbol{e}_{H}^{\perp}\cdot\boldsymbol{p}_{\perp}^{\chi_{c}}}{p_{\chi_{c}}^{-}},\,0,\boldsymbol{e}_{H}^{\perp}\right),\qquad\boldsymbol{e}_{H}^{\perp}=\frac{1}{\sqrt{2}}\left(\begin{array}{c}
1\\
iH
\end{array}\right),\\
E_{H=0}^{\mu} & =\frac{p_{\chi_{c}}^{\mu}}{M_{\chi_{c}}}-\frac{M_{\chi_{c}}}{p_{\chi_{c}}^{-}}n_{-}^{\mu}=\frac{1}{M_{\chi_{c}}}\left(\frac{\left(\boldsymbol{p}_{\perp}^{\chi_{c}}\right)^{2}-M_{\chi_{c}}^{2}}{2p_{\chi_{c}}^{-}},\,p_{\chi_{c}}^{-},\,\boldsymbol{p}_{\perp}^{\chi_{c}}\right),
\end{align}
\begin{equation}
E^{\mu\nu}=\left\{ \begin{array}{cc}
E_{\pm1}^{\mu}E_{\pm1}^{\nu}, & \qquad H=\pm2\\
\frac{1}{\sqrt{2}}\left(E_{\pm1}^{\mu}E_{0}^{\nu}+E_{\pm1}^{\nu}E_{0}^{\mu}\right), & \qquad H=\pm1\\
\frac{1}{\sqrt{6}}\left(E_{+1}^{\mu}E_{-1}^{\nu}+E_{-1}^{\mu}E_{+1}^{\nu}+2E_{0}^{\mu}E_{0}^{\nu}\right), & \qquad H=0
\end{array}\right.
\end{equation}
and $k^{\mu}\equiv k_{Q}^{\mu}-k_{\bar{Q}}^{\mu}\approx p_{\chi_{c}}^{-}\left(2z-1\right)n_{-}^{\mu}+\boldsymbol{k}_{\perp}^{\mu}$
is the relative momentum of the quark-antiquark motion. Note that
the vector $k^{\mu}$ always remains orthogonal to $p_{\chi_{c}}$,
namely $k\cdot p_{\chi_{c}}=0$. From~(\ref{eq:Gamma}) we may observe
that in the limit $k^{\mu}\to0$ the corresponding vertex with $J=2$
vanishes, as expected for the $P$-wave. Its is straightforward to
show that a similar property is valid for the states $J=0$ and $J=1$,
if we take into account that $\Gamma_{\chi_{cJ}}$ is plugged between
the spinors $\bar{u},\,v$ of the quark and antiquark (hidden in $\bar{\psi},\psi$)
and use the Dirac equation together with orthogonality of polarization
vector $E_{H}\cdot p_{\chi_{c}}=0$, namely 
\begin{align}
\bar{u}\left(k_{0}\right) & \Gamma_{\chi_{cJ}}v\left(k_{1}\right)=\frac{1}{2m}\bar{u}\left(k_{0}\right)\left(m\Gamma_{\chi_{cJ}}+\Gamma_{\chi_{cJ}}m\right)v\left(k_{1}\right)=\frac{1}{2m}\bar{u}\left(k_{0}\right)\left(\hat{k}_{0}\Gamma_{\chi_{cJ}}-\Gamma_{\chi_{cJ}}\hat{k}_{1}\right)v\left(k_{1}\right)\label{eq:Ga}\\
 & =\frac{1}{4m}\bar{u}\left(k_{0}\right)\left(\hat{p}_{\chi_{c}}\Gamma_{\chi_{cJ}}-\Gamma_{\chi_{cJ}}\hat{p}_{\chi_{c}}\right)v\left(k_{1}\right)+\frac{1}{4m}\bar{u}\left(k_{0}\right)\left(\hat{k}\Gamma_{\chi_{cJ}}+\Gamma_{\chi_{cJ}}\hat{k}\right)v\left(k_{1}\right)=\frac{k^{\mu}}{4m}\bar{u}\left(k_{0}\right)\left(\gamma^{\mu}\Gamma_{\chi_{cJ}}+\Gamma_{\chi_{cJ}}\gamma^{\mu}\right)v\left(k_{1}\right)\nonumber 
\end{align}
In view of this, we can replace the matrix $\Gamma_{\chi_{cJ}}$ with
an equivalent $\Gamma_{\chi_{cJ}}^{({\rm eff})}=\left(\gamma^{\mu}\Gamma_{\chi_{cJ}}+\Gamma_{\chi_{cJ}}\gamma^{\mu}\right)/4m=\gamma_{\chi_{cJ}}^{\mu}k_{\mu}$,
where $\gamma_{\chi_{c0}}^{\mu}\equiv\gamma^{\mu}/2m$, $\gamma_{\chi_{c1}}^{\mu}\equiv\left[\gamma^{\mu},\,\hat{E}_{H}\right]\gamma^{5}/4m$
and $\gamma_{\chi_{c2}}^{\mu}=E^{\mu\nu}\gamma_{\nu}$. In the configuration
space, the corresponding relative momentum $k_{\mu}$ can be replaced
with $i\overleftrightarrow{\partial}_{\mu}$. In the hard processes,
whose scale significantly exceeds the inverse charmonium radius, it
is possible to disregard completely the dependence on $z,\boldsymbol{k}_{\perp}$
in the partonic-level amplitude, so the final result will depend only
the wave function integrated over $z,\boldsymbol{k}_{\perp}$, namely
\begin{align}
\int dz & \int\frac{\,d^{2}\boldsymbol{k}_{\perp}}{\left(2\pi\right)^{2}}\Phi_{\chi_{c}}\left(z,\,\boldsymbol{k}_{\perp}\right)=\label{eq:DAdefEta-1}\\
 & =\lim_{\lambda,\,\boldsymbol{r}_{\perp}\to0}\left\langle 0\left|\bar{\Psi}\left(-\frac{\lambda}{2}n_{+}-\frac{\boldsymbol{r}_{\perp}}{2}\right)\gamma_{\chi_{cJ}}^{\mu}i\overleftrightarrow{\partial}_{\mu}\,\mathcal{L}\left(-\frac{\lambda}{2}n_{+}-\frac{\boldsymbol{r}_{\perp}}{2},\,\frac{\lambda}{2}n_{+}+\frac{\boldsymbol{r}_{\perp}}{2}\right)\Psi\left(\frac{\lambda}{2}n_{+}+\frac{\boldsymbol{r}_{\perp}}{2}\right)\right|\chi_{c}(p)\right\rangle \nonumber \\
 & =\left\langle 0\left|\bar{\Psi}\left(0\right)\gamma_{\chi_{cJ}}^{\mu}\frac{i}{2}\overleftrightarrow{D}_{\mu}\,\Psi\left(0\right)\right|\chi_{c}(p)\right\rangle \nonumber 
\end{align}
where $\overleftrightarrow{D}_{\mu}=\overrightarrow{\partial_{\mu}}-\overleftarrow{\partial_{\mu}}+2igA_{\mu}^{a}t_{a}+2ieA_{\mu}$
is the covariant derivative. The square of the matrix element which
appears in the last line of~(\ref{eq:DAdefEta-1}) in the nonrelativistic
limit reduces to NRQCD LDMEs. Indeed, using explicit form of $\gamma_{\chi_{cJ}}^{\mu}$
for $J=0,1,2$, we may recover the familiar NRQCD operators from~\cite{Ma:2006hc,Wang:2013ywc,Wang:2017bgv}
\begin{equation}
\left|\left\langle 0\left|\bar{\psi}\left(0\right)\gamma_{\chi_{cJ}}^{\mu}i\overleftrightarrow{D}_{\mu}\,\psi\left(0\right)\right|\chi_{c}(p)\right\rangle \right|^{2}=\frac{1}{4m^{2}}\times\left\{ \begin{array}{cc}
\left|\left\langle 0\left|\bar{\psi}\left(\frac{i}{2}\overleftrightarrow{D}_{\top}^{\mu}\gamma_{\mu}^{\top}\right)\,\chi\right|\chi_{c}(p)\right\rangle \right|^{2}=\left\langle \mathcal{\mathcal{O}}_{\chi_{c}}^{[1]}\left(^{3}P_{0}^{[1]}\right)\right\rangle , & \qquad J=0\\
\left|\left\langle 0\left|\psi^{\dagger}\left(-\frac{i}{4}\overleftrightarrow{D}_{\top}^{\nu}\left[\gamma_{\nu}^{\top},\,\gamma_{\mu}^{\top}\right]\right)\gamma_{5}\,\chi\right|\chi_{c}(p)\right\rangle \right|^{2}=\left\langle \mathcal{\mathcal{O}}_{\chi_{c}}^{[1]}\left(^{3}P_{1}^{[1]}\right)\right\rangle , & \qquad J=1\\
\left|\left\langle 0\left|\psi^{\dagger}\left(-\frac{i}{2}\overleftrightarrow{D_{\top}^{(\mu}}\gamma_{\top}^{\nu)}\right)\,\chi\right|\chi_{c}(p)\right\rangle \right|^{2}=\left\langle \mathcal{\mathcal{O}}_{\chi_{c}}^{[1]}\left(^{3}P_{2}^{[1]}\right)\right\rangle , & \qquad J=2
\end{array}\right.
\end{equation}
where $\top$ implies a part of the vector which is transverse to
$p_{\chi_{c}}^{\mu}$, namely $v_{\top}^{\mu}=\left(g^{\mu\nu}-p_{\chi_{c}}^{\mu}p_{\chi_{c}}^{\nu}/M_{\chi_{c}}^{2}\right)v^{\nu}$;
$a_{\top}^{(\mu}b_{\top}^{\nu)}=(a_{\top}^{\mu}b_{\top}^{\nu}+a_{\top}^{\nu}b_{\top}^{\mu})/2-a_{\top}\cdot b_{\top}\left(g^{\mu\nu}-p_{\chi_{c}}^{\mu}p_{\chi_{c}}^{\nu}/M_{\chi_{c}}^{2}\right)/3$,
the operators $\psi^{\dagger},\,\chi$ are the Pauli spinor operators
which create a quark and an antiquark, respectively. The presence
of the gauge fields in covariant derivative $\overleftrightarrow{D}_{\mu}$
implies that we should consider the $\chi_{c}$ meson as a mixture
of $|\bar{Q}Q\rangle$, $|\bar{Q}Qg\rangle$ and $|\bar{Q}Q\gamma\rangle$
Fock components, as shown in the Figure~\ref{fig:ChicWF}. Only in
the special gauge $A^{-}=0$, $A_{a}^{-}=0$ we may disregard the
contributions of $|\bar{Q}Qg\rangle$ and $|\bar{Q}Q\gamma\rangle$.
This choice of the gauge condition remains invariant under transverse
boosts~(\ref{eq:TransverseBoost}) and for this reason in what follows
we will use it for the electromagnetic field. We also need to mention
that in the leading order there is no diagrams with gluons connected
directly to$\chi_{c}$, so we do not impose any gauge conditions for
the gluonic field.

The explicit expression for the NRQCD projectors of $P$-wave include
differential operators which eventually require taking derivatives
of the partonic amplitude of the whole process with respect to components
of the momentum $k^{\mu}$, and eventually lead to sophisticated expressions.
For this reason, instead of applying NRQCD directly, we will assume
that the nonperturbative wave function $\Phi_{\chi_{cJ}}\left(z,\,\boldsymbol{k}_{\perp}\right)$
is given by
\begin{equation}
\Phi_{\chi_{cJ}}\left(z,\,\boldsymbol{k}_{\perp}\right)=\bar{u}\left(\frac{p_{\chi_{cJ}}}{2}+\frac{\boldsymbol{k}}{2}\right)\Gamma_{\chi_{cJ}}v\left(\frac{p_{\chi_{cJ}}}{2}-\frac{\boldsymbol{k}}{2}\right)\phi\left(z,\,\boldsymbol{k}_{\perp}\right)\label{eq:WF}
\end{equation}
where the scalar light-cone wave function $\phi\left(z,\,\boldsymbol{k}_{\perp}\right)$
is related to the rest frame wave function of $\chi_{c}$ via~\cite{Babiarz:2020jkh}
\begin{equation}
\frac{dz\,d^{2}\boldsymbol{k}}{z(1-z)16\pi^{3}}\phi\left(z,\,\boldsymbol{k}_{\perp}\right)=\frac{1}{4\pi^{2}\sqrt{M_{\chi_{c}}}}\frac{1}{2\sqrt{2}}d^{3}k\,\frac{u(k)}{k^{2}},\qquad k_{z}\equiv M_{\chi_{c}}\left(z-\frac{1}{2}\right)
\end{equation}
and $u(k)/k$ is the radial part of the rest frame wave function in
the momentum space. Using the identity
\begin{equation}
\int_{0}^{\infty}dk\,k^{2}dk\,u(k)=3\sqrt{\frac{\pi}{2}}R'(0)=\sqrt{2M_{\chi_{c}}}\int\frac{dz\,d^{2}\boldsymbol{k}}{z(1-z)8\pi^{2}}\phi\left(z,\,\boldsymbol{k}_{\perp}\right)\left[M_{\chi_{c}}^{2}\left(z-\frac{1}{2}\right)^{2}+\boldsymbol{k}_{\perp}^{2}\right]\label{eq:Rel}
\end{equation}
where $R'_{\chi_{c}}(0)$ is the slope of the radial wave function
of $\chi_{c}$, we may relate the light-cone wave function with NRQCD
matrix elements
\begin{equation}
\left\langle \mathcal{\mathcal{O}}_{\chi_{c}}^{[1]}\left(^{3}P_{J}^{[1]}\right)\right\rangle =6N_{c}\left(2J+1\right)\left|R'_{\chi_{c}}(0)\right|^{2}/4\pi\label{eq:LDME}
\end{equation}
 Using the explicit parametrization from~\cite{Benic:2024} rewritten
in momentum space,
\begin{equation}
\phi\left(z,\,\boldsymbol{k}_{\perp}\right)=\mathcal{N}\exp\left(-\frac{\mathcal{R}^{2}}{8}\left(\frac{\boldsymbol{k}_{\perp}^{2}+m^{2}}{z(1-z)}-4m^{2}\right)\right)\label{eq:Explicit}
\end{equation}
where $\mathcal{N},\mathcal{R}$ are some numerical constants, we
found that the results for $\left|R'(0)\right|$ agree within 10 per
cent with value $\left|R'(0)\right|^{2}=0.075\,{\rm GeV}^{5}$ calculated
in potential models and widely used in the literature for numerical
estimates.

\section{Evaluation of the coefficient functions}

\label{sec:CoefFunction}The evaluation of the coefficient functions
can be performed using the standard light--cone rules from~\cite{Lepage:1980fj,Brodsky:1997de,Diehl:2000xz,Diehl:2003ny,Diehl:1999cg,Ji:1998pc}.
As discussed in Section~\ref{subsec:Kinematics}, we disregard the
proton mass $m_{N}$ and the invariant momentum transfer to the proton
$t=\Delta^{2}$, treating all the other variables as parametrically
large quantities of order $m_{c}$. The evaluation of the partonic
amplitudes shown in the Figure~\ref{fig:Photoproduction-A} is straightforward.

In general, the coefficient functions $C_{\mathfrak{\gamma\chi_{c}}},\,\tilde{C}_{\mathfrak{\gamma\chi_{c}}}$
which appear in~(\ref{eq:Ha}-\ref{eq:ETildeA}) should be understood
as convolutions of partonic-level amplitudes with wave function $\Phi_{\chi_{c}}$,
namely, 
\begin{align}
C_{\mathfrak{\gamma\chi_{c}}}^{(\lambda,\sigma,H)}\left(x,\,\xi\right) & =\int\,\frac{dz\,d^{2}\boldsymbol{k}}{z(1-z)16\pi^{3}}\,c^{(\lambda,\sigma)}\left(x,\,\xi,\,z,\boldsymbol{k}_{\perp}\right)\Phi_{\chi_{c}}^{(H)}\left(z,\boldsymbol{k}_{\perp}\right),\label{eq:ca}\\
\tilde{C}_{\gamma\chi_{c}}^{(\lambda,\sigma,H)}\left(x,\,\xi\right) & =\int_{0}^{1}\,\frac{dz\,d^{2}\boldsymbol{k}}{z(1-z)16\pi^{3}}\,\tilde{c}^{(\lambda,\sigma)}\left(x,\,\xi,\,z,\boldsymbol{k}_{\perp}\right)\Phi_{\chi_{c}}^{(H)}\left(z,\boldsymbol{k}_{\perp}\right).
\end{align}
The evaluation of the partonic amplitudes $c\left(x,\,\xi,\,z,\boldsymbol{k}_{\perp}\right),\,\tilde{c}\left(x,\,\xi,\,z,\boldsymbol{k}_{\perp}\right)$
may be done perturbatively for all diagrams shown in the Figure~\ref{fig:Photoproduction-A}.
The (implicit) summation over helicities of the quark and antiquark
in the wave function~(\ref{eq:WF}) and the partonic amplitude using
identities $\sum_{h}u_{h}(k)\bar{u}_{h}(k)=\hat{k}+m$, $\sum_{\bar{h}}v_{\bar{h}}(k)\bar{v}_{\bar{h}}(k)=\hat{k}-m$
yields an effective vertex (projector)~\cite{Cho:1995ce,Cho:1995vh,DVMPcc1},
\begin{align}
\hat{P}_{\chi_{cJ}} & =-\sqrt{\frac{\left\langle \mathcal{O}_{\chi_{c}}\right\rangle }{m_{Q}}}\,\frac{1}{8m_{Q}}\left(\frac{\hat{p}_{\chi_{c}}}{2}-\frac{\hat{k}}{2}-m_{c}\right)\Gamma_{\chi_{cJ}}\left(\frac{\hat{p}_{\chi_{c}}}{2}+\frac{\hat{k}}{2}+m_{c}\right)\otimes\frac{\delta_{ij}}{\sqrt{N_{c}}},\label{eq:NRQCDProjector}
\end{align}
where the vector of relative momentum $k=k_{Q}-k_{\bar{Q}}$ was defined
in~(\ref{eq:VecK}), the helicity-dependent matrix $\Gamma_{\chi_{cJ}}$
was defined in~(\ref{eq:Gamma}). $\left\langle \mathcal{O}_{\chi_{c}}\right\rangle \equiv\left\langle \mathcal{\mathcal{O}}_{\chi_{c}}\left(^{3}P_{J}^{[1]}\right)\right\rangle $
is the corresponding color singlet LDME of $\chi_{c}$ meson~\cite{Braaten:2002fi}.
We need to mention that in the heavy quark mass limit the internal
motion is negligible, so the wave function $\Phi_{\chi_{c}}^{(H)}\left(z,\boldsymbol{k}_{\perp}\right)$
is strongly peaked near the point $z=1/2,\,\boldsymbol{k}_{\perp}=0$.
For this reason, making expansion near this point and using the relations~(\ref{eq:Rel},\ref{eq:LDME}),
we obtain a result which includes an LDME multiplied by the functions
of variables $x,\xi$ which do not depend on $z,\,\boldsymbol{k}_{\perp}$.
We need to mention that $C$-conjugate diagrams (with inverted direction
of quark lines) mutually cancel odd terms in the series expansion
over $\left(z-1/2\right),\,\boldsymbol{k}_{\perp}$ and double the
coefficients in front of even powers.

\begin{figure}
\includegraphics[scale=0.6]{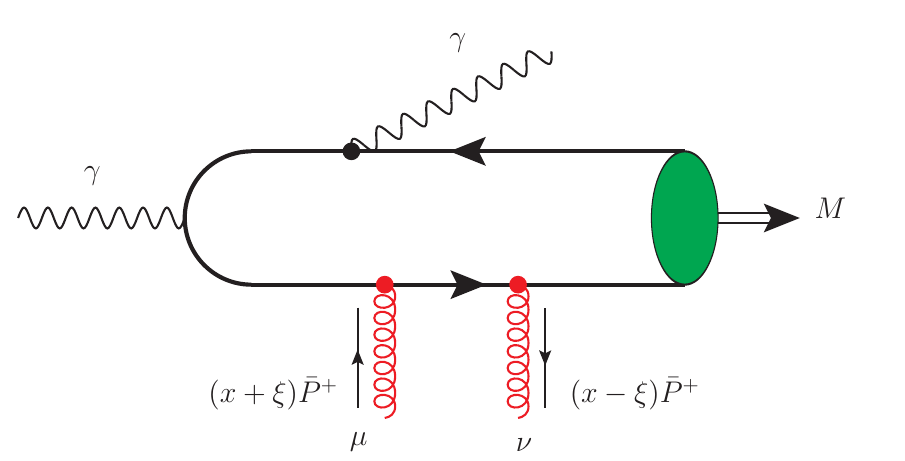}\includegraphics[scale=0.6]{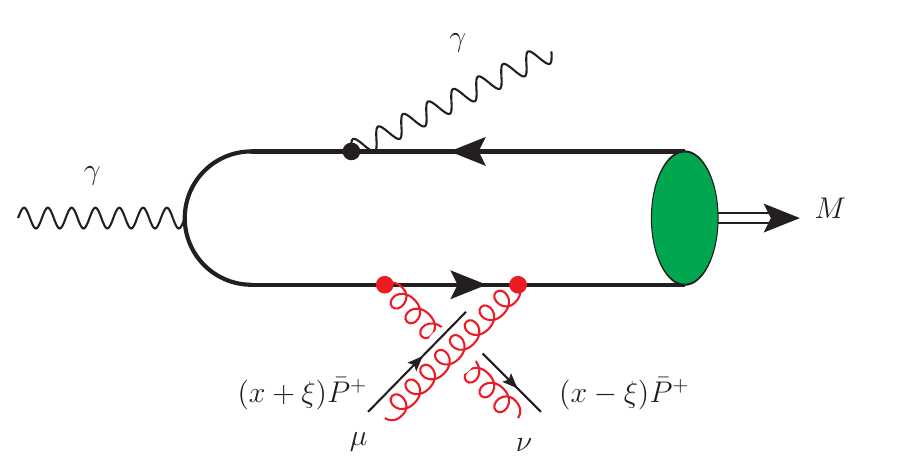}

\caption{\protect\label{fig:Photoproduction-Permute}Example of the diagrams
with direct and permuted $t$-channel gluons, which are related to
each other by inversion of sign in front of light-cone fraction $x\leftrightarrow-x$,
and permutation of the Lorentz indices $\mu\leftrightarrow\nu$.}
\end{figure}

The interaction of the heavy quarks with the target proceeds via exchange
of gluons in $t$-channel . At high energies this interaction is parametrized
by the leading twist gluon GPDs, which are defined as~\cite{Diehl:2003ny,DVMPcc1}
\begin{align}
F^{g}\left(x,\xi,t\right) & =\frac{1}{\bar{P}^{+}}\int\frac{dz}{2\pi}\,e^{ix\bar{P}^{+}}\left\langle P'\left|G^{+\mu\,a}\left(-\frac{z}{2}n\right)\mathcal{L}\left(-\frac{z}{2},\,\frac{z}{2}\right)G_{\,\,\mu}^{+\,a}\left(\frac{z}{2}n\right)\right|P\right\rangle =\label{eq:defF}\\
 & =\left(\bar{U}\left(P'\right)\gamma_{+}U\left(P\right)H^{g}\left(x,\xi,t\right)+\bar{U}\left(P'\right)\frac{i\sigma^{+\alpha}\Delta_{\alpha}}{2m_{N}}U\left(P\right)E^{g}\left(x,\xi,t\right)\right),\nonumber \\
\tilde{F}^{g}\left(x,\xi,t\right) & =\frac{-i}{\bar{P}^{+}}\int\frac{dz}{2\pi}\,e^{ix\bar{P}^{+}}\left\langle P'\left|G^{+\mu\,a}\left(-\frac{z}{2}n\right)\mathcal{L}\left(-\frac{z}{2},\,\frac{z}{2}\right)\tilde{G}_{\,\,\mu}^{+\,a}\left(\frac{z}{2}n\right)\right|P\right\rangle =\label{eq:defFTilde}\\
 & =\left(\bar{U}\left(P'\right)\gamma_{+}\gamma_{5}U\left(P\right)\tilde{H}^{g}\left(x,\xi,t\right)+\bar{U}\left(P'\right)\frac{\Delta^{+}\gamma_{5}}{2m_{N}}U\left(P\right)\tilde{E}^{g}\left(x,\xi,t\right)\right).\nonumber \\
 & \tilde{G}^{\mu\nu,\,a}\equiv\frac{1}{2}\varepsilon^{\mu\nu\alpha\beta}G_{\alpha\beta}^{a},\quad\mathcal{L}\left(-\frac{z}{2},\,\frac{z}{2}\right)\equiv\mathcal{P}{\rm exp}\left(i\int_{-z/2}^{z/2}d\zeta\,A^{+}\left(\zeta\right)\right).
\end{align}
where we introduced notations $U,\bar{U}$ for the spinors of the
incoming and outgoing proton. In the light-cone gauge $A^{+}=0$ the
Wilson link $\mathcal{L}$ can be disregarded, so the gluon operators
in~(\ref{eq:defF},~\ref{eq:defFTilde}) can be rewritten as 
\begin{align}
 & G^{+\mu_{\perp}\,a}\left(z_{1}\right)G_{\,\,\mu_{\perp}}^{+\,a}\left(z_{2}\right)=g_{\mu\nu}^{\perp}\left(\partial^{+}A^{\mu_{\perp},a}(z_{1})\right)\left(\partial^{+}A^{\nu_{\perp}a}(z_{2})\right),\\
 & G^{+\mu_{\perp}\,a}\left(z_{1}\right)\tilde{G}_{\,\,\mu_{\perp}}^{+\,a}\left(z_{2}\right)=G^{+\mu_{\perp}\,a}\left(z_{1}\right)\tilde{G}_{-\mu_{\perp}}^{\,a}\left(z_{2}\right)=\frac{1}{2}\varepsilon_{-\mu_{\perp}\alpha\nu}G^{+\mu_{\perp}\,a}\left(z_{1}\right)G^{\alpha\nu,\,a}\left(z_{2}\right)=\\
 & =\varepsilon_{-\mu_{\perp}+\nu_{\perp}}G^{+\mu_{\perp}\,a}\left(z_{1}\right)G^{+\nu_{\perp},\,a}\left(z_{2}\right)=\varepsilon_{\mu\nu}^{\perp}G^{+\mu_{\perp}\,a}\left(z_{1}\right)G^{+\nu_{\perp},\,a}\left(z_{2}\right)=\varepsilon_{\mu\nu}^{\perp}\left(\partial^{+}A^{\mu,a}(z_{1})\right)\left(\partial^{+}A^{\nu,\,a}(z_{2})\right).\nonumber 
\end{align}
The integration over the light-cone separation $z$ in~(\ref{eq:defF},~\ref{eq:defFTilde})
effectively corresponds to Fourier transformation to the momentum
space, where the derivatives $\partial_{z_{1}}^{+},\,\partial_{z_{2}}^{+}$
turn into the multiplicative factors $k_{1,2}^{+}\sim\left(x\pm\xi\right)\bar{P}^{+}$.
This allows to rewrite the Eqs.~(\ref{eq:defF},~\ref{eq:defFTilde})
as~\cite{DVMPcc1}
\begin{align}
\frac{1}{\bar{P}^{+}}\int\frac{dz}{2\pi}\,e^{ix\bar{P}^{+}}\left.\left\langle P'\left|A_{\mu}^{a}\left(-\frac{z}{2}n\right)A_{\nu}^{b}\left(\frac{z}{2}n\right)\right|P\right\rangle \right|_{A^{+}=0\,{\rm gauge}} & =\frac{\delta^{ab}}{N_{c}^{2}-1}\left(\frac{-g_{\mu\nu}^{\perp}F^{g}\left(x,\xi,t\right)-\varepsilon_{\mu\nu}^{\perp}\tilde{F}^{g}\left(x,\xi,t\right)}{2\,\left(x-\xi+i0\right)\left(x+\xi-i0\right)}\right).\label{eq:defF-1}
\end{align}
The Eq.~(\ref{eq:defF-1}) suggests that the coefficient functions
$C_{\gamma\chi_{c}}$ and $\tilde{C}_{\gamma\chi_{c}}$can be obtained
making convolutions of the Lorentz indices of $t$-channel gluons
in diagrams of Figure~\ref{fig:Photoproduction-A} with $g_{\mu\nu}^{\perp}$
and $\varepsilon_{\mu\nu}^{\perp}$ respectively. The remaining steps
in evaluation of the diagrams shown in Figure~\ref{fig:Photoproduction-A}
were performed using Mathematica and yield
\begin{align}
c\left(x,\,\xi\right) & =2\left[\mathcal{C}\left(x,\,\xi\right)+\mathcal{C}\left(-x,\,\xi\right)\right],\label{eq:C}\\
\tilde{c}\left(x,\,\xi\right) & =2\left[\tilde{\mathcal{C}}\left(x,\,\xi\right)-\tilde{\mathcal{C}}\left(-x,\,\xi\right)\right],\label{eq:CTilde}
\end{align}
where the terms $\mathcal{C}\left(x,\,\xi\right)$ and $\mathcal{\tilde{C}}\left(x,\,\xi\right)$
in the right-hand sides of~(\ref{eq:C},~\ref{eq:CTilde}) correspond
to the contributions of the six diagrams shown explicitly in the Figure~$~\ref{fig:Photoproduction-A}$,
the last two terms (with $x\to-x$ substitution) are the contributions
of diagrams with permuted gluons in the $t$-channel, as shown in
the Figure~\ref{fig:Photoproduction-Permute}, and the prefactor
2 in front of parenthesis in the right-hand side takes into account
that $C$-conjugate diagrams (with inverted direction of quark lines)
eventually give exactly the same contribution. From the right-hand
side of~(\ref{eq:C},~\ref{eq:CTilde}) we may infer that $c$ and
$\tilde{c}$ are even and odd functions of their first argument $x$.
Explicitly, the components $\mathcal{C}^{(\lambda,\sigma,H)}\left(x,\,\xi,\,z\right)$
and $\tilde{\mathcal{C}}^{(\lambda,\sigma,H)}\left(x,\,\xi,\,z\right)$
are given in Appendix~\ref{sec:CoefFunction}.

\section{Final results for the coefficient functions}

In this Appendix we provide explicit expressions for the coefficient
functions (partonic amplitudes) of the $\chi_{cJ}\gamma$ production
in helicity basis. Due to the $P$-parity the cross-sections which
differ by global inversion of the sign of all helicities should coincide,
$\mathcal{A}_{\gamma p\to\chi_{c}\gamma p}^{(-\lambda,-\sigma,-H)}=\mathcal{A}_{\gamma p\to\chi_{c}\gamma p}^{(\lambda,\sigma,H)}$,
which implies that only half of all possible combinations are independent.
For the sake of definiteness we will assume that for each helicity
$H$, the incoming photon always has helicity ``+'', the emitted
photon may have helicity +'' or ``-'' (we'll refer to them as helicity
non-flip and helicity flip components, $C_{\gamma\chi_{c}}^{(+,+,H)}$
and $C_{\gamma\chi_{c}}^{(+.-,H)}$ respectively). We'll assume that
the remaining components can be recovered changing the sign of all
helicities. For the sake of definiteness, we'll choose the direction
of the axis $\hat{\boldsymbol{x}}$ in direction of the vector $\boldsymbol{p}_{\perp}$.
In any other frame, which differs by azimuthal rotation in transverse
plane $Oxy$, the amplitudes in helicity basis are multiplied by a
trivial phase factors $\sim e^{i\Lambda\alpha}$, where $\Lambda$
is a linear combination of helicities of the incoming and outgoing
bosons. In the cross-section, all such phase factors eventually cancel.

\subsection{Coefficient functions for $\chi_{c0}$}

\begin{align}
c_{\gamma\chi_{c0}}^{(++)} & \left(r,\,\alpha,\,\zeta=x/\xi\right)=-\frac{4\mathfrak{C}_{\chi_{c}}}{\alpha^{2}\left(1+\bar{\alpha}\right)^{2}\left(\bar{\alpha}+1/r^{2}\right)\left(r^{2}-1\right)^{2}\left(\zeta+1-i0\right)\left(\zeta-1+i0\right)\left(\zeta-\kappa+i0\right)^{2}}\times\\
\times & \left\{ \mathinner{\color{white}\frac{\frac{}{}}{\frac{}{}}}\left(\alpha^{6}-3\alpha^{5}-13\alpha^{4}+68\alpha^{3}-68\alpha^{2}+16\alpha\right)\zeta^{4}+\alpha^{7}+10\alpha^{6}+84\alpha^{5}-227\alpha^{4}+122\alpha^{3}+120\alpha^{2}\right.\nonumber \\
 & \qquad+\left(43\alpha^{7}-273\alpha^{6}+665\alpha^{5}-834\alpha^{4}+754\alpha^{3}-584\alpha^{2}+328\alpha-96\right)\zeta^{3}-144\alpha+32\nonumber \\
 & \qquad+\left(-149\alpha^{7}+713\alpha^{6}-1473\alpha^{5}+1844\alpha^{4}-1474\alpha^{3}+628\alpha^{2}+8\alpha-96\right)\zeta^{2}\nonumber \\
 & \qquad+\left(105\alpha^{7}-323\alpha^{6}+487\alpha^{5}-290\alpha^{4}-286\alpha^{3}+560\alpha^{2}-256\alpha\right)\zeta\nonumber \\
 & +\frac{1}{r^{2}}\left[\left(4\alpha^{5}-14\alpha^{4}+2\alpha^{3}+16\alpha^{2}-8\alpha\right)\zeta^{4}+\left(-48\alpha^{6}+252\alpha^{5}-468\alpha^{4}+306\alpha^{3}-56\alpha^{2}+40\alpha-32\right)\zeta^{3}+\right.\nonumber \\
 & \qquad+\left(264\alpha^{6}-1044\alpha^{5}+1624\alpha^{4}-1602\alpha^{3}+1256\alpha^{2}-664\alpha+160\right)\zeta^{2}-8\alpha^{6}-52\alpha^{5}-150\alpha^{4}\nonumber \\
 & \qquad-\left.\left(272\alpha^{6}-696\alpha^{5}+960\alpha^{4}-782\alpha^{3}+48\alpha^{2}+384\alpha-192\right)\zeta+640\alpha^{3}-624\alpha^{2}+168\alpha+32\right]\nonumber \\
 & +\frac{1}{r^{4}}\left[\left(-5\alpha^{4}+24\alpha^{3}-36\alpha^{2}+16\alpha\right)\zeta^{4}+\left(31\alpha^{5}-134\alpha^{4}+222\alpha^{3}-144\alpha^{2}-8\alpha+32\right)\zeta^{3}\right.\nonumber \\
 & \qquad+\left(-249\alpha^{5}+780\alpha^{4}-710\alpha^{3}+192\alpha^{2}-104\alpha+96\right)\zeta^{2}+25\alpha^{5}+125\alpha^{4}-18\alpha^{3}\nonumber \\
 & \qquad+\left.\left(369\alpha^{5}-670\alpha^{4}+642\alpha^{3}-656\alpha^{2}+344\alpha-32\right)\zeta-524\alpha^{2}+552\alpha-160\right]\nonumber \\
 & +\frac{1}{r^{6}}\left[\left(-10\alpha^{4}+38\alpha^{3}-64\alpha^{2}+72\alpha-32\right)\zeta^{3}+\left(142\alpha^{4}-410\alpha^{3}+320\alpha^{2}+48\alpha-96\right)\zeta^{2}\right.+\nonumber \\
 & \qquad\left.+\left(-302\alpha^{4}+338\alpha^{3}+40\alpha^{2}+24\alpha-96\right)\zeta-38\alpha^{4}-126\alpha^{3}+216\alpha^{2}+32\alpha-96\right]\nonumber \\
 & \left.+\frac{4\left(7\alpha^{3}+9\alpha^{2}-\left(10\alpha^{3}-33\alpha^{2}+40\alpha-16\right)\zeta^{2}+\left(39\alpha^{3}-42\alpha^{2}-22\alpha+24\right)\zeta-30\alpha+16\right)}{r^{8}}+\frac{8\bar{\alpha}\left((5\alpha-4)\zeta+\alpha\right)}{r^{10}}\right\} \nonumber \\
 & +\left(\mathinner{\color{white}\frac{}{}}\zeta\to-\zeta\right)\nonumber 
\end{align}
where we introduced shorthand notations $\alpha\equiv\alpha_{\chi_{c}}=(-u')/M_{\gamma\chi_{c}}$,
$\zeta=x/\xi$, the constant $\mathfrak{C}_{\chi_{c}}$ in the prefactor
is given explicitly by
\begin{equation}
\mathfrak{C}_{\chi_{c}}=\frac{4}{9N_{c}}\pi^{2}\alpha_{{\rm em}}\alpha_{s}\left(\mu\right)\pi\sqrt{\frac{3\left\langle \mathcal{\mathcal{O}}_{\chi_{c}}^{[1]}\left(^{3}P_{J}^{[1]}\right)\right\rangle }{2m_{c}^{5}N_{c}(2J+1)}},\label{eq:CDef-1}
\end{equation}
$\left\langle \mathcal{O}_{\chi_{c}}\right\rangle \equiv\left\langle \mathcal{O}_{\chi_{c}}\left[^{3}P_{J}^{[1]}\right]\right\rangle $
is the color singlet long-distance matrix element (LDME) of $\chi_{c}$~\cite{Braaten:2002fi},
and the notations $\kappa,\,r$ were defined in~(\ref{eq:kappa}).
Similarly, we may obtain for the constants $c_{\gamma\chi_{c0}}^{(\lambda,\sigma)},\,d_{\gamma\chi_{c0}}^{(\lambda,\sigma)}$
of other helicity states
\begin{align}
d_{\gamma\chi_{c0}}^{(++)} & \left(r,\,\alpha,\,\zeta=x/\xi\right)=\frac{4\mathfrak{C}_{\chi_{c}}}{\alpha^{4}\left(1+\bar{\alpha}\right)^{2}\left(\bar{\alpha}+1/r^{2}\right)^{2}\left(r^{2}-1\right)^{2}\left(\zeta+1-i0\right)\left(\zeta-1+i0\right)\left(\zeta-\kappa+i0\right)^{2}}\times\\
\times & \left\{ \mathinner{\color{white}\frac{\frac{}{}}{\frac{}{}}}\frac{\bar{\alpha}\alpha^{2}}{8}\left[-5\left(1+\bar{\alpha}\right)^{2}\left(\alpha^{2}+2\bar{\alpha}\right)\zeta^{4}+\left(1+\bar{\alpha}\right)\left(8\alpha^{5}-45\alpha^{4}+100\alpha^{3}-126\alpha^{2}+92\alpha-40\right)\zeta^{3}\right.\right.\nonumber \\
 & \qquad+\left(16\alpha^{6}-91\alpha^{5}+212\alpha^{4}-254\alpha^{3}+156\alpha^{2}-24\alpha-24\right)\zeta^{2}-\alpha\left(8\alpha^{5}-31\alpha^{4}+42\alpha^{3}+2\alpha^{2}-56\alpha+40\right)\zeta\nonumber \\
 & \qquad-\left.\left(\alpha^{5}+7\alpha^{4}-12\alpha^{3}-26\alpha^{2}+64\alpha-32\right)\right]+\nonumber \\
 & +\frac{1}{8r^{2}}\left[\left(-\alpha^{7}-\alpha^{6}+29\alpha^{5}-88\alpha^{4}+124\alpha^{3}-96\alpha^{2}+32\alpha\right)\zeta^{4}+\right.\nonumber \\
 & \qquad+\left(-19\alpha^{8}+157\alpha^{7}-513\alpha^{6}+914\alpha^{5}-1054\alpha^{4}+872\alpha^{3}-520\alpha^{2}+160\alpha\right)\zeta^{3}\nonumber \\
 & \qquad+\left(69\alpha^{8}-449\alpha^{7}+1209\alpha^{6}-1772\alpha^{5}+1514\alpha^{4}-628\alpha^{3}-72\alpha^{2}+128\alpha\right)\zeta^{2}\nonumber \\
 & \qquad+\left(-49\alpha^{8}+239\alpha^{7}-463\alpha^{6}+370\alpha^{5}+82\alpha^{4}-352\alpha^{3}+176\alpha^{2}\right)\zeta\nonumber \\
 & \qquad\left.-\alpha^{8}-10\alpha^{7}-24\alpha^{6}+187\alpha^{5}-182\alpha^{4}-160\alpha^{3}+288\alpha^{2}-96\alpha\right]\nonumber \\
 & +\frac{1}{4r^{4}}\left[\left(3\alpha^{5}-15\alpha^{4}+24\alpha^{3}-12\alpha^{2}\right)\zeta^{4}+\left(8\alpha^{7}-70\alpha^{6}+214\alpha^{5}-317\alpha^{4}+276\alpha^{3}-172\alpha^{2}+96\alpha-32\right)\zeta^{3}+\right.\nonumber \\
 & \qquad+\left(-52\alpha^{7}+338\alpha^{6}-840\alpha^{5}+1105\alpha^{4}-928\alpha^{3}+508\alpha^{2}-96\alpha-32\right)\zeta^{2}\nonumber \\
 & \qquad+\left(56\alpha^{7}-276\alpha^{6}+508\alpha^{5}-419\alpha^{4}+8\alpha^{3}+280\alpha^{2}-160\alpha\right)\zeta\nonumber \\
 & \qquad+\left.4\alpha^{7}+16\alpha^{6}-5\alpha^{5}-242\alpha^{4}+404\alpha^{3}-116\alpha^{2}-96\alpha+32\right]\nonumber \\
 & +\frac{1}{8r^{6}}\left[\left(\alpha^{5}-12\alpha^{4}+44\alpha^{3}-64\alpha^{2}+32\alpha\right)\zeta^{4}+\left(-7\alpha^{6}+74\alpha^{5}-242\alpha^{4}+360\alpha^{3}-280\alpha^{2}+96\alpha\right)\zeta^{3}\right.+\nonumber \\
 & \qquad+\left(73\alpha^{6}-532\alpha^{5}+1238\alpha^{4}-1256\alpha^{3}+696\alpha^{2}-352\alpha+128\right)\zeta^{2}\nonumber \\
 & \qquad+\left(-121\alpha^{6}+642\alpha^{5}-1030\alpha^{4}+760\alpha^{3}-280\alpha^{2}-96\alpha+128\right)\zeta\nonumber \\
 & \qquad\left.-25\alpha^{6}-41\alpha^{5}+142\alpha^{4}+444\alpha^{3}-1032\alpha^{2}+512\alpha\right]\nonumber \\
 & +\frac{1}{4r^{8}}\left[\left(\alpha^{5}-11\alpha^{4}+36\alpha^{3}-60\alpha^{2}+64\alpha-32\right)\zeta^{3}+\left(-15\alpha^{5}+141\alpha^{4}-352\alpha^{3}+320\alpha^{2}-64\alpha-32\right)\zeta^{2}\right.+\nonumber \\
 & \qquad\left.+\left(35\alpha^{5}-233\alpha^{4}+296\alpha^{3}-68\alpha^{2}-32\right)\zeta+19\alpha^{5}+7\alpha^{4}-108\alpha^{3}-8\alpha^{2}+192\alpha-96\right]\nonumber \\
 & +\frac{\left(2\alpha^{4}-21\alpha^{3}+60\alpha^{2}-72\alpha+32\right)\zeta^{2}+\left(-7\alpha^{4}+62\alpha^{3}-78\alpha^{2}-8\alpha+32\right)\zeta-7\alpha^{4}+3\alpha^{3}+26\alpha^{2}-24\alpha}{2r^{10}}\nonumber \\
 & \left.-\bar{\alpha}\frac{\left(\alpha^{2}+8\bar{\alpha}\right)\zeta+\alpha^{2}}{r^{12}}\right\} +\left(\mathinner{\color{white}\frac{}{}}\zeta\to-\zeta\right)\nonumber 
\end{align}
\begin{align}
c_{\gamma\chi_{c0}}^{(+-)} & \left(r,\,\alpha,\,\zeta=x/\xi\right)=\frac{4\mathfrak{C}_{\chi_{c}}\left(1-\alpha\,r^{2}\right)}{\alpha\left(1+\bar{\alpha}\right)^{2}\left(\bar{\alpha}+1/r^{2}\right)\left(r^{2}-1\right)^{2}\left(\zeta+1-i0\right)\left(\zeta-1+i0\right)\left(\zeta-\kappa+i0\right)^{2}}\times\\
\times & \left\{ \mathinner{\color{white}\frac{\frac{}{}}{\frac{}{}}}-\bar{\alpha}\left[8\alpha\left(2-\alpha\right)^{2}\zeta^{4}+\left(16\alpha^{5}-87\alpha^{4}+174\alpha^{3}-192\alpha^{2}+112\alpha\right)\zeta^{3}-\alpha^{2}\left(32\alpha^{3}-109\alpha^{2}+142\alpha\right)\zeta^{2}\right.\right.\nonumber \\
 & \qquad+\left.92\alpha^{2}\zeta^{2}+\left(16\alpha^{5}-21\alpha^{4}+18\alpha^{3}+12\alpha^{2}+8\alpha-16\right)\zeta-\alpha^{4}+6\alpha^{3}+24\alpha^{2}-24\alpha\right]+\nonumber \\
 & +\frac{1}{r^{2}}\left[\left(4\alpha^{3}-16\alpha^{2}+16\alpha\right)\zeta^{4}+\left(-25\alpha^{5}+130\alpha^{4}-208\alpha^{3}+128\alpha^{2}-56\alpha+32\right)\zeta^{3}+\right.\nonumber \\
 & \qquad+\left(115\alpha^{5}-446\alpha^{4}+672\alpha^{3}-560\alpha^{2}+184\alpha+32\right)\zeta^{2}+\left(-91\alpha^{5}+182\alpha^{4}-172\alpha^{3}+16\alpha^{2}+72\alpha\right)\zeta+\nonumber \\
 & \qquad+\left.\alpha^{5}+6\alpha^{4}-72\alpha^{3}+48\alpha^{2}+56\alpha-32\right]\nonumber \\
 & +\frac{1}{r^{4}}\left[\left(-4\alpha^{3}+16\alpha^{2}-16\alpha\right)\zeta^{4}+\left(17\alpha^{4}-66\alpha^{3}+72\alpha^{2}-8\alpha-32\right)\zeta^{3}+\right.\nonumber \\
 & \qquad+\left(-135\alpha^{4}+386\alpha^{3}-308\alpha^{2}+152\alpha-96\right)\zeta^{2}\nonumber \\
 & \qquad+\left.\left(187\alpha^{4}-290\alpha^{3}+248\alpha^{2}-56\alpha-96\right)\zeta-5\alpha^{4}+6\alpha^{3}+132\alpha^{2}-152\alpha\right]\nonumber \\
 & +\frac{4}{r^{6}}\left[(-2\alpha^{3}+6\alpha^{2}-8\alpha+8)\zeta^{3}+(21\alpha^{3}-44\alpha^{2}+4\alpha+24)\zeta^{2}+(-45\alpha^{3}+34\alpha^{2}-12\alpha+24)\zeta\right.+\nonumber \\
 & \qquad\left.+2\alpha^{3}-12\alpha^{2}-12\alpha+24\right]\nonumber \\
 & \left.-\frac{4\left((8\alpha^{2}-20\,\alpha+16)\zeta^{2}+(6-25\alpha)\,\alpha\,\zeta+(\alpha-10)\,\alpha+20\,\zeta+8\right)}{r^{8}}+\frac{32\bar{\alpha}\zeta}{r^{10}}\right\} +\left(\mathinner{\color{white}\frac{}{}}\zeta\to-\zeta\right)\nonumber 
\end{align}
\begin{align}
d_{\gamma\chi_{c0}}^{(+-)} & \left(r,\,\alpha,\,\zeta=x/\xi\right)=-\frac{4\mathfrak{C}_{\chi_{c}}\left(1/r^{2}-\alpha\right)}{\alpha^{2}r^{2}\left(1+\bar{\alpha}\right)^{2}\left(\bar{\alpha}+1/r^{2}\right)^{2}\left(r^{2}-1\right)^{2}\left(\zeta+1-i0\right)\left(\zeta-1+i0\right)\left(\zeta-\kappa+i0\right)^{2}}\times\\
 & \times\left\{ \frac{\bar{\alpha}}{8}\left[4(\alpha-2)^{2}\alpha\zeta^{4}+\left(8\alpha^{5}-47\alpha^{4}+106\alpha^{3}-132\alpha^{2}+96\alpha-16\right)\zeta^{3}+\alpha^{2}\left(-16\alpha^{3}+61\alpha^{2}-94\alpha\right)\zeta^{2}\right.\right.\nonumber \\
 & \qquad\left.+68\alpha^{2}\zeta^{2}+\left(8\alpha^{5}-13\alpha^{4}+14\alpha^{3}-8\alpha^{2}+24\alpha-16\right)\zeta-\alpha\left(\alpha^{3}-2\alpha^{2}-24\alpha+24\right)\right]+\nonumber \\
 & +\frac{1}{8r^{2}}\left[\left(17\alpha^{5}-94\alpha^{4}+180\alpha^{3}-160\alpha^{2}+96\alpha-32\right)\zeta^{3}-\alpha^{5}-6\alpha^{4}+60\alpha^{3}-24\alpha^{2}-64\alpha+32\right.\nonumber \\
 & \qquad\left.+\left(-67\alpha^{5}+278\alpha^{4}-480\alpha^{3}+488\alpha^{2}-240\alpha+32\right)\zeta^{2}+\alpha\left(51\alpha^{4}-114\alpha^{3}+112\alpha^{2}-48\right)\zeta\right]\nonumber \\
 & +\frac{1}{8r^{4}}\left[4(\alpha-2)^{2}\alpha\zeta^{4}-\left(17\alpha^{4}-78\alpha^{3}+124\alpha^{2}-80\alpha+16\right)\zeta^{3}+\left(103\alpha^{4}-330\alpha^{3}+348\alpha^{2}-208\alpha\right)\zeta^{2}+\right.\nonumber \\
 & \qquad+\left.64\zeta^{2}+\left(-123\alpha^{4}+222\alpha^{3}-220\alpha^{2}+80+16\right)\zeta+\alpha\left(5\alpha^{3}-6\alpha^{2}-132\alpha+128\right)\right]\nonumber \\
 & +\frac{1}{2r^{6}}\left[2(2-\alpha)\left(\alpha^{2}-\alpha+2\right)\zeta^{3}-\alpha\left(21\alpha^{2}-56\alpha+36\right)\zeta^{2}\right.+\left(37\alpha^{3}-42\alpha^{2}+20\alpha-8\right)\zeta+\nonumber \\
 & \qquad\left.-\left(2\alpha^{3}-12\alpha^{2}-16\alpha+16\right)\right]\nonumber \\
 & \qquad\left.-\frac{4\left(2\alpha^{2}-5\alpha+4\right)\zeta^{2}-\left(25\alpha^{2}-18\alpha+8\right)\zeta+\left(\alpha^{2}-10\alpha\right)}{2r^{8}}-\frac{4\bar{\alpha}\zeta}{r^{10}}\right\} +\left(\mathinner{\color{white}\frac{}{}}\zeta\to-\zeta\right)\nonumber 
\end{align}
\begin{align}
\tilde{c}_{\gamma\chi_{c0}}^{(++)} & \left(r,\,\alpha,\,\zeta=x/\xi\right)=-\frac{4\mathfrak{C}_{\chi_{c}}}{\alpha^{2}\left(1+\bar{\alpha}\right)^{2}\left(\bar{\alpha}+1/r^{2}\right)\left(r^{2}-1\right)^{2}\left(\zeta+1-i0\right)\left(\zeta-1+i0\right)\left(\zeta-\kappa+i0\right)^{2}}\times\\
\times & \left\{ \mathinner{\color{white}\frac{\frac{}{}}{\frac{}{}}}-\alpha\left(\alpha^{5}-29\alpha^{4}+137\alpha^{3}-262\alpha^{2}+212\alpha-56\right)\zeta^{4}-\alpha\left(3\alpha^{6}-33\alpha^{5}+141\alpha^{4}-256\alpha^{3}+116\alpha^{2}+24\alpha\right)\zeta^{3}\right.\nonumber \\
 & \qquad-\alpha\left(3\alpha^{6}+5\alpha^{5}-5\alpha^{4}-178\alpha^{3}+280\alpha^{2}-236\alpha+120\right)\zeta^{2}+\nonumber \\
 & \qquad+\alpha\left(15\alpha^{6}-117\alpha^{5}+421\alpha^{4}-876\alpha^{3}+1088\alpha^{2}-648\alpha+112\right)\zeta\nonumber \\
 & \qquad-\alpha\left(9\alpha^{6}-90\alpha^{5}+330\alpha^{4}-435\alpha^{3}+186\alpha^{2}+24\alpha-16\right)\nonumber \\
 & +\frac{1}{r^{2}}\left[\left(-4\alpha^{5}+12\alpha^{4}+4\alpha^{3}-40\alpha^{2}+32\alpha\right)\zeta^{4}+\left(6\alpha^{6}-60\alpha^{5}+296\alpha^{4}-712\alpha^{3}+792\alpha^{2}-448\alpha+128\right)\zeta^{3}+\right.\nonumber \\
 & \qquad+\left(6\alpha^{6}+24\alpha^{5}-140\alpha^{4}-20\alpha^{3}+176\alpha^{2}-208\alpha+128\right)\zeta^{2}\nonumber \\
 & \qquad+\left(-78\alpha^{6}+320\alpha^{5}-508\alpha^{4}+488\alpha^{3}-600\alpha^{2}+368\alpha\right)\zeta\nonumber \\
 & \qquad+\left.66\alpha^{6}-200\alpha^{5}+484\alpha^{4}-704\alpha^{3}+408\alpha^{2}-32\alpha\right]\nonumber \\
 & +\frac{1}{r^{4}}\left[\left(5\alpha^{4}-26\alpha^{3}+44\alpha^{2}-24\alpha\right)\zeta^{4}+\left(-3\alpha^{5}-20\alpha^{3}+144\alpha^{2}-192\alpha+64\right)\zeta^{3}\right.\nonumber \\
 & -\left(3\alpha^{5}+30\alpha^{4}-320\alpha^{3}+616\alpha^{2}-560\alpha+192\right)\zeta^{2}\nonumber \\
 & \qquad+\left.\left(147\alpha^{5}-524\alpha^{4}+564\alpha^{3}-240\alpha^{2}+384\alpha-256\right)\zeta-189\alpha^{5}+309\alpha^{4}-246\alpha^{3}+428\alpha^{2}-280\alpha\right]\nonumber \\
\nonumber \\ & -\frac{4}{r^{6}}\left[8(2-\alpha)\bar{\alpha}^{2}\zeta^{3}+\left(13\alpha^{3}-2\alpha^{2}-52\alpha+48\right)\zeta^{2}+\left(30\alpha^{4}-102\alpha^{3}+56\alpha^{2}+48\alpha\right)\zeta\right.+\nonumber \\
 & \qquad\left.-66\alpha^{4}+93\alpha^{3}-18\alpha^{2}+56\alpha-32\right]\nonumber \\
 & \left.+\frac{4\left[\left(17\alpha^{2}-46\alpha+32\right)\zeta^{2}+\left(9\alpha^{3}-40\alpha^{2}+4\alpha+48\right)\zeta-45\alpha^{3}+59\alpha^{2}+10\alpha+16\right]}{r^{8}}+16\frac{3\alpha^{2}+(1+3\bar{\alpha})\bar{\zeta}}{r^{10}}\right\} \nonumber \\
 & +\left(\mathinner{\color{white}\frac{}{}}\zeta\to-\zeta\right)\nonumber 
\end{align}
\begin{align}
\tilde{d}_{\gamma\chi_{c0}}^{(++)} & \left(r,\,\alpha,\,\zeta=x/\xi\right)=\frac{4\mathfrak{C}_{\chi_{c}}}{\alpha^{4}\left(1+\bar{\alpha}\right)^{2}\left(\bar{\alpha}+1/r^{2}\right)^{2}\left(r^{2}-1\right)^{2}\left(\zeta+1-i0\right)\left(\zeta-1+i0\right)\left(\zeta-\kappa+i0\right)^{2}}\times\\
\times & \left\{ \mathinner{\color{white}\frac{\frac{}{}}{\frac{}{}}}\frac{\bar{\alpha}\alpha^{2}}{8}\left[(\alpha-2)^{2}\alpha(5\alpha-4)\zeta^{4}+\left(3\alpha^{5}-28\alpha^{4}+72\alpha^{3}-72\alpha^{2}+32\alpha\right)\zeta^{3}\right.\right.\nonumber \\
 & \qquad+\left(3\alpha^{5}-34\alpha^{4}+132\alpha^{3}-184\alpha^{2}+112\alpha-16\right)\zeta^{2}+\left(-15\alpha^{5}+84\alpha^{4}-188\alpha^{3}+224\alpha^{2}-96\alpha\right)\zeta\nonumber \\
 & \qquad+\left.9\alpha^{5}-27\alpha^{4}+24\alpha^{3}-4\alpha^{2}+32\alpha-32\right]+\nonumber \\
 & +\frac{1}{8r^{2}}\left[(\alpha-2)^{2}\alpha^{2}\left(\alpha^{3}-23\alpha^{2}+45\alpha-22\right)\zeta^{4}+\right.\left(\alpha^{8}-19\alpha^{7}+127\alpha^{6}-284\alpha^{5}+164\alpha^{4}+120\alpha^{3}-112\alpha^{2}\right)\zeta^{3}\nonumber \\
 & \qquad+\left(\alpha^{8}-19\alpha^{7}+127\alpha^{6}-284\alpha^{5}+164\alpha^{4}+120\alpha^{3}-112\alpha^{2}\right)\zeta^{3}\nonumber \\
 & \qquad+\left(\alpha^{8}+15\alpha^{7}-47\alpha^{6}-150\alpha^{5}+444\alpha^{4}-356\alpha^{3}+88\alpha^{2}\right)\zeta^{2}\nonumber \\
 & \qquad+\left(-5\alpha^{8}+55\alpha^{7}-263\alpha^{6}+592\alpha^{5}-816\alpha^{4}+568\alpha^{3}-128\alpha^{2}\right)\zeta\nonumber \\
 & \qquad\left.+3\alpha^{8}-52\alpha^{7}+194\alpha^{6}-299\alpha^{5}+86\alpha^{4}+104\alpha^{3}-32\alpha\right]\nonumber \\
 & +\frac{1}{4r^{4}}\left[(\alpha-2)^{2}\alpha\left(\alpha^{3}+5\alpha^{2}-16\alpha+8\right)\zeta^{4}+\right.\nonumber \\
 & \qquad-\left(\alpha^{7}-20\alpha^{6}+150\alpha^{5}-430\alpha^{4}+520\alpha^{3}-248\alpha^{2}+32\alpha\right)\zeta^{3}\\
 & \qquad+\left(-5\alpha^{7}-8\alpha^{6}+82\alpha^{5}+72\alpha^{4}-424\alpha^{3}+384\alpha^{2}-96\alpha\right)\zeta^{2}+\nonumber \\
 & \qquad+\left(21\alpha^{7}-90\alpha^{6}+228\alpha^{5}-342\alpha^{4}+376\alpha^{3}-184\alpha^{2}\right)\zeta\nonumber \\
 & \qquad+\left.-15\alpha^{7}+69\alpha^{6}-177\alpha^{5}+328\alpha^{4}-140\alpha^{3}-96\alpha^{2}+32\alpha\right]\nonumber \\
 & +\frac{1}{8r^{6}}\left[3\alpha^{2}\left(2-\alpha\right)^{3}\zeta^{4}+\left(\alpha^{6}-4\alpha^{5}+92\alpha^{4}-440\alpha^{3}+752\alpha^{2}-512\alpha+128\right)\zeta^{3}\right.+\nonumber \\
 & \qquad+\left(17\alpha^{6}+10\alpha^{5}-404\alpha^{4}+784\alpha^{3}-400\alpha^{2}-128\alpha+128\right)\zeta^{2}\nonumber \\
 & \qquad+\left(-113\alpha^{6}+416\alpha^{5}-804\alpha^{4}+1080\alpha^{3}-1168\alpha^{2}+512\alpha\right)\zeta\nonumber \\
 & \qquad\left.+111\alpha^{6}-291\alpha^{5}+378\alpha^{4}-876\alpha^{3}+504\alpha^{2}+64\alpha\right]\nonumber \\
 & +\frac{1}{2r^{8}}\left[+\left(-5\alpha^{4}+24\alpha^{3}-36\alpha^{2}+16\alpha\right)\zeta^{3}+\left(-2\alpha^{5}+8\alpha^{4}+32\alpha^{3}-148\alpha^{2}+176\alpha-64\right)\zeta^{2}\right.+\nonumber \\
 & \qquad\left.\left(30\alpha^{5}-97\alpha^{4}+112\alpha^{3}-68\alpha^{2}+112\alpha-64\right)\zeta-48\alpha^{5}+106\alpha^{4}-76\alpha^{3}+200\alpha^{2}-112\alpha\right]\nonumber \\
 & +\frac{\left(-11\alpha^{3}+38\alpha^{2}-32\alpha\right)\zeta^{2}+\left(-11\alpha^{4}+36\alpha^{3}-12\alpha^{2}-64\alpha+32\right)\zeta+39\alpha^{4}-73\alpha^{3}+10\alpha^{2}-72\alpha+32}{2r^{10}}\nonumber \\
 & \left.+\frac{-6\alpha^{3}+8\alpha^{2}+8\alpha+\left(8\alpha-4\alpha^{2}\right)\zeta}{r^{12}}\right\} +\left(\mathinner{\color{white}\frac{}{}}\zeta\to-\zeta\right)\nonumber 
\end{align}
\begin{align}
\tilde{c}_{\gamma\chi_{c0}}^{(+-)} & \left(r,\,\alpha,\,\zeta=x/\xi\right)=-\frac{4\mathfrak{C}_{\chi_{c}}\left(1-\alpha\,r^{2}\right)}{\alpha\left(1+\bar{\alpha}\right)^{2}\left(\bar{\alpha}+1/r^{2}\right)\left(r^{2}-1\right)^{2}\left(\zeta+1-i0\right)\left(\zeta-1+i0\right)\left(\zeta-\kappa+i0\right)^{2}}\times\\
\times & \left\{ \mathinner{\color{white}\frac{\frac{}{}}{\frac{}{}}}\bar{\alpha}\left[\alpha\left(2\alpha^{2}-16\right)\zeta^{4}+\alpha\left(5\alpha^{3}-34\alpha^{2}+52\alpha-48\right)\zeta^{3}+\alpha\left(5\alpha^{3}-46\alpha^{2}+104\alpha-72\right)\zeta^{2}\right.\right.\nonumber \\
 & \qquad+\left.\alpha\left(-25\alpha^{3}+106\alpha^{2}-88\alpha+8\right)\zeta+\left(15\alpha^{4}-28\alpha^{3}+76\alpha^{2}-112\alpha+48\right)\right]+\nonumber \\
 & +\frac{1}{r^{2}}\left[\left(-2\alpha^{4}+14\alpha^{3}-40\alpha^{2}+32\alpha\right)\zeta^{4}+\left(-3\alpha^{5}+30\alpha^{4}-148\alpha^{3}+268\alpha^{2}-248\alpha+112\right)\zeta^{3}+\right.\nonumber \\
 & \qquad+\left(-3\alpha^{5}-14\alpha^{4}+76\alpha^{3}-44\alpha^{2}-112\alpha+112\right)\zeta^{2}+\left(15\alpha^{5}-86\alpha^{4}+360\alpha^{3}-524\alpha^{2}+248\alpha\right)\zeta+\nonumber \\
 & \qquad+\left.-9\alpha^{5}+72\alpha^{4}-62\alpha^{3}+20\alpha^{2}+\zeta^{2}+\zeta-80\alpha+64\right]\nonumber \\
 & +\frac{1}{r^{4}}\left[-4\alpha\left(1+\bar{\alpha}\right)^{2}\zeta^{4}+\left(3\alpha^{4}-6\alpha^{3}+36\alpha^{2}-56\alpha\right)\zeta^{3}+\left(3\alpha^{4}+42\alpha^{3}-272\alpha^{2}+416\alpha-224\right)\zeta^{2}\right.\nonumber \\
 & \qquad+\left.\left(-63\alpha^{4}+98\alpha^{3}-212\alpha^{2}+360\alpha-224\right)\zeta+57\alpha^{4}-178\alpha^{3}+16\alpha^{2}+176\alpha-96\right]\nonumber \\
 & +\frac{4\left[\left(-5\alpha^{2}+12\alpha-4\right)\zeta^{3}+\left(7\alpha^{2}+2\alpha-4\right)\zeta^{2}+\left(21\alpha^{3}-7\alpha^{2}-36\alpha+28\right)\zeta-33\alpha^{3}+69\alpha^{2}-22\alpha-4\right]}{r^{6}}\nonumber \\
 & \left.+\frac{4\left(8\bar{\alpha}\zeta^{2}+\left(-9\alpha^{2}+6\alpha+8\right)\zeta+\left(33\alpha^{2}-50\alpha+12\right)\right)}{r^{8}}-\frac{16\,\left(\zeta+3\bar{\alpha}\right)}{r^{10}}\right\} +\left(\mathinner{\color{white}\frac{}{}}\zeta\to-\zeta\right)\nonumber 
\end{align}
\begin{align}
\tilde{d}_{\gamma\chi_{c0}}^{(+-)} & \left(r,\,\alpha,\,\zeta=x/\xi\right)=-\frac{4\mathfrak{C}_{\chi_{c}}\left(1/r^{2}-\alpha\right)}{\alpha^{3}\left(1+\bar{\alpha}\right)^{2}\left(\bar{\alpha}+1/r^{2}\right)^{2}\left(r^{2}-1\right)^{2}\left(\zeta+1-i0\right)\left(\zeta-1+i0\right)\left(\zeta-\kappa+i0\right)^{2}}\times\\
 & \times\left\{ \mathinner{\color{white}\frac{\frac{}{}}{\frac{}{}}}-\frac{\alpha\bar{\alpha}}{8}\left[\left(6\alpha^{3}-16\alpha^{2}+8\alpha\right)\zeta^{4}+\left(3\alpha^{4}-30\alpha^{3}+52\alpha^{2}-32\alpha\right)\zeta^{3}+\left(3\alpha^{4}-30\alpha^{3}+76\alpha^{2}-40\alpha-16\right)\zeta^{2}\right.\right.\nonumber \\
 & \qquad\left.+\left(-15\alpha^{4}+78\alpha^{3}-88\alpha^{2}+24\alpha\right)\zeta+\left(9\alpha^{4}-24\alpha^{3}+56\alpha^{2}-88\alpha+48\right)\right]+\nonumber \\
 & +\frac{1}{8r^{2}}\left[16\alpha\bar{\alpha}^{2}\left(2-\alpha\right)\zeta^{4}+\left(\alpha^{6}-14\alpha^{5}+128\alpha^{4}-304\alpha^{3}+288\alpha^{2}-96\alpha\right)\zeta^{3}\right.\nonumber \\
 & \qquad+\left(\alpha^{6}+22\alpha^{5}-88\alpha^{4}+192\alpha^{2}-128\alpha\right)\zeta^{2}+\left(-5\alpha^{6}+30\alpha^{5}-228\alpha^{4}+416\alpha^{3}-192\alpha^{2}-32\alpha\right)\zeta\nonumber \\
 & \qquad\left.+3\alpha^{6}-38\alpha^{5}+60\alpha^{4}+64\alpha^{2}-160\alpha+64\right]\nonumber \\
 & +\frac{1}{8r^{4}}\left[2\alpha^{2}\left(2-\alpha\right)^{2}\zeta^{4}+\left(-\alpha^{5}+2\alpha^{4}-84\alpha^{3}+240\alpha^{2}-224\alpha+64\right)\zeta^{3}+\left(-9\alpha^{5}-22\alpha^{4}+284\alpha^{3}-432\alpha^{2}+112\alpha+64\right)\zeta^{2}+\right.\nonumber \\
 & \qquad+\left.\left(37\alpha^{5}-30\alpha^{4}+140\alpha^{3}-464\alpha^{2}+320\alpha\right)\zeta-27\alpha^{5}+96\alpha^{4}-124\alpha^{3}-184\alpha^{2}+240\alpha\right]\nonumber \\
 & +\frac{1}{2r^{6}}\left[4\alpha\bar{\alpha}\left(2-\alpha\right)\zeta^{3}+\left(2\alpha^{4}-8\alpha^{3}-28\alpha^{2}+72\alpha-32\right)\zeta^{2}\right.\left(-19\alpha^{4}+16\alpha^{3}+28\alpha^{2}+16\alpha-32\right)\zeta\nonumber \\
 & \qquad\left.+\left(21\alpha^{4}-40\alpha^{3}+76\alpha^{2}-32\right)\right]\nonumber \\
 & \left.+\frac{\left(10\alpha^{2}-16\alpha\right)\zeta^{2}+\left(11\alpha^{3}-18\alpha^{2}-24\alpha+16\right)\zeta-27\alpha^{3}+32\alpha^{2}-64\alpha+16}{2r^{8}}+\frac{2\left(3\alpha^{2}-2\alpha+4\right)+4\alpha\zeta}{r^{10}}\right\} +\left(\mathinner{\color{white}\frac{}{}}\zeta\to-\zeta\right)\nonumber 
\end{align}

\subsection{Coefficient functions for $\chi_{c1}$}

\begin{align}
c_{\gamma\chi_{c1}}^{(++,-1)} & \left(r,\,\alpha,\,\zeta=x/\xi\right)=\frac{4i\mathfrak{C}_{\chi_{c}}\sqrt{\left(1-\alpha\right)\left(\alpha\,r^{2}-1\right)}}{\alpha\left(1+\bar{\alpha}\right)^{2}\left(\bar{\alpha}+1/r^{2}\right)\left(r^{2}-1\right)^{2}\left(\zeta+1-i0\right)\left(\zeta-1+i0\right)\left(\zeta-\kappa+i0\right)^{2}}\times\\
\times & \left\{ \mathinner{\color{white}\frac{\frac{}{}}{\frac{}{}}}-4\alpha^{7}\bar{\zeta}^{2}\zeta+\alpha^{6}\left(25\zeta^{3}-23\zeta^{2}+3\zeta-5\right)-\alpha^{5}\left(67\zeta^{3}-9\zeta^{2}-7\zeta-3\right)+\alpha^{4}\left(15\zeta^{4}+49\zeta^{3}+103\zeta^{2}-17\zeta-22\right)\right.\nonumber \\
 & \qquad+\alpha^{3}\left(-90\zeta^{4}+16\zeta^{3}-208\zeta^{2}+4\zeta+54\right)+4\alpha^{2}\left(33\zeta^{4}-21\zeta^{3}+20\zeta^{2}+12\zeta+4\right)\nonumber \\
 & \qquad-8\alpha\left(7\zeta^{4}-20\zeta^{3}-15\zeta^{2}+6\zeta+10\right)-32\left(3\zeta^{3}+3\zeta^{2}-1\right)\nonumber \\
 & \qquad-\alpha\bar{\alpha}r^{2}\left(4\alpha^{5}\bar{\zeta}^{2}\zeta+\alpha^{4}\left(-23\zeta^{3}+33\zeta^{2}-13\zeta+3\right)+\alpha^{3}\left(\zeta^{4}+58\zeta^{3}-68\zeta^{2}+30\zeta-5\right)\right.\nonumber \\
 & \qquad\left.-2\alpha^{2}\left(7\zeta^{4}+31\zeta^{3}-35\zeta^{2}+23\zeta-10\right)+4\alpha\left(13\zeta^{4}+12\zeta^{3}-9\zeta^{2}+8\zeta-8\right)-8\left(5\zeta^{4}+3\zeta^{3}-2\zeta^{2}-2\right)\right)\nonumber \\
 & +\frac{1}{r^{2}}\left[2\alpha^{6}\left(5\zeta^{3}-17\zeta^{2}+11\zeta+1\right)-\alpha^{5}\left(\zeta^{4}+54\zeta^{3}-136\zeta^{2}+50\zeta-17\right)-\alpha^{4}\left(\zeta^{4}-138\zeta^{3}+248\zeta^{2}-74\zeta+27\right)\right.\nonumber \\
 & \qquad+2\alpha^{3}\left(15\zeta^{4}-50\zeta^{3}+72\zeta^{2}-44\zeta+47\right)-4\alpha^{2}\left(13\zeta^{4}+11\zeta^{3}-41\zeta^{2}-31\zeta+64\right)\nonumber \\
 & \qquad+\left.8\alpha\left(3\zeta^{4}-48\zeta^{2}-31\zeta+22\right)+48\zeta\left(\zeta^{2}+5\zeta+4\right)\right]\nonumber \\
 & +\frac{1}{r^{4}}\left[-4\alpha^{5}\left(2\zeta^{3}-13\zeta^{2}+12\zeta+3\right)+\alpha^{4}\left(\zeta^{4}+35\zeta^{3}-167\zeta^{2}+69\zeta-2\right)-2\alpha^{3}\left(3\zeta^{4}+40\zeta^{3}-112\zeta^{2}+34\zeta-5\right)\right.\nonumber \\
 & \qquad\left.+4\alpha^{2}\left(3\zeta^{4}+21\zeta^{3}-27\zeta^{2}+15\zeta-8\right)-8\alpha\left(\zeta^{4}-14\zeta^{2}-18\zeta-25\right)-32\left(\zeta^{3}+4\zeta^{2}+6\zeta+5\right)\right]\nonumber \\
 & +\frac{2}{r^{6}}\left[\alpha^{4}\left(\zeta^{3}-17\zeta^{2}+27\zeta+13\right)-\alpha^{3}\left(5\zeta^{3}-49\zeta^{2}+11\zeta+25\right)+2\alpha^{2}\left(5\zeta^{3}-22\zeta^{2}-17\zeta+6\right)\right.\nonumber \\
 & \qquad\left.-12\alpha\left(\zeta^{3}+2\zeta^{2}+2\zeta+3\right)+8(\zeta+1)^{2}(\zeta+3)\right]\nonumber \\
 & \left.+\frac{4\left(2\alpha^{3}\left(\zeta^{2}-4\zeta-3\right)+\alpha^{2}\left(-9\zeta^{2}+5\zeta+16\right)+2\alpha\left(7\zeta^{2}+8\zeta-3\right)-8\zeta(\zeta+2)\right)}{r^{8}}+\frac{8\left(\alpha^{2}-3\alpha+2\right)(\zeta+1)}{r^{10}}\right\} \nonumber \\
 & +\left(\mathinner{\color{white}\frac{}{}}\zeta\to-\zeta\right)\nonumber 
\end{align}
\begin{align}
d_{\gamma\chi_{c1}}^{(++,-1)} & \left(r,\,\alpha,\,\zeta=x/\xi\right)=\frac{4i\mathfrak{C}_{\chi_{c}}\sqrt{\left(1-\alpha\right)\left(\alpha\,r^{2}-1\right)}}{\alpha^{4}\left(1+\bar{\alpha}\right)^{2}\left(\bar{\alpha}+1/r^{2}\right)^{2}\left(r^{2}-1\right)^{2}\left(\zeta+1-i0\right)\left(\zeta-1+i0\right)\left(\zeta-\kappa+i0\right)^{2}}\times\\
\times & \left\{ \mathinner{\color{white}\frac{\frac{}{}}{\frac{}{}}}\frac{\bar{\alpha}\alpha^{2}}{8}\left[4\alpha^{5}\bar{\zeta}^{2}\zeta+\alpha^{4}\bar{\zeta}\left(27\zeta^{2}-14\zeta+3\right)+\alpha^{3}\left(\zeta^{4}+82\zeta^{3}-100\zeta^{2}+38\zeta-5\right)+\right.\right.\nonumber \\
 & \qquad-\left.2\alpha^{2}\left(7\zeta^{4}+59\zeta^{3}-59\zeta^{2}+27\zeta-10\right)+4\alpha\left(9\zeta^{4}+24\zeta^{3}-13\zeta^{2}+12\zeta-8\right)-8\left(3\zeta^{4}+5\zeta^{3}+2\zeta-2\right)\right]+\nonumber \\
 & +\frac{\alpha}{8r^{2}}\left[-4\alpha^{7}\bar{\zeta}^{2}\zeta-\alpha^{6}\bar{\zeta}(3\zeta+1)(3\zeta+5)+\alpha^{5}\left(37\zeta^{3}-151\zeta^{2}+63\zeta+3\right)+\alpha^{4}\left(7\zeta^{4}-215\zeta^{3}+439\zeta^{2}-105\zeta-30\right)\right.\nonumber \\
 & \qquad+\alpha^{3}\left(-10\zeta^{4}+404\zeta^{3}-632\zeta^{2}+72\zeta+6\right)-4\alpha^{2}\left(9\zeta^{4}+113\zeta^{3}-96\zeta^{2}+4\zeta-38\right)\nonumber \\
 & \qquad\left.+8\alpha\left(9\zeta^{4}+44\zeta^{3}+5\zeta^{2}+4\zeta-24\right)-32\left((\zeta+1)(\zeta+3)\zeta^{2}+\zeta-2\right)\right]\nonumber \\
 & +\frac{1}{8r^{4}}\left[-2\alpha^{7}\bar{\zeta}\left(5\zeta^{2}-12\zeta-1\right)-\alpha^{6}(\zeta+1)\left(\zeta^{3}+21\zeta^{2}-29\zeta-17\right)-\alpha^{5}\left(\zeta^{4}+46\zeta^{3}-280\zeta^{2}+158\zeta+27\right)+\right.\nonumber \\
 & \qquad+2\alpha^{4}\left(7\zeta^{4}+118\zeta^{3}-336\zeta^{2}+68\zeta+111\right)-4\alpha^{3}\left(5\zeta^{4}+71\zeta^{3}-233\zeta^{2}-31\zeta+100\right)+\nonumber \\
 & \qquad+\left.8\alpha^{2}\left(\zeta^{4}+22\zeta^{3}-96\zeta^{2}-53\zeta+8\right)+16\alpha\left(-9\zeta^{3}+11\zeta^{2}+16\zeta+12\right)+64\left(\zeta^{3}+\zeta^{2}-1\right)\right]\nonumber \\
 & +\frac{1}{8r^{6}}\left[-4\alpha^{6}\left(2\zeta^{3}-13\zeta^{2}+12\zeta+3\right)+\alpha^{5}\left(\zeta^{4}+3\zeta^{3}+25\zeta^{2}-155\zeta-2\right)+2\alpha^{4}\left(\zeta^{4}+42\zeta^{3}-228\zeta^{2}+192\zeta-15\right)\right.+\nonumber \\
 & \qquad-4\alpha^{3}\left(7\zeta^{4}+55\zeta^{3}-153\zeta^{2}+73\zeta+70\right)+8\alpha^{2}\left(7\zeta^{4}+26\zeta^{3}-46\zeta^{2}+20\zeta+93\right)\nonumber \\
 & \qquad\left.-32\alpha\left(\zeta^{4}+2\zeta^{3}-9\zeta^{2}-3\zeta+13\right)-128\zeta(\zeta+1)\right]\nonumber \\
 & +\frac{1}{4r^{8}}\left[\alpha^{5}\left(\zeta^{3}-17\zeta^{2}+27\zeta+13\right)+\alpha^{4}\left(3\zeta^{3}-31\zeta^{2}+125\zeta-25\right)+\alpha^{3}(76-2\zeta(11(\zeta-10)\zeta+125))\right.+\nonumber \\
 & \qquad\left.+4\alpha^{2}\left(11\zeta^{3}-62\zeta^{2}+16\zeta-1\right)-8\alpha\left(7\zeta^{3}-5\zeta^{2}+\zeta+21\right)+32\left(\zeta^{3}+\zeta^{2}+\zeta+3\right)\right]\nonumber \\
 & +\frac{2\alpha^{4}((\zeta-4)\zeta-3)+\alpha^{3}(\zeta(7\zeta-43)+16)-6\alpha^{2}(7(\zeta-2)\zeta+5)+8\alpha\left(8\zeta^{2}+3\right)-32\zeta(\zeta+1)}{2r^{10}}\nonumber \\
 & \left.-\frac{\bar{\alpha}(\alpha((\alpha+6)\zeta+\alpha-2)-8\zeta)}{r^{12}}\right\} +\left(\mathinner{\color{white}\frac{}{}}\zeta\to-\zeta\right)\nonumber 
\end{align}
\begin{align}
c_{\gamma\chi_{c1}}^{(+-,-1)} & \left(r,\,\alpha,\,\zeta=x/\xi\right)=\frac{4i\mathfrak{C}_{\chi_{c}}\sqrt{\left(1-\alpha\right)\left(\alpha\,r^{2}-1\right)}\left(\alpha r^{2}-1\right)}{\alpha\left(1+\bar{\alpha}\right)^{2}\left(r^{2}-1\right)^{2}\left(\zeta+1-i0\right)\left(\zeta-1+i0\right)\left(\zeta-\kappa+i0\right)^{2}}\times\\
\times & \left\{ \mathinner{\color{white}\frac{\frac{}{}}{\frac{}{}}}\alpha^{3}\left(-(\zeta-1)^{2}\right)(3\zeta+1)+4\alpha^{2}\left(5\zeta^{3}-7\zeta^{2}+3\zeta-1\right)-4\alpha\left(11\zeta^{3}-9\zeta^{2}+4\zeta-2\right)+16\left(2\zeta^{3}+\zeta\right)\right.\nonumber \\
 & -\frac{\alpha^{3}\bar{\zeta}^{2}(5\zeta-1)-2\alpha^{2}\bar{\zeta}^{2}(9\zeta+5)+4\alpha\zeta\left(3\zeta^{2}-11\zeta+4\right)+8\left(\zeta^{3}+9\zeta^{2}+2\zeta+2\right)}{r^{2}}\nonumber \\
 & \qquad+\frac{2\alpha^{2}\left(\zeta^{3}-13\zeta^{2}+15\zeta-3\right)+4\alpha\left(-2\zeta^{3}+10\zeta^{2}+\zeta-5\right)+8\left(\zeta^{3}+3\zeta^{2}+4\zeta+2\right)}{r^{4}}\nonumber \\
 & \qquad\left.+\frac{4\left(\alpha\left(2\zeta^{2}-9\zeta+3\right)-2\left(2\zeta^{2}+\zeta-1\right)\right)}{r^{6}}-\frac{8\bar{\zeta}}{r^{8}}\right\} +\left(\mathinner{\color{white}\frac{}{}}\zeta\to-\zeta\right)\nonumber 
\end{align}
\begin{align}
d_{\gamma\chi_{c1}}^{(+-,-1)} & \left(r,\,\alpha,\,\zeta=x/\xi\right)=\frac{4i\mathfrak{C}_{\chi_{c}}\sqrt{\left(1-\alpha\right)\left(\alpha\,r^{2}-1\right)}\left(\alpha-1/r^{2}\right)}{\alpha^{2}\left(1+\bar{\alpha}\right)\left(\bar{\alpha}+1/r^{2}\right)\left(r^{2}-1\right)^{2}\left(\zeta+1-i0\right)\left(\zeta-1+i0\right)\left(\zeta-\kappa+i0\right)}\times\\
 & \times\mathinner{\color{white}\frac{\frac{}{}}{\frac{}{}}}\frac{1}{8}\left[(\alpha-2)\zeta^{2}\left(3\alpha-4+\frac{\alpha-2}{r^{2}}-\frac{2}{r^{4}}\right)-2\zeta\left(\alpha-\frac{2}{r^{2}}\right)\left(\alpha-3+\frac{\alpha}{r^{2}}-\frac{1}{r^{4}}\right)\right.-\nonumber \\
 & \qquad\left.-\left(1-\frac{1}{r^{2}}\right)\left(\alpha(\alpha+4)-8-\frac{4\alpha}{r^{2}}+\frac{4}{r^{4}}\right)\right]+\left(\mathinner{\color{white}\frac{}{}}\zeta\to-\zeta\right)\nonumber 
\end{align}
\begin{align}
c_{\gamma\chi_{c1}}^{(++,0)} & \left(r,\,\alpha,\,\zeta=x/\xi\right)=-\frac{4i\mathfrak{C}_{\chi_{c}}}{\alpha^{2}\left(1+\bar{\alpha}\right)^{2}\left(\bar{\alpha}+1/r^{2}\right)\left(r^{2}-1\right)^{2}\left(\zeta+1-i0\right)\left(\zeta-1+i0\right)\left(\zeta-\kappa+i0\right)^{2}}\times\\
\times & \left\{ \mathinner{\color{white}\frac{\frac{}{}}{\frac{}{}}}\left(\alpha^{6}-9\alpha^{5}+5\alpha^{4}+70\alpha^{3}-140\alpha^{2}+72\alpha\right)\zeta^{4}+\alpha^{7}+10\alpha^{6}+72\alpha^{5}-195\alpha^{4}+10\alpha^{3}+168\alpha^{2}-64\alpha\right.\nonumber \\
 & \qquad+\left(35\alpha^{7}-227\alpha^{6}+557\alpha^{5}-650\alpha^{4}+508\alpha^{3}-448\alpha^{2}+224\alpha\right)\zeta^{3}+\zeta^{2}\left(-117\alpha^{7}+597\alpha^{6}-1283\alpha^{5}\right.\nonumber \\
 & \qquad+\left.1614\alpha^{4}-1220\alpha^{3}+324\alpha^{2}+88\alpha\right)+\left(81\alpha^{7}-301\alpha^{6}+495\alpha^{5}-350\alpha^{4}-120\alpha^{3}+216\alpha^{2}-16\alpha\right)\zeta\nonumber \\
 & +\frac{1}{r^{2}}\left[\left(4\alpha^{5}-12\alpha^{4}-8\alpha^{3}+48\alpha^{2}-32\alpha\right)\zeta^{4}+\left(-44\alpha^{6}+242\alpha^{5}-474\alpha^{4}+314\alpha^{3}+52\alpha^{2}-8\alpha-80\right)\zeta^{3}\right.\nonumber \\
 & \qquad+\left(224\alpha^{6}-916\alpha^{5}+1404\alpha^{4}-1286\alpha^{3}+1052\alpha^{2}-408\alpha-80\right)\zeta^{2}-8\alpha^{6}-48\alpha^{5}-120\alpha^{4}\nonumber \\
 & \qquad+\left.\left(-220\alpha^{6}+614\alpha^{5}-838\alpha^{4}+686\alpha^{3}+60\alpha^{2}-320\alpha\right)\zeta+550\alpha^{3}-332\alpha^{2}-80\alpha+32\right]\nonumber \\
 & +\frac{1}{r^{4}}\left[\left(-5\alpha^{4}+26\alpha^{3}-44\alpha^{2}+24\alpha\right)\zeta^{4}+\left(31\alpha^{5}-148\alpha^{4}+272\alpha^{3}-216\alpha^{2}+64\right)\zeta^{3}\right.+\nonumber \\
 & \qquad+\left(-233\alpha^{5}+792\alpha^{4}-760\alpha^{3}+104\alpha^{2}-112\alpha+224\right)\zeta^{2}\nonumber \\
 & \qquad\left.+\left(321\alpha^{5}-620\alpha^{4}+520\alpha^{3}-608\alpha^{2}+256\alpha+160\right)\zeta+25\alpha^{5}+101\alpha^{4}-50\alpha^{3}-484\alpha^{2}+408\alpha\right]\nonumber \\
 & +\frac{1}{r^{6}}\left[\left(-10\alpha^{4}+42\alpha^{3}-76\alpha^{2}+88\alpha-48\right)\zeta^{3}+\left(142\alpha^{4}-470\alpha^{3}+436\alpha^{2}+56\alpha-176\right)\zeta^{2}\right.\nonumber \\
 & \qquad+\left.\left(-286\alpha^{4}+382\alpha^{3}+76\alpha^{2}+8\alpha-208\right)\zeta-38\alpha^{4}-74\alpha^{3}+212\alpha^{2}+56\alpha-144\right]\nonumber \\
 & \qquad+\frac{\left(-40\alpha^{3}+148\alpha^{2}-200\alpha+96\right)\zeta^{2}+\left(156\alpha^{3}-224\alpha^{2}-80\alpha+160\right)\zeta+28\alpha^{3}-12\alpha^{2}-88\alpha+64}{r^{8}}\nonumber \\
 & \left.+\frac{\left(-40\alpha^{2}+88\alpha-48\right)\zeta-8\alpha^{2}+24\alpha-16}{r^{10}}\right\} +\left(\mathinner{\color{white}\frac{}{}}\zeta\to-\zeta\right)\nonumber 
\end{align}
\begin{align}
d_{\gamma\chi_{c1}}^{(++,0)} & \left(r,\,\alpha,\,\zeta=x/\xi\right)=\frac{4i\mathfrak{C}_{\chi_{c}}r^{2}}{\alpha^{3}\left(1+\bar{\alpha}\right)^{2}\left(\bar{\alpha}+1/r^{2}\right)^{2}\left(r^{2}-1\right)^{2}\left(\zeta+1-i0\right)\left(\zeta-1+i0\right)\left(\zeta-\kappa+i0\right)^{2}}\times\\
\times & \left\{ \mathinner{\color{white}\frac{\frac{}{}}{\frac{}{}}}\frac{\bar{\alpha}\alpha^{2}}{4}\left[\alpha\bar{\alpha}\left(2-\alpha\right)\left(\zeta^{4}+1\right)+\left(2\alpha^{5}-13\alpha^{4}+31\alpha^{3}-36\alpha^{2}+20\alpha-4\right)\zeta^{3}\right.\right.\nonumber \\
 & \qquad+\left.\left(-4\alpha^{5}+18\alpha^{4}-28\alpha^{3}+18\alpha^{2}-4\alpha\right)\zeta^{2}+\left(2\alpha^{5}-5\alpha^{4}+3\alpha^{3}+4\alpha^{2}-8\alpha+4\right)\zeta\right]+\nonumber \\
 & +\frac{\alpha}{8r^{2}}\left[3(2-\alpha)\alpha\bar{\alpha}\left(\alpha^{2}-8\alpha+8\right)\zeta^{4}+\left(4\alpha^{7}-47\alpha^{6}+209\alpha^{5}-456\alpha^{4}+562\alpha^{3}-392\alpha^{2}+120\alpha\right)\zeta^{3}\right.\nonumber \\
 & \qquad+\left(-8\alpha^{7}+95\alpha^{6}-359\alpha^{5}+604\alpha^{4}-468\alpha^{3}+120\alpha^{2}+16\alpha\right)\zeta^{2}\nonumber \\
 & \qquad\left.+\left(4\alpha^{7}-49\alpha^{6}+127\alpha^{5}-104\alpha^{4}-66\alpha^{3}+200\alpha^{2}-160\alpha+48\right)\zeta+\alpha^{2}\bar{\alpha}^{2}\left(\alpha^{2}+6\alpha-8\right)\right]\nonumber \\
 & +\frac{1}{8r^{4}}\left[\left(-\alpha^{6}+5\alpha^{5}+19\alpha^{4}-106\alpha^{3}+140\alpha^{2}-56\alpha\right)\zeta^{4}+\right.\nonumber \\
 & \qquad+\left(-11\alpha^{7}+99\alpha^{6}-345\alpha^{5}+574\alpha^{4}-576\alpha^{3}+408\alpha^{2}-144\alpha\right)\zeta^{3}-\alpha^{7}-10\alpha^{6}\nonumber \\
 & \qquad+\left(37\alpha^{7}-309\alpha^{6}+963\alpha^{5}-1438\alpha^{4}+1020\alpha^{3}-196\alpha^{2}-72\alpha\right)\zeta^{2}-12\alpha^{5}+99\alpha^{4}\nonumber \\
 & \qquad+\left.\left(-25\alpha^{7}+205\alpha^{6}-475\alpha^{5}+450\alpha^{4}-76\alpha^{3}-160\alpha^{2}+112\alpha-32\right)\zeta-102\alpha^{3}+24\alpha^{2}\right]\nonumber \\
 & +\frac{1}{4r^{6}}\left[-2\alpha^{2}\bar{\alpha}\left(2-\alpha\right)\zeta^{4}+\left(6\alpha^{6}-45\alpha^{5}+125\alpha^{4}-133\alpha^{3}+62\alpha^{2}-44\alpha+24\right)\zeta^{3}\right.+\nonumber \\
 & \qquad+\left(-32\alpha^{6}+210\alpha^{5}-530\alpha^{4}+683\alpha^{3}-474\alpha^{2}+108\alpha+24\right)\zeta^{2}+4\alpha^{6}+14\alpha^{5}\nonumber \\
 & \qquad\left.+\left(30\alpha^{6}-191\alpha^{5}+387\alpha^{4}-375\alpha^{3}+130\alpha^{2}+16\alpha\right)\zeta-8\alpha^{4}-77\alpha^{3}+78\alpha^{2}-8\alpha\right]\nonumber \\
 & +\frac{1}{4r^{8}}\left[-\alpha\left(2-\alpha\right)^{3}\zeta^{4}+\left(-7\alpha^{5}+44\alpha^{4}-108\alpha^{3}+104\alpha^{2}-32\alpha\right)\zeta^{3}-25\alpha^{5}+\right.\nonumber \\
 & \qquad+\left(57\alpha^{5}-296\alpha^{4}+536\alpha^{3}-464\alpha^{2}+288\alpha-96\right)\zeta^{2}-17\alpha^{4}+54\alpha^{3}\nonumber \\
 & \qquad\left.+\left(-73\alpha^{5}+340\alpha^{4}-500\alpha^{3}+448\alpha^{2}-160\alpha-32\right)\zeta+108\alpha^{2}-120\alpha\right]\nonumber \\
 & +\frac{1}{4r^{10}}\left[\left(2-\alpha\right)^{2}\left(2-\alpha\bar{\alpha}\right)\zeta^{3}+\left(-15\alpha^{4}+67\alpha^{3}-90\alpha^{2}+28\alpha+8\right)\zeta^{2}\right.\nonumber \\
 & \qquad\left.\left(27\alpha^{4}-79\alpha^{3}+22\alpha^{2}-4\alpha+24\right)\zeta+19\alpha^{4}-19\alpha^{3}-10\alpha^{2}-20\alpha+24\right]\nonumber \\
 & \left.+\frac{\left(2\alpha^{3}-9\alpha^{2}+14\alpha-8\right)\zeta^{2}+\left(-7\alpha^{3}+16\alpha^{2}+8\alpha-16\right)\zeta-7\alpha^{3}+15\alpha^{2}-6\alpha}{2r^{12}}\right\} +\left(\mathinner{\color{white}\frac{}{}}\zeta\to-\zeta\right)\nonumber 
\end{align}
\begin{align}
c_{\gamma\chi_{c1}}^{(+-,0)} & \left(r,\,\alpha,\,\zeta=x/\xi\right)=\frac{4i\mathfrak{C}_{\chi_{c}}\left(1-\alpha\,r^{2}\right)}{\alpha\left(1+\bar{\alpha}\right)^{2}\left(\bar{\alpha}+1/r^{2}\right)\left(r^{2}-1\right)^{2}\left(\zeta+1-i0\right)\left(\zeta-1+i0\right)\left(\zeta-\kappa+i0\right)^{2}}\times\\
\times & \left\{ \mathinner{\color{white}\frac{\frac{}{}}{\frac{}{}}}-\bar{\alpha}\left[-10\alpha\left(2-\alpha\right)^{2}\zeta^{4}+\left(-12\alpha^{5}+73\alpha^{4}-172\alpha^{3}+236\alpha^{2}-176\alpha+32\right)\zeta^{3}+\alpha^{4}-8\alpha^{3}\right.\right.\nonumber \\
 & \qquad+\left.\left(24\alpha^{5}-113\alpha^{4}+222\alpha^{3}-196\alpha^{2}+48\alpha\right)\zeta^{2}+\left(-12\alpha^{5}+39\alpha^{4}-48\alpha^{3}-8\alpha^{2}+16\alpha\right)\zeta+16\alpha\bar{\alpha}\right]+\nonumber \\
 & +\frac{1}{r^{2}}\left[\left(2\alpha^{4}-12\alpha^{3}+24\alpha^{2}-16\alpha\right)\zeta^{4}+\left(17\alpha^{5}-98\alpha^{4}+178\alpha^{3}-164\alpha^{2}+144\alpha-80\right)\zeta^{3}+\right.\nonumber \\
 & \qquad+\left(-83\alpha^{5}+370\alpha^{4}-708\alpha^{3}+760\alpha^{2}-360\alpha+16\right)\zeta^{2}+\left(67\alpha^{5}-190\alpha^{4}+278\alpha^{3}-148\alpha^{2}-16\alpha\right)\zeta+\nonumber \\
 & \qquad-\left.\alpha^{5}-4\alpha^{4}+64\alpha^{3}-56\alpha^{2}-24\alpha+16\right]\nonumber \\
 & +\frac{1}{r^{4}}\left[2\alpha\left(\alpha-2\right)^{2}\zeta^{4}+\left(-9\alpha^{4}+40\alpha^{3}-44\alpha^{2}-8\alpha+32\right)\zeta^{3}+\left(87\alpha^{4}-282\alpha^{3}+340\alpha^{2}-320\alpha+192\right)\zeta^{2}\right.\nonumber \\
 & \qquad+\left.\left(-131\alpha^{4}+272\alpha^{3}-348\alpha^{2}+168\alpha+64\right)\zeta+5\alpha^{4}-8\alpha^{3}-100\alpha^{2}+120\alpha\right]\nonumber \\
 & +\frac{4}{r^{6}}\left[\left(\alpha^{3}-3\alpha^{2}+4\alpha-4\right)\zeta^{3}+\left(-11\alpha^{3}+22\alpha^{2}+4\alpha-20\right)\zeta^{2}+\left(28\alpha^{3}-29\alpha^{2}+30\alpha-36\right)\zeta\right.+\nonumber \\
 & \qquad\left.-2\alpha^{3}+10\alpha^{2}+8\alpha-20\right]\nonumber \\
 & \left.+\frac{4\left(\left(4\alpha^{2}-10\alpha+8\right)\zeta^{2}+\left(16-13\alpha^{2}\right)\zeta+\alpha^{2}-8\alpha+8\right)}{r^{8}}-\frac{4\bar{\alpha}\zeta}{r^{10}}\right\} +\left(\mathinner{\color{white}\frac{}{}}\zeta\to-\zeta\right)\nonumber 
\end{align}
\begin{align}
d_{\gamma\chi_{c1}}^{(+-,0)} & \left(r,\,\alpha,\,\zeta=x/\xi\right)=-\frac{4i\mathfrak{C}_{\chi_{c}}}{\alpha^{2}\left(1+\bar{\alpha}\right)^{2}\left(\bar{\alpha}+1/r^{2}\right)^{2}\left(r^{2}-1\right)^{2}\left(\zeta+1-i0\right)\left(\zeta-1+i0\right)\left(\zeta-\kappa+i0\right)^{2}}\times\\
 & \times\left\{ \mathinner{\color{white}\frac{\frac{}{}}{\frac{}{}}}-\frac{\bar{\alpha}\alpha}{8}\left[6\alpha\left(\alpha-2\right)^{2}\zeta^{4}+\left(4\alpha^{5}-37\alpha^{4}+122\alpha^{3}-194\alpha^{2}+144\alpha-24\right)\zeta^{3}+\right.\right.\nonumber \\
 & \qquad\left.+\left(-8\alpha^{5}+73\alpha^{4}-194\alpha^{3}+168\alpha^{2}-24\alpha\right)\zeta^{2}+\left(4\alpha^{5}-35\alpha^{4}+46\alpha^{3}-6\alpha^{2}\right)\zeta-\alpha^{4}+4\alpha^{3}-16\alpha\bar{\alpha}\right]+\nonumber \\
 & +\frac{1}{8r^{2}}\left[2\alpha\left(2-\alpha\right)^{2}\left(\alpha^{2}+3\alpha-3\right)\zeta^{4}+\left(-13\alpha^{6}+95\alpha^{5}-259\alpha^{4}+392\alpha^{3}-384\alpha^{2}+176\alpha-16\right)\zeta^{3}\right.\nonumber \\
 & +\left(43\alpha^{6}-267\alpha^{5}+699\alpha^{4}-866\alpha^{3}+412\alpha^{2}-32\alpha\right)\zeta^{2}+\left(-31\alpha^{6}+153\alpha^{5}-265\alpha^{4}+120\alpha^{3}+20\alpha^{2}\right)\zeta\nonumber \\
 & \qquad\left.+\alpha^{6}+5\alpha^{5}-57\alpha^{4}+20\alpha^{3}+64\alpha^{2}-32\alpha\right]\nonumber \\
 & +\frac{1}{4r^{4}}\left[\left(9\alpha^{5}-53\alpha^{4}+103\alpha^{3}-94\alpha^{2}+68\alpha-24\right)\zeta^{3}+\left(-45\alpha^{5}+198\alpha^{4}-384\alpha^{3}+420\alpha^{2}-176\alpha+8\right)\zeta^{2}+\right.\nonumber \\
 & \qquad+\left.\left(47\alpha^{5}-151\alpha^{4}+233\alpha^{3}-94\alpha^{2}-20\alpha\right)\zeta-3\alpha^{5}+2\alpha^{4}+76\alpha^{3}-64\alpha^{2}-16\alpha+8\right]\nonumber \\
 & +\frac{1}{8r^{6}}\left[2\alpha\left(2-\alpha\right)^{2}\zeta^{4}+\left(-13\alpha^{4}+60\alpha^{3}-96\alpha^{2}+64\alpha-16\right)\zeta^{3}+\left(99\alpha^{4}-322\alpha^{3}+380\alpha^{2}-296\alpha+96\right)\zeta^{2}\right.+\nonumber \\
 & \qquad\left.+\left(-147\alpha^{4}+300\alpha^{3}-392\alpha^{2}+160\alpha+16\right)\zeta+13\alpha^{4}-48\alpha^{3}-148\alpha^{2}+144\alpha\right]\nonumber \\
 & \qquad+\frac{(\alpha-2)\left(2-\alpha\bar{\alpha}\right)\zeta^{3}+\left(-15\alpha^{3}+40\alpha^{2}-28\alpha+4\right)\zeta^{2}+\left(33\alpha^{3}-37\alpha^{2}+26\alpha-12\right)\zeta-3\alpha^{3}+18\alpha^{2}-12\bar{\alpha}}{2r^{8}}\nonumber \\
 & \left.+\frac{\left(4\alpha^{2}-10\alpha+8\right)\zeta^{2}+\left(-17\alpha^{2}+12\alpha+4\right)\zeta+\alpha^{2}-8\alpha}{2r^{10}}-\frac{2\bar{\alpha}\zeta}{r^{12}}\right\} +\left(\mathinner{\color{white}\frac{}{}}\zeta\to-\zeta\right)\nonumber 
\end{align}

\begin{align}
c_{\gamma\chi_{c1}}^{(++,+1)} & \left(r,\,\alpha,\,\zeta=x/\xi\right)=-\frac{4i\mathfrak{C}_{\chi_{c}}r^{2}\sqrt{\left(1-\alpha\right)\left(\alpha r^{2}-1\right)}}{\alpha^{2}\left(1+\bar{\alpha}\right)^{2}\left(\bar{\alpha}+1/r^{2}\right)\left(r^{2}-1\right)^{2}\left(\zeta+1-i0\right)\left(\zeta-1+i0\right)\left(\zeta-\kappa+i0\right)^{2}}\times\\
\times & \left\{ \mathinner{\color{white}\frac{\frac{}{}}{\frac{}{}}}\alpha\,\bar{\alpha}^{2}\left[\left(\alpha^{3}-22\alpha^{2}+68\alpha-40\right)\zeta^{4}-\left(9\alpha^{4}-62\alpha^{3}+118\alpha^{2}-128\alpha+56\right)\zeta^{3}-\alpha^{3}\bar{\alpha}\right.\right.\nonumber \\
 & \qquad\left.+\left(19\alpha^{4}-96\alpha^{3}+142\alpha^{2}-84\alpha+16\right)\zeta^{2}-\alpha\left(11\alpha^{3}-34\alpha^{2}+54\alpha-32\right)\zeta+\left(20\alpha^{2}-32\alpha+16\right)\right]+\nonumber \\
 & +\frac{\bar{\alpha}}{r^{2}}\left[-\alpha\left(\alpha^{4}-15\alpha^{3}+26\alpha^{2}+60\alpha-88\right)\zeta^{4}-\left(15\alpha^{6}-93\alpha^{5}+293\alpha^{4}-444\alpha^{3}+524\alpha^{2}-432\alpha+96\right)\zeta^{3}+\right.\nonumber \\
 & \qquad-\left(29\alpha^{6}-153\alpha^{5}+449\alpha^{4}-688\alpha^{3}+392\alpha^{2}+120\alpha-96\right)\zeta^{2}\nonumber \\
 & \qquad\left.+\alpha\left(13\alpha^{5}-63\alpha^{4}+107\alpha^{3}-24\alpha^{2}-160\alpha+80\right)\zeta+\left(\alpha^{6}+2\alpha^{5}+64\alpha^{4}-70\alpha^{3}-128\alpha^{2}+144\alpha-32\right)\right]\nonumber \\
 & +\frac{1}{r^{4}}\left[-\left(8\alpha^{4}-31\alpha^{3}+10\alpha^{2}+52\alpha-40\right)\alpha\zeta^{4}-4\bar{\alpha}\left(11\alpha^{5}-44\alpha^{4}+73\alpha^{3}-19\alpha^{2}-16\alpha-28\right)\zeta^{3}\right.\nonumber \\
 & \qquad-2\left(74\alpha^{6}-306\alpha^{5}+589\alpha^{4}-854\alpha^{3}+946\alpha^{2}-488\alpha+40\right)\zeta^{2}\nonumber \\
 & \qquad-\left.4\bar{\alpha}\left(25\alpha^{5}-55\alpha^{4}+64\alpha^{3}-7\alpha^{2}-134\alpha+48\right)\zeta+4\alpha^{6}+32\alpha^{5}+219\alpha^{4}-790\alpha^{3}+616\alpha^{2}-16\alpha-64\right]\nonumber \\
 & +\frac{1}{r^{6}}\left[(\alpha-2)^{2}(9\alpha-10)\alpha\zeta^{4}+\left(-47\alpha^{5}+199\alpha^{4}-336\alpha^{3}+228\alpha^{2}+48\alpha-96\right)\zeta^{3}\right.+\nonumber \\
 & \qquad\,+\left(273\alpha^{5}-883\alpha^{4}+848\alpha^{3}-204\alpha^{2}+288\alpha-320\right)\zeta^{2}\nonumber \\
 & \qquad\left.+\left(-281\alpha^{5}+621\alpha^{4}-684\alpha^{3}+924\alpha^{2}-448\alpha-128\right)\zeta-9\alpha^{5}-154\alpha^{4}-38\alpha^{3}+912\alpha^{2}-872\alpha+160\right]\nonumber \\
 & +\frac{2}{r^{8}}\left[(\alpha-2)(9\alpha-10)\left(\alpha^{2}-\alpha+2\right)\zeta^{3}-\left(113\alpha^{4}-351\alpha^{3}+292\alpha^{2}+80\alpha-136\right)\zeta^{2}\right.+\nonumber \\
 & \qquad\left.\left(191\alpha^{4}-291\alpha^{3}+10\alpha^{2}-64\alpha+152\right)\zeta+\left(9\alpha^{4}+105\alpha^{3}-184\alpha^{2}-52\alpha+120\right)\right]\nonumber \\
 & \left.+\frac{4\left[(9\alpha-10)\left(2\alpha^{2}-5\alpha+4\right)\zeta^{2}+\bar{\alpha}(5\alpha+4)(13\alpha-16)\zeta-\left(5\alpha^{3}+16\alpha^{2}-54\alpha+32\right)\right]}{r^{10}}-\frac{8\bar{\alpha}\left[(9\alpha-10)\zeta+(2-\alpha)\right]}{r^{12}}\right\} \nonumber \\
 & +\left(\mathinner{\color{white}\frac{}{}}\zeta\to-\zeta\right)\nonumber 
\end{align}
\begin{align}
d_{\gamma\chi_{c1}}^{(++,+1)} & \left(r,\,\alpha,\,\zeta=x/\xi\right)=\frac{4i\mathfrak{C}_{\chi_{c}}\sqrt{\left(1-\alpha\right)\left(\alpha r^{2}-1\right)}}{\bar{\alpha}\alpha^{4}\left(1+\bar{\alpha}\right)^{2}\left(\bar{\alpha}+1/r^{2}\right)^{2}\left(r^{2}-1\right)^{2}\left(\zeta+1-i0\right)\left(\zeta-1+i0\right)\left(\zeta-\kappa+i0\right)^{2}}\times\\
\times & \left\{ \mathinner{\color{white}\frac{\frac{}{}}{\frac{}{}}}\frac{\bar{\alpha}^{2}\alpha^{2}}{8}\left[(2-\alpha)\left(\alpha^{2}+12\bar{\alpha}\right)\zeta^{4}+\left(9\alpha^{4}-46\alpha^{3}+78\alpha^{2}-80\alpha+40\right)\zeta^{3}-\alpha\left(19\alpha^{3}-64\alpha^{2}+78\alpha-36\right)\zeta^{2}\right.\right.\nonumber \\
 & \qquad+\left.\left(11\alpha^{4}-18\alpha^{3}+14\alpha^{2}+16\bar{\alpha}\right)\zeta+\left(\alpha^{3}\bar{\alpha}+4\alpha^{2}-16\bar{\alpha}\right)\right]+\nonumber \\
 & +\frac{\bar{\alpha}\alpha}{8r^{2}}\left[(\alpha-2)\bar{\alpha}\left(\alpha^{3}-4\alpha^{2}-20\alpha+16\right)\zeta^{4}-\left(3\alpha^{6}-57\alpha^{5}+261\alpha^{4}-472\alpha^{3}+508\alpha^{2}-368\alpha+128\right)\zeta^{3}\right.\nonumber \\
 & \qquad+\left(5\alpha^{6}-129\alpha^{5}+449\alpha^{4}-632\alpha^{3}+360\alpha^{2}+40\alpha-96\right)\zeta^{2}\nonumber \\
 & \qquad\left.-\left(\alpha^{6}-75\alpha^{5}+139\alpha^{4}-68\alpha^{3}-128\alpha^{2}+160\alpha-32\right)\zeta-\left(\alpha^{6}+2\alpha^{5}+8\alpha^{4}-86\alpha^{3}+8\alpha^{2}+128\alpha-64\right)\right]\nonumber \\
 & +\frac{1}{8r^{4}}\left[(\alpha-2)\alpha^{2}\left(\alpha^{2}+4\alpha-4\right)\zeta^{4}+4\left(-\alpha^{7}+18\alpha^{6}-83\alpha^{5}+148\alpha^{4}-143\alpha^{3}+104\alpha^{2}-60\alpha+16\right)\zeta^{3}+\right.\nonumber \\
 & \qquad+2\left(10\alpha^{7}-166\alpha^{6}+605\alpha^{5}-1022\alpha^{4}+954\alpha^{3}-456\alpha^{2}+40\alpha+32\right)\zeta^{2}\nonumber \\
 & \qquad-4\alpha\left(3\alpha^{6}-67\alpha^{5}+165\alpha^{4}-151\alpha^{3}-15\alpha^{2}+130\alpha-64\right)\zeta\nonumber \\
 & \qquad+\left.\left(-4\alpha^{7}-8\alpha^{6}-11\alpha^{5}+382\alpha^{4}-584\alpha^{3}+96\alpha^{2}+192\alpha-64\right)\right]\nonumber \\
 & +\frac{1}{8r^{6}}\left[-(\alpha-2)^{2}\alpha\left(\alpha^{2}-10\alpha+8\right)\zeta^{4}+\alpha\left(3\alpha^{5}-51\alpha^{4}+204\alpha^{3}-308\alpha^{2}+224\alpha-64\right)\zeta^{3}\right.+\nonumber \\
 & \qquad+\left(-25\alpha^{6}+347\alpha^{5}-1000\alpha^{4}+1244\alpha^{3}-896\alpha^{2}+480\alpha-128\right)\zeta^{2}\nonumber \\
 & \qquad+\left(29\alpha^{6}-441\alpha^{5}+920\alpha^{4}-860\alpha^{3}+336\alpha^{2}+160\alpha-128\right)\zeta\nonumber \\
 & \qquad\left.+\alpha\left(9\alpha^{5}+34\alpha^{4}-106\alpha^{3}-456\alpha^{2}+936\alpha-416\right)\right]\nonumber \\
 & +\frac{1}{4r^{8}}\left[(2-\alpha)\left(\alpha^{2}-10\alpha+8\right)\left(\alpha^{2}-\alpha+2\right)\zeta^{3}+\left(9\alpha^{5}-119\alpha^{4}+300\alpha^{3}-256\alpha^{2}+24\alpha+32\right)\zeta^{2}\right.+\nonumber \\
 & \qquad\left.+\left(-15\alpha^{5}+187\alpha^{4}-250\alpha^{3}+88\alpha^{2}-56\alpha+32\right)\zeta+\left(-9\alpha^{5}-9\alpha^{4}+104\alpha^{3}+36\alpha^{2}-216\alpha+96\right)\right]\nonumber \\
 & +\frac{\left(-\alpha^{2}+10\alpha-8\right)\left(2\alpha^{2}-5\alpha+4\right)\zeta^{2}+\left(5\alpha^{4}-57\alpha^{3}+56\alpha^{2}+32\alpha-32\right)\zeta+\alpha\left(5\alpha^{3}-8\alpha^{2}-22\alpha+24\right)}{2r^{10}}\nonumber \\
 & \left.+\frac{\bar{\alpha}\left[\left(\alpha^{2}-10\alpha+8\right)\zeta-(2-\alpha)\alpha\right]}{r^{12}}\right\} +\left(\mathinner{\color{white}\frac{}{}}\zeta\to-\zeta\right)\nonumber 
\end{align}
\begin{align}
c_{\gamma\chi_{c1}}^{(+-,+1)} & \left(r,\,\alpha,\,\zeta=x/\xi\right)=-\frac{4i\mathfrak{C}_{\chi_{c}}r^{2}\sqrt{\left(1-\alpha\right)\left(\alpha r^{2}-1\right)}}{\alpha\left(1+\bar{\alpha}\right)^{2}\left(\bar{\alpha}+1/r^{2}\right)\left(r^{2}-1\right)^{2}\left(\zeta+1-i0\right)\left(\zeta-1+i0\right)\left(\zeta-\kappa+i0\right)^{2}}\times\\
\times & \left\{ \mathinner{\color{white}\frac{\frac{}{}}{\frac{}{}}}\bar{\alpha}^{2}\left[(2-\alpha)^{2}\alpha\bar{\alpha}^{2}\left(4\alpha^{2}+3\alpha+4\right)\zeta^{3}+(2-\alpha)\alpha^{2}\bar{\alpha}^{2}\left(8\alpha^{2}+5\alpha+6\right)\zeta^{2}\right.\right.\nonumber \\
 & \qquad+\left.\alpha^{2}\bar{\alpha}^{2}\left(4\alpha^{3}+\alpha^{2}+8\alpha-8\right)\zeta+(4-\alpha)\bar{\alpha}^{2}\alpha^{3}\right]+\nonumber \\
 & +\frac{1}{r^{2}}\left[(2-\alpha)^{2}\bar{\alpha}\alpha(\alpha+22)\zeta^{4}-\bar{\alpha}\left(4\alpha^{6}-31\alpha^{5}+111\alpha^{4}-229\alpha^{3}+336\alpha^{2}-260\alpha+32\right)\zeta^{3}+\right.\nonumber \\
 & \qquad+\bar{\alpha}\alpha\left(8\alpha^{5}-33\alpha^{4}+95\alpha^{3}-223\alpha^{2}+172\alpha+12\right)\zeta^{2}+\nonumber \\
 & \qquad-\left.\bar{\alpha}\left(4\alpha^{6}-5\alpha^{5}+37\alpha^{4}-35\alpha^{3}-92\alpha^{2}+64\alpha-16\right)-\bar{\alpha}\alpha\left(3\alpha^{4}-4\alpha^{3}-37\alpha^{2}-12\alpha+24\right)\right]\nonumber \\
 & +\frac{1}{r^{4}}\left[(2-\alpha)^{2}\left(\alpha^{2}+6\alpha-8\right)\alpha\zeta^{4}+2\bar{\alpha}\left(5\alpha^{5}-31\alpha^{4}+76\alpha^{3}-53\alpha^{2}+28\alpha-68\right)\zeta^{3}+\right.\nonumber \\
 & \qquad+2\left(17\alpha^{6}-107\alpha^{5}+281\alpha^{4}-495\alpha^{3}+561\alpha^{2}-236\alpha-20\right)\zeta^{2}-2\alpha^{6}-11\alpha^{5}\nonumber \\
 & \qquad+\left.2\bar{\alpha}\left(11\alpha^{4}-41\alpha^{3}+86\alpha^{2}-129\alpha-48\right)\alpha\zeta+48\alpha^{4}+106\alpha^{3}-150\alpha^{2}-8\alpha+16\right]\nonumber \\
 & +\frac{1}{r^{6}}\left[\left(8\alpha^{5}-45\alpha^{4}+101\alpha^{3}-64\alpha^{2}-60\alpha+64\right)\zeta^{3}+(2-\alpha)\left(52\alpha^{4}-165\alpha^{3}+117\alpha^{2}-174\alpha+168\right)\zeta^{2}\right.+\nonumber \\
 & \qquad\left.+\alpha\left(12\alpha^{4}+4\alpha^{3}-107\alpha^{2}-120\alpha+212\right)+\left(48\alpha^{5}-211\alpha^{4}+351\alpha^{3}-612\alpha^{2}+276\alpha+144\right)\zeta+(2-\alpha)^{3}\alpha\zeta^{4}\right]\nonumber \\
 & -\frac{2}{r^{8}}\left[(2-\alpha)^{2}\left(\alpha^{2}-2\alpha+3\right)\zeta^{3}+(2-\alpha)\left(17\alpha^{3}-40\alpha^{2}-13\alpha+38\right)\zeta^{2}\right.+\nonumber \\
 & \qquad\left.\left(27\alpha^{4}-86\alpha^{3}+25\alpha^{2}-100\alpha+132\right)\zeta+\left(13\alpha^{4}-14\alpha^{3}-61\alpha^{2}-8\alpha+68\right)\right]\nonumber \\
 & \left.-\frac{4\left[(\alpha-2)^{2}(2\alpha-3)\zeta^{2}+\bar{\alpha}\left(8\alpha^{2}-13\alpha-32\right)\zeta-\left(6\alpha^{3}-10\alpha^{2}-13\alpha+16\right)\right]}{r^{10}}+\frac{8(\alpha-3)\bar{\alpha}\zeta-8\,\bar{\alpha}^{2}}{r^{12}}\right\} +\left(\mathinner{\color{white}\frac{}{}}\zeta\to-\zeta\right)\nonumber 
\end{align}
\begin{align}
d_{\gamma\chi_{c1}}^{(+-,+1)} & \left(r,\,\alpha,\,\zeta=x/\xi\right)=\frac{4i\mathfrak{C}_{\chi_{c}}\sqrt{\left(1-\alpha\right)\left(\alpha r^{2}-1\right)}}{\alpha^{2}\bar{\alpha}\left(1+\bar{\alpha}\right)^{2}\left(\bar{\alpha}+1/r^{2}\right)\left(r^{2}-1\right)^{2}\left(\zeta+1-i0\right)\left(\zeta-1+i0\right)\left(\zeta-\kappa+i0\right)}\times\\
 & \times\left\{ \mathinner{\color{white}\frac{\frac{}{}}{\frac{}{}}}\frac{\bar{\alpha}}{8}\left[(2-\alpha)\left(4\alpha^{2}-9\alpha+8\right)\alpha\zeta^{2}-2\bar{\alpha}(2\alpha-3)\alpha^{2}\zeta-(4-\alpha)\alpha^{2}\right]\right.\nonumber \\
 & +\frac{1}{8r^{2}}\left[(\alpha-2)^{2}\alpha\zeta^{3}+(2-\alpha)\left(4\alpha^{4}-3\alpha^{3}-9\alpha^{2}+15\alpha-4\right)\zeta^{2}+\alpha\left(4\alpha^{4}+2\alpha^{3}-37\alpha^{2}+52\alpha-18\right)\zeta\right.\nonumber \\
 & \qquad\left.\left(3\alpha^{4}-7\alpha^{3}+\alpha^{2}-4\alpha+8\right)\right]\nonumber \\
 & +\frac{1}{8r^{4}}\left[-(\alpha-2)^{2}\alpha\zeta^{3}+(\alpha-2)\left(6\alpha^{3}-8\alpha^{2}-3\alpha+10\right)\zeta^{2}+\left(-12\alpha^{4}+17\alpha^{3}+20\alpha^{2}-32\alpha-4\right)\zeta\right.\nonumber \\
 & \qquad+\left.\left(-2\alpha^{4}-4\alpha^{3}+11\alpha^{2}+4\alpha-8\right)\right]\nonumber \\
 & +\frac{(2-\alpha)\left(\alpha^{2}-3\alpha+3\right)\zeta^{2}+\left(6\alpha^{3}-11\alpha^{2}-7\alpha+16\right)\zeta+\left(3\alpha^{3}-2\alpha^{2}-4\alpha+2\right)}{4r^{6}}\nonumber \\
 & \left.-\frac{(3-\alpha)\bar{\alpha}\zeta+\bar{\alpha}^{2}}{2r^{8}}\right\} +\left(\mathinner{\color{white}\frac{}{}}\zeta\to-\zeta\right)\nonumber 
\end{align}

\subsection{Coefficient functions for $\chi_{c2}$}

\begin{align}
c_{\gamma\chi_{c2}}^{(++,-2)} & \left(r,\,\alpha,\,\zeta=x/\xi\right)=\frac{4\mathfrak{C}_{\chi_{c}}\left(1-\alpha\right)\left(\alpha\,r^{2}-1\right)r^{2}}{\alpha^{3}\left(1+\bar{\alpha}\right)^{2}\left(\bar{\alpha}+1/r^{2}\right)\left(r^{2}-1\right)^{2}\left(\zeta+1-i0\right)\left(\zeta-1+i0\right)\left(\zeta-\kappa+i0\right)^{2}}\times\\
\times & \left\{ \mathinner{\color{white}\frac{\frac{}{}}{\frac{}{}}}\bar{\alpha}\alpha^{2}\left[-\left(2\alpha^{4}-11\alpha^{3}+20\alpha^{2}-16\alpha+4\right)\zeta^{3}-\left(2\alpha^{4}-3\alpha^{3}+4\alpha-4\right)\zeta+(2-\alpha)\alpha\left(\zeta^{4}+2\bar{\alpha}(2\alpha-1)\zeta^{2}+1\right)\right]\right.\nonumber \\
 & +\frac{1}{r^{2}}\left[(\alpha-4)(2-\alpha)\bar{\alpha}\alpha(3\alpha-2)\zeta^{4}+\alpha\left(6\alpha^{6}-41\alpha^{5}+113\alpha^{4}-171\alpha^{3}+176\alpha^{2}-116\alpha+32\right)\zeta^{3}\right.\nonumber \\
 & \qquad-2\alpha\left(6\alpha^{6}-33\alpha^{5}+76\alpha^{4}-92\alpha^{3}+54\alpha^{2}-6\alpha-4\right)\zeta^{2}+\alpha^{2}\left(6\alpha^{5}-25\alpha^{4}+41\alpha^{3}-23\alpha^{2}-8\alpha+8\right)\zeta\nonumber \\
 & \qquad-\left.\bar{\alpha}\alpha\left(3\alpha^{3}+8\alpha^{2}-16\alpha+8\right)\right]\nonumber \\
 & +\frac{1}{r^{4}}\left[4(\alpha-2)\bar{\alpha}^{2}\alpha\zeta^{4}-2\left(5\alpha^{6}-28\alpha^{5}+52\alpha^{4}-38\alpha^{3}+36\alpha^{2}-52\alpha+24\right)\zeta^{3}\right.\nonumber \\
 & \qquad+4\left(11\alpha^{6}-47\alpha^{5}+81\alpha^{4}-87\alpha^{3}+54\alpha^{2}+2\alpha-12\right)\zeta^{2}\nonumber \\
 & \qquad-\left.2\alpha\left(17\alpha^{5}-46\alpha^{4}+56\alpha^{3}-18\alpha^{2}-48\alpha+36\right)\zeta+4\bar{\alpha}\left(9\alpha^{3}-7\alpha^{2}-8\alpha+4\right)\right]\nonumber \\
 & +\frac{1}{r^{6}}\left[\left(8\alpha^{5}-37\alpha^{4}+56\alpha^{3}-4\alpha^{2}-56\alpha+32\right)\zeta^{3}-2\left(28\alpha^{5}-93\alpha^{4}+92\alpha^{3}-74\alpha^{2}+116\alpha-64\right)\zeta^{2}\right.\nonumber \\
 & \qquad\left.+\left(72\alpha^{5}-145\alpha^{4}+144\alpha^{3}-108\alpha^{2}-72\alpha+96\right)\zeta-2\bar{\alpha}\alpha\left(5\alpha^{2}+34\alpha-56\right)+2(\alpha-2)^{2}\bar{\alpha}\alpha\zeta^{4}\right]\nonumber \\
 & +\frac{1}{r^{8}}\left[-4(2-\alpha)\bar{\alpha}\left(\alpha^{2}-\alpha+2\right)\zeta^{3}+4\left(10\alpha^{4}-29\alpha^{3}+12\alpha^{2}+28\alpha-20\right)\zeta^{2}\right.\nonumber \\
 & \qquad\left.-4\left(19\alpha^{4}-25\alpha^{3}+7\alpha^{2}-32\alpha+28\right)\zeta+16\bar{\alpha}\left(2\alpha^{2}+\alpha-5\right)\right]\nonumber \\
 & \left.\frac{8\bar{\alpha}\left(2\alpha^{2}-5\alpha+4\right)\zeta^{2}+4(3\alpha-4)\left(4\alpha^{2}+\alpha-4\right)\zeta-8\bar{\alpha}(3\alpha-4)}{r^{10}}-\frac{16\bar{\alpha}^{2}\zeta}{r^{12}}\right\} +\left(\mathinner{\color{white}\frac{}{}}\zeta\to-\zeta\right)\nonumber 
\end{align}
\begin{align}
d_{\gamma\chi_{c2}}^{(++,-2)} & \left(r,\,\alpha,\,\zeta=x/\xi\right)=\frac{4\mathfrak{C}_{\chi_{c}}\bar{\alpha}^{2}\left(\alpha\,r^{2}-1\right)}{\alpha^{4}\left(1+\bar{\alpha}\right)^{2}r^{2}\left(\bar{\alpha}+1/r^{2}\right)^{2}\left(r^{2}-1\right)^{2}\left(\zeta+1-i0\right)\left(\zeta-1+i0\right)\left(\zeta-\kappa+i0\right)^{2}}\times\\
\times & \left\{ \mathinner{\color{white}\frac{\frac{}{}}{\frac{}{}}}\frac{\bar{\alpha}\alpha}{4}\left[-(\alpha-2)^{2}\zeta^{4}-2(2-\alpha)\left(\alpha^{3}-3\alpha^{2}+3\alpha-3\right)\zeta^{3}-2\left(2\alpha^{4}-6\alpha^{3}+7\alpha^{2}-3\alpha-4\right)\zeta^{2}\right.\right.\nonumber \\
 & \qquad\left.+\left(\alpha^{2}+6\alpha-4\right)+2\left(\alpha^{4}-\alpha^{3}+\alpha^{2}+\alpha+2\right)\zeta\right]+\nonumber \\
 & +\frac{1}{2r^{2}}\left[(\alpha-2)\left(2\alpha^{4}-5\alpha^{3}+\alpha^{2}-\alpha+2\right)\zeta^{3}+\alpha\left(6\alpha^{4}-9\alpha^{3}+7\alpha^{2}+9\alpha-12\right)\zeta\right.+\nonumber \\
 & \qquad\left.+2\left(-4\alpha^{5}+13\alpha^{4}-17\alpha^{3}+17\alpha^{2}-6\alpha-2\right)\zeta^{2}-4\bar{\alpha}(\alpha+1)(2\alpha-1)\right]\nonumber \\
 & +\frac{1}{4r^{4}}\left[(\alpha-2)^{2}\alpha\zeta^{4}+2(2-\alpha)\alpha\left(2\alpha^{2}-4\alpha+1\right)\zeta^{3}+2(\alpha-2)\left(12\alpha^{3}-4\alpha^{2}+3\alpha-4\right)\zeta^{2}\right.+\nonumber \\
 & \qquad\left.+2\left(-14\alpha^{4}+14\alpha^{3}-15\alpha^{2}+2\alpha+8\right)\zeta-\alpha\left(5\alpha^{2}+34\alpha-40\right)\right]\nonumber \\
 & +\frac{(\alpha-2)\left(\alpha^{2}-\alpha+2\right)\zeta^{3}+4\left(2\alpha^{2}+2\alpha-3\right)+2\left(-5\alpha^{3}+11\alpha^{2}-2\alpha-2\right)\zeta^{2}+\left(17\alpha^{3}-7\alpha^{2}-4\right)\zeta}{2r^{6}}\nonumber \\
 & \left.\frac{\left(2\alpha^{2}-5\alpha+4\right)\zeta^{2}+2\bar{\alpha}(3\alpha+2)\zeta-3\alpha}{r^{8}}-\frac{2\bar{\alpha}\zeta}{r^{10}}\right\} +\left(\mathinner{\color{white}\frac{}{}}\zeta\to-\zeta\right)\nonumber 
\end{align}
\begin{align}
c_{\gamma\chi_{c2}}^{(+-,-2)} & \left(r,\,\alpha,\,\zeta=x/\xi\right)=-\frac{4\mathfrak{C}_{\chi_{c}}\left(1-\alpha\right)\left(\alpha\,r^{2}-1\right)^{2}}{\alpha^{2}\left(1+\bar{\alpha}\right)^{2}r^{2}\left(\bar{\alpha}+1/r^{2}\right)\left(r^{2}-1\right)\left(\zeta+1-i0\right)\left(\zeta-1+i0\right)\left(\zeta-\kappa+i0\right)}\times\\
\times & \left\{ \mathinner{\color{white}\frac{\frac{}{}}{\frac{}{}}}\bar{\alpha}\alpha(2\alpha-1)\left[\alpha\zeta+(2-\alpha)\zeta^{2}\right]+\frac{(\alpha-2)\alpha\zeta^{3}-2(2-\alpha)\bar{\alpha}\alpha\zeta^{2}-2\bar{\alpha}\alpha(3\alpha-1)\zeta-(\alpha-2)^{2}}{r^{2}}\right.\nonumber \\
 & \left.\frac{2\left(\alpha^{2}-2\alpha+2\right)\zeta^{2}-2\left(3\alpha^{2}-2\alpha-2\right)\zeta-2\alpha}{r^{4}}-\frac{4\bar{\alpha}\zeta}{r^{6}}\right\} +\left(\mathinner{\color{white}\frac{}{}}\zeta\to-\zeta\right)\nonumber 
\end{align}
\begin{align}
d_{\gamma\chi_{c2}}^{(+-,-2)} & \left(r,\,\alpha,\,\zeta=x/\xi\right)=0
\end{align}

\begin{align}
c_{\gamma\chi_{c2}}^{(++,-1)} & \left(r,\,\alpha,\,\zeta=x/\xi\right)=\frac{4\mathfrak{C}_{\chi_{c}}r^{2}\sqrt{\left(1-\alpha\right)\left(\alpha r^{2}-1\right)}}{\alpha^{3}\left(1+\bar{\alpha}\right)^{2}\left(\bar{\alpha}+1/r^{2}\right)\left(r^{2}-1\right)^{2}\left(\zeta+1-i0\right)\left(\zeta-1+i0\right)\left(\zeta-\kappa+i0\right)^{2}}\times\\
\times & \left\{ \mathinner{\color{white}\frac{\frac{}{}}{\frac{}{}}}\bar{\alpha}\alpha^{2}\left[-2\left(2\alpha^{2}-7\alpha+4\right)\zeta^{4}-\left(3\alpha^{4}-20\alpha^{3}+38\alpha^{2}-36\alpha+12\right)\zeta^{3}+2\left(3\alpha^{4}-14\alpha^{3}+18\alpha^{2}-8\alpha+2\right)\zeta^{2}\right.\right.\nonumber \\
 & \qquad\left.-\left(3\alpha^{4}-8\alpha^{3}+10\alpha^{2}-4\alpha-4\right)\zeta+2\left(2\alpha^{2}-3\alpha+2\right)\right]+\nonumber \\
 & +\frac{\alpha}{r^{2}}\left[2(4-\alpha)\bar{\alpha}\left(\alpha^{2}-6\alpha+4\right)\zeta^{4}+\left(11\alpha^{6}-71\alpha^{5}+203\alpha^{4}-322\alpha^{3}+362\alpha^{2}-272\alpha+88\right)\zeta^{3}\right.\nonumber \\
 & \qquad\,\,-2\left(11\alpha^{6}-55\alpha^{5}+133\alpha^{4}-180\alpha^{3}+108\alpha^{2}+4\alpha-20\right)\zeta^{2}\nonumber \\
 & \qquad\left.+\alpha\left(11\alpha^{5}-39\alpha^{4}+63\alpha^{3}-26\alpha^{2}-42\alpha+32\right)\zeta-2\bar{\alpha}\left(9\alpha^{3}+4\alpha^{2}-24\alpha+12\right)\right]\nonumber \\
 & +\frac{2}{r^{4}}\left[\alpha\bar{\alpha}^{2}\left(\alpha^{2}-8\right)\zeta^{4}-\left(11\alpha^{6}-55\alpha^{5}+103\alpha^{4}-80\alpha^{3}+68\alpha^{2}-96\alpha+48\right)\zeta^{3}\right.\nonumber \\
 & \qquad+2\left(22\alpha^{6}-85\alpha^{5}+145\alpha^{4}-176\alpha^{3}+132\alpha^{2}-12\alpha-24\right)\zeta^{2}\nonumber \\
 & \qquad-\left.\alpha\left(33\alpha^{5}-79\alpha^{4}+95\alpha^{3}-20\alpha^{2}-128\alpha+96\right)\zeta+\bar{\alpha}\left(3\alpha^{4}+51\alpha^{3}-52\alpha^{2}-32\alpha+16\right)\right]\nonumber \\
 & +\frac{1}{r^{6}}\left[4(\alpha-2)^{2}\alpha\bar{\alpha}\zeta^{4}+\left(19\alpha^{5}-78\alpha^{4}+110\alpha^{3}-12\alpha^{2}-104\alpha+64\right)\zeta^{3}-4\alpha\bar{\alpha}\left(10\alpha^{2}+37\alpha-66\right)\right.+\nonumber \\
 & \qquad-\left.2\left(63\alpha^{5}-190\alpha^{4}+170\alpha^{3}-122\alpha^{2}+212\alpha-128\right)\zeta^{2}+\left(151\alpha^{5}+270\alpha^{3}\bar{\alpha}-268\alpha^{2}-88\alpha+192\right)\zeta\right]\nonumber \\
 & +\frac{4}{r^{8}}\left[2(\alpha-2)\bar{\alpha}\left(\alpha^{2}-\alpha+2\right)\zeta^{3}+\left(23\alpha^{4}-64\alpha^{3}+30\alpha^{2}+52\alpha-40\right)\zeta^{2}\right.+\nonumber \\
 & \qquad\left.-\left(43\alpha^{4}-52\alpha^{3}+6\alpha^{2}-56\alpha+56\right)\zeta+4\bar{\alpha}\left(5\alpha^{2}+\alpha-10\right)\right]\nonumber \\
 & \left.+\frac{16\bar{\alpha}\left(2\alpha^{2}-5\alpha+4\right)\zeta^{2}+4\left(27\alpha^{3}-30\alpha^{2}-30\alpha+32\right)\zeta-16\bar{\alpha}(3\alpha-4)}{r^{10}}-\frac{32\bar{\alpha}^{2}\zeta}{r^{12}}\right\} +\left(\mathinner{\color{white}\frac{}{}}\zeta\to-\zeta\right)\nonumber 
\end{align}
\begin{align}
d_{\gamma\chi_{c2}}^{(++,-1)} & \left(r,\,\alpha,\,\zeta=x/\xi\right)=\frac{4\mathfrak{C}_{\chi_{c}}\bar{\alpha}\sqrt{\left(1-\alpha\right)\left(\alpha r^{2}-1\right)}}{\alpha^{4}\left(1+\bar{\alpha}\right)^{2}\left(\bar{\alpha}+1/r^{2}\right)^{2}\left(r^{2}-1\right)^{2}\left(\zeta+1-i0\right)\left(\zeta-1+i0\right)\left(\zeta-\kappa+i0\right)^{2}}\times\\
\times & \left\{ \mathinner{\color{white}\frac{\frac{}{}}{\frac{}{}}}\frac{\bar{\alpha}^{2}\alpha^{2}}{4}\left[2\alpha(2\alpha-3)\zeta^{2}-(2-\alpha)\zeta^{4}-2\left(\alpha^{2}-2\alpha+2\right)\zeta^{3}-2\left(\alpha^{2}-2\alpha+2\right)\zeta-(3\alpha-2)\right]\right.\nonumber \\
 & +\frac{\bar{\alpha}\alpha}{8r^{2}}\left[\left(3\alpha^{4}-9\alpha^{3}+22\alpha^{2}-24\alpha+32\right)\zeta^{3}-2\left(3\alpha^{4}-3\alpha^{3}+14\alpha^{2}-18\alpha-12\right)\zeta^{2}\right.\nonumber \\
 & \qquad\left.+\left(3\alpha^{4}+3\alpha^{3}+6\alpha^{2}+12\alpha+8\right)\zeta-2\left(\alpha^{2}-4\right)\zeta^{4}-2(\alpha+2)(7\alpha-4)\right]\nonumber \\
 & +\frac{1}{4r^{4}}\left[\left(5\alpha^{5}-17\alpha^{4}+24\alpha^{3}-10\alpha^{2}+8\alpha-8\right)\zeta^{3}-4\left(-4\alpha^{5}+8\alpha^{4}-9\alpha^{3}+16\alpha^{2}-8\alpha-2\right)\zeta^{2}\right.\nonumber \\
 & \qquad+\left.\alpha\left(11\alpha^{4}-5\alpha^{3}+6\alpha^{2}+22\alpha-32\right)\zeta-\bar{\alpha}\left(3\alpha^{3}+34\alpha^{2}+8\alpha-8\right)-(2-\alpha)\bar{\alpha}\alpha^{2}\zeta^{4}\right]\nonumber \\
 & +\frac{1}{8r^{6}}\left[2(\alpha-2)^{2}\alpha\zeta^{4}-\alpha\left(11\alpha^{3}-34\alpha^{2}+40\alpha-16\right)\zeta^{3}+2\left(31\alpha^{4}-58\alpha^{3}+18\alpha^{2}-16\alpha+16\right)\zeta^{2}\right.+\nonumber \\
 & \qquad\left.+\left(-63\alpha^{4}+30\alpha^{3}-40\alpha^{2}+16\alpha+32\right)\zeta-2\alpha\left(15\alpha^{2}+42\alpha-52\right)\right]\nonumber \\
 & +\frac{(\alpha-2)\left(2-\alpha\bar{\alpha}\right)\zeta^{3}+\left(-13\alpha^{3}+26\alpha^{2}-8\alpha-4\right)\zeta^{2}+\left(22\alpha^{3}-7\alpha^{2}-4\alpha-4\right)\zeta+4\left(3\alpha^{2}+2\alpha-3\right)}{2r^{8}}\nonumber \\
 & \left.+\frac{2\left(2\alpha^{2}-5\alpha+4\right)\zeta^{2}+\left(-15\alpha^{2}+6\alpha+8\right)\zeta-6\alpha}{2r^{10}}-\frac{2\bar{\alpha}\zeta}{r^{12}}\right\} +\left(\mathinner{\color{white}\frac{}{}}\zeta\to-\zeta\right)\nonumber 
\end{align}
\begin{align}
c_{\gamma\chi_{c2}}^{(+-,-1)} & \left(r,\,\alpha,\,\zeta=x/\xi\right)=-\frac{4\mathfrak{C}_{\chi_{c}}\sqrt{\left(1-\alpha\right)^{3}\left(\alpha r^{2}-1\right)^{3}}}{\bar{\alpha}\alpha^{2}\left(1+\bar{\alpha}\right)^{2}\left(\bar{\alpha}+1/r^{2}\right)\left(r^{2}-1\right)^{2}\left(\zeta+1-i0\right)\left(\zeta-1+i0\right)\left(\zeta-\kappa+i0\right)}\times\\
\times & \left\{ \mathinner{\color{white}\frac{\frac{}{}}{\frac{}{}}}3\bar{\alpha}\alpha^{2}\left(\alpha\zeta+(2-\alpha)\zeta^{2}\right)+\frac{(2-\alpha)\alpha^{2}(5\alpha-4)\zeta^{2}+\alpha^{2}\left(5\alpha^{2}+2\alpha-6\right)\zeta+2(\alpha-2)\alpha\zeta^{3}-2(\alpha-2)^{2}}{r^{2}}\right.\nonumber \\
 & \qquad+\frac{2(2-\alpha)\alpha\zeta^{3}+\left(5\alpha^{3}-10\alpha^{2}+8\right)\zeta^{2}+\left(-15\alpha^{3}+8\alpha^{2}+8\right)\zeta+2\left(\alpha^{2}-6\alpha+4\right)}{r^{4}}\nonumber \\
 & \qquad\left.+\frac{4\alpha-4\left(\alpha^{2}-2\alpha+2\right)\zeta^{2}+2\left(7\alpha^{2}-8\right)\zeta}{r^{6}}+\frac{8\bar{\alpha}\zeta}{r^{8}}\right\} +\left(\mathinner{\color{white}\frac{}{}}\zeta\to-\zeta\right)\nonumber 
\end{align}
\begin{align}
d_{\gamma\chi_{c2}}^{(+-,-1)} & \left(r,\,\alpha,\,\zeta=x/\xi\right)=\frac{\mathfrak{C}_{\chi_{c}}\zeta\sqrt{\left(1-\alpha\right)\left(\alpha r^{2}-1\right)^{3/2}}}{2\alpha^{2}r^{2}\left(r^{2}-1\right)^{2}\left(\zeta+1-i0\right)\left(\zeta-1+i0\right)}+\left(\mathinner{\color{white}\frac{}{}}\zeta\to-\zeta\right)=0
\end{align}
\begin{align}
c_{\gamma\chi_{c2}}^{(++,\,0)} & \left(r,\,\alpha,\,\zeta=x/\xi\right)=\frac{8\bar{\alpha}r^{2}\mathfrak{C}_{\chi_{c}}}{\sqrt{6}\alpha^{3}\left(1+\bar{\alpha}\right)^{2}\left(\bar{\alpha}+1/r^{2}\right)\left(r^{2}-1\right)^{2}\left(\zeta+1-i0\right)\left(\zeta-1+i0\right)\left(\zeta-\kappa+i0\right)^{2}}\times\\
\times & \left\{ \mathinner{\color{white}\frac{\frac{}{}}{\frac{}{}}}2\alpha\,\bar{\alpha}^{2}\left[\left(\alpha^{3}-5\alpha^{2}+12\alpha-16\right)\zeta^{4}+\left(2\alpha^{5}-11\alpha^{4}+24\alpha^{3}-37\alpha^{2}+36\alpha-24\right)\zeta^{3}\right.\right.\nonumber \\
 & \qquad\left.-\alpha\left(4\alpha^{4}-14\alpha^{3}+22\alpha^{2}-31\alpha+18\right)\zeta^{2}+\alpha\left(2\alpha^{4}-3\alpha^{3}+4\alpha^{2}-3\alpha+4\right)\zeta+\left(\alpha^{3}+2\alpha^{2}-10\alpha+8\right)\right]+\nonumber \\
 & +\frac{\alpha}{r^{2}}\left[\left(\alpha^{4}-4\alpha^{3}+16\alpha^{2}-64\alpha+48\right)\zeta^{4}+2\left(5\alpha^{6}-36\alpha^{5}+95\alpha^{4}-153\alpha^{3}+177\alpha^{2}-180\alpha+96\right)\zeta^{3}\right.\nonumber \\
 & \qquad-2\left(18\alpha^{6}-88\alpha^{5}+169\alpha^{4}-228\alpha^{3}+199\alpha^{2}-26\alpha-60\right)\zeta^{2}\nonumber \\
 & \qquad\left.+2\alpha\left(13\alpha^{5}-36\alpha^{4}+43\alpha^{3}-21\alpha^{2}-45\alpha+64\right)\zeta+\left(21\alpha^{4}-72\alpha^{3}-2\alpha^{2}+124\alpha-56\right)\right]\nonumber \\
 & +\frac{1}{r^{4}}\left[\left(\alpha^{4}-9\alpha^{3}+18\alpha^{2}+12\alpha-24\right)\alpha\zeta^{4}-2\left(5\alpha^{6}-41\alpha^{5}+96\alpha^{4}-98\alpha^{3}+40\alpha^{2}-64\alpha+72\right)\zeta^{3}\right.\nonumber \\
 & \qquad+2\left(30\alpha^{6}-166\alpha^{5}+281\alpha^{4}-315\alpha^{3}+368\alpha^{2}-176\alpha-72\right)\zeta^{2}\nonumber \\
 & \qquad-\left.2\left(33\alpha^{5}-107\alpha^{4}+108\alpha^{3}-84\alpha^{2}-84\alpha+204\right)\alpha\zeta-5\alpha^{5}-33\alpha^{4}+296\alpha^{3}-276\alpha^{2}-104\alpha+48\right]\nonumber \\
 & +\frac{1}{r^{6}}\left[(6-\alpha)(\alpha-2)^{2}\alpha\zeta^{4}+2\left(\alpha^{2}-2\alpha+2\right)\left(3\alpha^{3}-21\alpha^{2}+10\alpha+24\right)\zeta^{3}+\alpha\left(21\alpha^{3}-54\alpha^{2}-336\alpha+496\right)\right.+\nonumber \\
 & \qquad\left.-4\left(13\alpha^{5}-85\alpha^{4}+111\alpha^{3}-27\alpha^{2}+54\alpha-96\right)\zeta^{2}+2\left(43\alpha^{5}-169\alpha^{4}+106\alpha^{3}-150\alpha^{2}+124\alpha+144\right)\zeta\right]\nonumber \\
 & +\frac{2}{r^{8}}\left[(\alpha-6)(2-\alpha)\left(2-\alpha\bar{\alpha}\right)\zeta^{3}+2\left(7\alpha^{4}-54\alpha^{3}+71\alpha^{2}+18\alpha-60\right)\zeta^{2}\right.+\nonumber \\
 & \qquad\left.-\left(33\alpha^{4}-161\alpha^{3}+10\alpha^{2}+12\alpha+168\right)\zeta-2\left(7\alpha^{3}-39\alpha^{2}-6\alpha+60\right)\right]\nonumber \\
 & \left.-\frac{4(\alpha-6)\left(2\alpha^{2}-5\alpha+4\right)\zeta^{2}-32\left(\alpha^{3}-6\alpha^{2}+6\right)\zeta-4(\alpha-6)(3\alpha-4)}{r^{10}}+\frac{8(\alpha-6)\bar{\alpha}\zeta}{r^{12}}\right\} +\left(\mathinner{\color{white}\frac{}{}}\zeta\to-\zeta\right)\nonumber 
\end{align}
\begin{align}
d_{\gamma\chi_{c2}}^{(++,\,0)} & \left(r,\,\alpha,\,\zeta=x/\xi\right)=-\frac{4\bar{\alpha}\mathfrak{C}_{\chi_{c}}}{\sqrt{6}\alpha^{4}\left(1+\bar{\alpha}\right)^{2}\left(\bar{\alpha}+1/r^{2}\right)^{2}\left(r^{2}-1\right)^{2}\left(\zeta+1-i0\right)\left(\zeta-1+i0\right)\left(\zeta-\kappa+i0\right)^{2}}\times\\
\times & \left\{ \mathinner{\color{white}\frac{\frac{}{}}{\frac{}{}}}\frac{(2-\alpha)\alpha^{2}}{4}\left[-2\left(\alpha^{5}-6\alpha^{4}+13\alpha^{3}-15\alpha^{2}+10\alpha-4\right)\zeta^{3}-2\alpha(2-\alpha)\left(2\alpha^{3}-4\alpha^{2}+3\alpha-2\right)\zeta^{2}\right.\right.\nonumber \\
 & \qquad+\left.(2-\alpha)\left(\alpha^{2}+2\alpha-2\right)-2\alpha\left(\alpha^{4}-2\alpha^{3}+\alpha^{2}+\alpha-2\right)\zeta+(2-\alpha)\left(\alpha^{2}+2\bar{\alpha}\right)\zeta^{4}\right]+\nonumber \\
 & +\frac{\alpha}{2r^{2}}\left[(2-\alpha)\left(2\alpha^{5}-13\alpha^{4}+24\alpha^{3}-25\alpha^{2}+18\alpha-10\right)\zeta^{3}-(\alpha-2)^{2}\left(1-\alpha\bar{\alpha}\right)\zeta^{4}-\alpha\left(\alpha^{2}+6\bar{\alpha}\right)\left(3\alpha^{2}-2\bar{\alpha}\right)\right.\nonumber \\
 & \qquad\left.+\left(8\alpha^{6}-50\alpha^{5}+108\alpha^{4}-109\alpha^{3}+59\alpha^{2}-4\alpha-16\right)\zeta^{2}-\alpha\left(6\alpha^{5}-25\alpha^{4}+30\alpha^{3}-11\alpha^{2}-14\alpha+18\right)\zeta\right]\nonumber \\
 & +\frac{1}{2r^{4}}\left[(\alpha-2)\left(\alpha^{5}-11\alpha^{4}+16\alpha^{3}-12\alpha^{2}+6\alpha-4\right)\zeta^{3}+2\bar{\alpha}\left(5\alpha^{5}-32\alpha^{4}+44\alpha^{3}-24\alpha^{2}+20\alpha+4\right)\zeta^{2}\right.\nonumber \\
 & \qquad+\left.\alpha\left(13\alpha^{5}-61\alpha^{4}+62\alpha^{3}-20\alpha^{2}-30\alpha+40\right)\zeta+\left(3\alpha^{4}-52\alpha^{3}+44\alpha^{2}+16\alpha-8\right)+(\alpha-2)^{2}\alpha^{2}\zeta^{4}\right]\nonumber \\
 & +\frac{1}{2r^{6}}\left[-(\alpha-2)^{2}\alpha\zeta^{4}+2(\alpha-2)\alpha\left(3\alpha^{2}-4\alpha+3\right)\zeta^{3}+\left(4\alpha^{5}-56\alpha^{4}+111\alpha^{3}-64\alpha^{2}+28\alpha-16\right)\zeta^{2}\right.+\nonumber \\
 & \qquad\left.+2\left(-6\alpha^{5}+39\alpha^{4}-29\alpha^{3}+11\alpha^{2}-2\alpha-8\right)\zeta+8\alpha\left(\alpha^{2}-8\bar{\alpha}\right)\right]\nonumber \\
 & +\frac{(2-\alpha)\left(2-\alpha\bar{\alpha}\right)\zeta^{3}-2\bar{\alpha}\left(7\alpha^{2}-8\alpha-2\right)\zeta^{2}-2\left(5\alpha^{2}-6\bar{\alpha}\right)+\left(2\alpha^{4}-31\alpha^{3}+13\alpha^{2}+4\alpha+4\right)\zeta}{r^{8}}\nonumber \\
 & \left.\frac{2\left(8\alpha^{2}-3\alpha-4\right)\zeta-2\left(2\alpha^{2}-5\alpha+4\right)\zeta^{2}+6\alpha}{r^{10}}+\frac{4\bar{\alpha}\zeta}{r^{12}}\right\} +\left(\mathinner{\color{white}\frac{}{}}\zeta\to-\zeta\right)\nonumber 
\end{align}
\begin{align}
c_{\gamma\chi_{c2}}^{(+-,\,0)} & \left(r,\,\alpha,\,\zeta=x/\xi\right)=-\frac{8\mathfrak{C}_{\chi_{c}}\left(\alpha r^{2}-1\right)}{\sqrt{6}\alpha^{2}\left(1+\bar{\alpha}\right)^{2}\left(\bar{\alpha}+1/r^{2}\right)\left(r^{2}-1\right)^{2}\left(\zeta+1-i0\right)\left(\zeta-1+i0\right)\left(\zeta-\kappa+i0\right)^{2}}\times\\
\times & \left\{ \mathinner{\color{white}\frac{\frac{}{}}{\frac{}{}}}2\bar{\alpha}\alpha\left[\bar{\alpha}\left(2\alpha^{4}-\alpha^{3}+\alpha+2\right)\zeta-(\alpha-2)^{2}\alpha\zeta^{4}-\alpha\left(\alpha^{2}+3\alpha-3\right)+\bar{\alpha}\left(2\alpha^{4}-9\alpha^{3}+12\alpha^{2}-9\alpha+6\right)\zeta^{3}\right.\right.\nonumber \\
 & \qquad+\left.\alpha\left(4\alpha^{4}-14\alpha^{3}+16\alpha^{2}-9\alpha+5\right)\zeta^{2}\right]+\nonumber \\
 & +\frac{1}{r^{2}}\left[2\alpha\bar{\alpha}\left(3\alpha^{4}-17\alpha^{3}+30\alpha^{2}-24\alpha+10\right)\zeta^{3}+2\alpha\left(14\alpha^{5}-64\alpha^{4}+107\alpha^{3}-84\alpha^{2}+37\alpha-8\right)\zeta^{2}\right.\nonumber \\
 & \qquad+\left.2\bar{\alpha}\left(11\alpha^{5}-17\alpha^{4}+6\alpha^{3}+2\alpha^{2}+6\alpha-12\right)\zeta-\alpha\left(19\alpha^{3}-11\alpha^{2}-26\alpha+20\right)+(\alpha-3)(\alpha-2)^{2}\alpha\zeta^{4}\right]\nonumber \\
 & +\frac{1}{r^{4}}\left[(3-\alpha)(\alpha-2)^{2}\alpha\zeta^{4}+4(\alpha-3)\left(\alpha^{2}-2\alpha+2\right)\left(\alpha^{2}-\alpha-1\right)\zeta^{3}-4\left(8\alpha^{5}-36\alpha^{4}+54\alpha^{3}-31\alpha^{2}+12\alpha-6\right)\zeta^{2}\right.\nonumber \\
 & \qquad+\left.4\left(11\alpha^{5}-29\alpha^{4}+23\alpha^{3}-\alpha^{2}-12\alpha+6\right)\zeta+5\alpha^{4}+37\alpha^{3}+-52\alpha^{2}-12\alpha+24\right]\nonumber \\
 & -\frac{2}{r^{6}}\left[(\alpha-3)(\alpha-2)\left(\alpha^{2}-\alpha+2\right)\zeta^{3}-2(\alpha-2)\left(5\alpha^{3}-12\alpha^{2}+9\right)\zeta^{2}+\left(21\alpha^{4}-54\alpha^{3}+29\alpha^{2}-12\alpha+12\right)\zeta\right.+\nonumber \\
 & \qquad\left.+2\left(4\alpha^{3}+3\alpha^{2}-15\alpha+6\right)\right]\nonumber \\
 & \left.-\frac{4(\alpha-3)\left(2\alpha^{2}-5\alpha+4\right)\zeta^{2}+8\bar{\alpha}\left(3\alpha^{2}-5\alpha-6\right)\zeta-4\alpha(3\alpha-5)}{r^{8}}+\frac{8(\alpha-3)\bar{\alpha}\zeta}{r^{10}}\right\} +\left(\mathinner{\color{white}\frac{}{}}\zeta\to-\zeta\right)\nonumber 
\end{align}
\begin{align}
d_{\gamma\chi_{c2}}^{(+-,\,0)} & \left(r,\,\alpha,\,\zeta=x/\xi\right)=\frac{4\mathfrak{C}_{\chi_{c}}\left(\alpha-1/r^{2}\right)}{\sqrt{6}\alpha^{2}\left(1+\bar{\alpha}\right)^{2}\left(\bar{\alpha}+1/r^{2}\right)^{2}\left(r^{2}-1\right)^{2}\left(\zeta+1-i0\right)\left(\zeta-1+i0\right)\left(\zeta-\kappa+i0\right)^{2}}\times\\
 & \times\left\{ \mathinner{\color{white}\frac{\frac{}{}}{\frac{}{}}}\frac{\bar{\alpha}}{4}\left[(\alpha-2)^{2}\alpha\zeta^{4}-2(2-\alpha)\left(\alpha^{4}-4\alpha^{3}+6\alpha^{2}-6\alpha+1\right)\zeta^{3}-2\alpha\left(2\alpha^{4}-8\alpha^{3}+13\alpha^{2}-9\alpha-1\right)\zeta^{2}\right.\right.\nonumber \\
 & \qquad\qquad\left.+2\left(\alpha^{5}-2\alpha^{4}+2\alpha^{3}+3\alpha-2\right)\zeta+\alpha\left(\alpha^{2}+6\alpha-6\right)\right]\nonumber \\
 & +\frac{1}{2r^{2}}\left[(\alpha-2)\left(2\alpha^{4}-7\alpha^{3}+7\alpha^{2}-5\alpha+2\right)\zeta^{3}+2\left(-4\alpha^{5}+17\alpha^{4}-31\alpha^{3}+33\alpha^{2}-16\alpha+2\right)\zeta^{2}\right.\nonumber \\
 & \qquad\left.\alpha\left(6\alpha^{4}-15\alpha^{3}+17\alpha^{2}+\alpha-8\right)\zeta-4\bar{\alpha}(\alpha+1)(2\alpha-1)\right]\nonumber \\
 & +\frac{1}{4r^{4}}\left[(\alpha-2)^{2}\alpha\zeta^{4}+2(2-\alpha)\bar{\alpha}^{2}(2\alpha-1)\zeta^{3}+2\left(12\alpha^{4}-38\alpha^{3}+41\alpha^{2}-27\alpha+8\right)\zeta^{2}\right.\nonumber \\
 & \qquad+\left.2\left(-14\alpha^{4}+27\alpha^{3}-32\alpha^{2}+11\alpha+2\right)\zeta-\alpha\left(5\alpha^{2}+34\alpha-34\right)\right]\nonumber \\
 & +\frac{(\alpha-2)\left(2-\alpha\bar{\alpha}\right)\zeta^{3}+2\bar{\alpha}\alpha(5\alpha-8)\zeta^{2}+\left(17\alpha^{3}-19\alpha^{2}+12\alpha-4\right)\zeta+8\left(\alpha^{2}+\alpha-1\right)}{2r^{6}}\nonumber \\
 & \left.\frac{\left(2\alpha^{2}-5\alpha+4\right)\zeta^{2}+2\bar{\alpha}(3\alpha+1)\zeta-3\alpha}{r^{8}}-\frac{2\bar{\alpha}\zeta}{r^{10}}\right\} +\left(\mathinner{\color{white}\frac{}{}}\zeta\to-\zeta\right)\nonumber 
\end{align}
\begin{align}
c_{\gamma\chi_{c2}}^{(++,+1)} & \left(r,\,\alpha,\,\zeta=x/\xi\right)=-\frac{4\mathfrak{C}_{\chi_{c}}r^{2}\sqrt{\left(1-\alpha\right)\left(\alpha r^{2}-1\right)}}{\alpha\left(1+\bar{\alpha}\right)^{2}\left(\bar{\alpha}+1/r^{2}\right)\left(r^{2}-1\right)^{2}\left(\zeta+1-i0\right)\left(\zeta-1+i0\right)\left(\zeta-\kappa+i0\right)^{2}}\times\\
\times & \left\{ \mathinner{\color{white}\frac{\frac{}{}}{\frac{}{}}}\,\bar{\alpha}\alpha^{2}\left[-2(\alpha-4)\bar{\alpha}\zeta^{4}-\left(\alpha^{4}-2\alpha^{3}+2\alpha^{2}+4\alpha-4\right)\zeta^{3}+2\left(\alpha^{4}-4\alpha^{2}+4\alpha-2\right)\zeta^{2}\right.\right.\nonumber \\
 & \qquad\qquad\left.-\left(\alpha^{4}+2\alpha^{3}-10\alpha^{2}+12\alpha-4\right)\zeta+2\bar{\alpha}(3\alpha-2)\right]+\nonumber \\
 & -\frac{1}{r^{2}}\left[4\alpha\left(5\alpha^{2}-12\alpha+8\right)\zeta^{4}+\alpha\left(7\alpha^{6}-41\alpha^{5}+113\alpha^{4}-188\alpha^{3}+230\alpha^{2}-200\alpha+88\right)\zeta^{3}\right.\nonumber \\
 & \qquad-2\alpha\left(7\alpha^{6}-25\alpha^{5}+49\alpha^{4}-74\alpha^{3}+54\alpha^{2}+4\alpha-20\right)\zeta^{2}\nonumber \\
 & \qquad\left.+\alpha^{2}\left(7\alpha^{5}-9\alpha^{4}+5\alpha^{3}+12\alpha^{2}-38\alpha+32\right)\zeta+4\alpha\left(-4\alpha^{3}\bar{\alpha}-9\alpha^{2}+16\alpha-6\right)\right]\nonumber \\
 & +\frac{2}{r^{4}}\left[-\alpha\left(\alpha^{4}-5\alpha^{3}+3\alpha^{2}+8\alpha-8\right)\zeta^{4}+\left(7\alpha^{6}-35\alpha^{5}+71\alpha^{4}-76\alpha^{3}+68\alpha^{2}-72\alpha+48\right)\zeta^{3}\right.\nonumber \\
 & \qquad-2\left(14\alpha^{6}-51\alpha^{5}+84\alpha^{4}-111\alpha^{3}+100\alpha^{2}-24\alpha-24\right)\zeta^{2}+\alpha\left(21\alpha^{5}+39\alpha^{3}\bar{\alpha}+4\alpha^{2}+96\bar{\alpha}\right)\zeta\nonumber \\
 & \qquad+\left.+\left(3\alpha^{5}+31\alpha^{4}-83\alpha^{3}+36\alpha^{2}+40\alpha-16\right)\right]\nonumber \\
 & +\frac{1}{r^{6}}\left[\left(-\alpha^{2}+2\alpha-2\right)\left(11\alpha^{3}-26\alpha^{2}-4\alpha+32\right)\zeta^{3}+2\left(39\alpha^{5}-128\alpha^{4}+130\alpha^{3}-90\alpha^{2}+148\alpha-128\right)\zeta^{2}\right.\nonumber \\
 & \qquad\left.+\left(176\alpha^{4}-95\alpha^{5}-166\alpha^{3}+188\alpha^{2}-8\alpha-192\right)\zeta+2(\alpha-2)^{3}\alpha\zeta^{4}-2\alpha\left(15\alpha^{3}+22\alpha^{2}-146\alpha+132\right)\right]\nonumber \\
 & +\frac{4}{r^{8}}\left[(\alpha-2)^{2}\left(2-\alpha\bar{\alpha}\right)\zeta^{3}-\left(13\alpha^{4}-44\alpha^{3}+32\alpha^{2}+32\alpha-40\right)\zeta^{2}+\left(26\alpha^{4}-43\alpha^{3}+6\alpha^{2}-28\alpha+56\right)\zeta\right.\nonumber \\
 & \qquad\left.+2\left(6\alpha^{3}-9\alpha^{2}-12\alpha+20\right)\right]\nonumber \\
 & \left.+\frac{8(\alpha-2)\left(2\alpha^{2}-5\alpha+4\right)\zeta^{2}-4\left(15\alpha^{3}-28\alpha^{2}-14\alpha+32\right)\zeta-8(\alpha-2)(3\alpha-4)}{r^{10}}+\frac{16(2-\alpha)\bar{\alpha}\zeta}{r^{12}}\right\} \nonumber \\
 & +\left(\mathinner{\color{white}\frac{}{}}\zeta\to-\zeta\right)\nonumber 
\end{align}
\begin{align}
d_{\gamma\chi_{c2}}^{(++,+1)} & \left(r,\,\alpha,\,\zeta=x/\xi\right)=\frac{4\mathfrak{C}_{\chi_{c}}\sqrt{\left(1-\alpha\right)\left(\alpha r^{2}-1\right)}}{\bar{\alpha}\alpha^{4}\left(1+\bar{\alpha}\right)^{2}\left(\bar{\alpha}+1/r^{2}\right)^{2}\left(r^{2}-1\right)^{2}\left(\zeta+1-i0\right)\left(\zeta-1+i0\right)\left(\zeta-\kappa+i0\right)^{2}}\times\\
\times & \left\{ \mathinner{\color{white}\frac{\frac{}{}}{\frac{}{}}}\frac{\bar{\alpha}^{2}\alpha^{2}}{8}\left[2(2-\alpha)\bar{\alpha}\zeta^{4}+\left(3\alpha^{4}-16\alpha^{3}+28\alpha^{2}-24\alpha+8\right)\zeta^{3}-2\alpha\left(3\alpha^{3}-10\alpha^{2}+10\alpha-2\right)\zeta^{2}+\right.\right.\nonumber \\
 & \qquad+\left.\left(3\alpha^{4}-4\alpha^{3}-8\bar{\alpha}\right)\zeta-2\bar{\alpha}(\alpha+2)\right]+\nonumber \\
 & +\frac{\bar{\alpha}\alpha}{8r^{2}}\left[\left(10\alpha^{5}-59\alpha^{4}+121\alpha^{3}-138\alpha^{2}+96\alpha-32\right)\zeta^{3}-2\left(16\alpha^{5}-65\alpha^{4}+95\alpha^{3}-50\alpha^{2}-6\alpha+12\right)\zeta^{2}\right.\nonumber \\
 & \qquad\left.+\left(22\alpha^{5}-47\alpha^{4}+29\alpha^{3}+22\alpha^{2}-36\alpha+8\right)\zeta+2(2-\alpha)\bar{\alpha}(3\alpha-2)\zeta^{4}-2\bar{\alpha}(3\alpha-2)(3\alpha+4)\right]\nonumber \\
 & +\frac{1}{8r^{4}}\left[2(2-\alpha)\bar{\alpha}\alpha^{2}\zeta^{4}+\left(11\alpha^{6}-66\alpha^{5}+134\alpha^{4}-148\alpha^{3}+116\alpha^{2}-64\alpha+16\right)\zeta^{3}-2\bar{\alpha}\left(37\alpha^{3}-26\alpha^{2}-16\alpha+8\right)\right.\nonumber \\
 & \qquad+\left.\alpha\left(63\alpha^{5}-178\alpha^{4}+174\alpha^{3}-124\alpha+64\right)\zeta+2\left(-31\alpha^{6}+142\alpha^{5}-264\alpha^{4}+248\alpha^{3}-112\alpha^{2}+8\alpha+8\right)\zeta^{2}\right]\nonumber \\
 & +\frac{1}{8r^{6}}\left[-\alpha\left(8\alpha^{4}-39\alpha^{3}+62\alpha^{2}-48\alpha+16\right)\zeta^{3}+2\left(30\alpha^{5}-111\alpha^{4}+162\alpha^{3}-126\alpha^{2}+64\alpha-16\right)\zeta^{2}\right.+\nonumber \\
 & \qquad+\left.\left(-92\alpha^{5}+219\alpha^{4}-202\alpha^{3}+64\alpha^{2}+48\alpha-32\right)\zeta-2(\alpha-2)^{2}\bar{\alpha}\alpha\zeta^{4}+2\alpha\bar{\alpha}\left(5\alpha^{2}+62\alpha-52\right)\right]\nonumber \\
 & +\frac{1}{2r^{8}}\left[(2-\alpha)\bar{\alpha}\left(\alpha^{2}-\alpha+2\right)\zeta^{3}+\left(-10\alpha^{4}+31\alpha^{3}-30\alpha^{2}+4\alpha+4\right)\zeta^{2}+\left(20\alpha^{4}-33\alpha^{3}+15\alpha^{2}-8\alpha+4\right)\zeta\right.\nonumber \\
 & \qquad\left.-4\bar{\alpha}\left(2\alpha^{2}+4\alpha-3\right)\right]\nonumber \\
 & \left.+\frac{6\alpha\bar{\alpha}-2\bar{\alpha}\left(2\alpha^{2}-5\alpha+4\right)\zeta^{2}-\left(12\alpha^{3}-15\alpha^{2}-6\alpha+8\right)\zeta}{2r^{10}}+\frac{2\bar{\alpha}^{2}\zeta}{r^{12}}\right\} +\left(\mathinner{\color{white}\frac{}{}}\zeta\to-\zeta\right)\nonumber 
\end{align}
\begin{align}
c_{\gamma\chi_{c2}}^{(+-,+1)} & \left(r,\,\alpha,\,\zeta=x/\xi\right)=-\frac{4\mathfrak{C}_{\chi_{c}}r^{2}\sqrt{\left(1-\alpha\right)\left(\alpha r^{2}-1\right)}}{\alpha\left(1+\bar{\alpha}\right)^{2}\left(\bar{\alpha}+1/r^{2}\right)\left(r^{2}-1\right)^{2}\left(\zeta+1-i0\right)\left(\zeta-1+i0\right)\left(\zeta-\kappa+i0\right)^{2}}\times\\
\times & \left\{ \mathinner{\color{white}\frac{\frac{}{}}{\frac{}{}}}\bar{\alpha}\alpha^{2}(2-\alpha)\left[\alpha^{2}\zeta+2(2-\alpha)\alpha\zeta^{2}+(2-\alpha)^{2}\zeta^{3}\right]\right.\nonumber \\
 & +\frac{\alpha}{r^{2}}\left[\left(9\alpha^{5}-51\alpha^{4}+109\alpha^{3}-130\alpha^{2}+88\alpha-16\right)\zeta^{3}-2\alpha\left(9\alpha^{4}-35\alpha^{3}+57\alpha^{2}-50\alpha+14\right)\zeta^{2}\right.\nonumber \\
 & \qquad+\left.4(\alpha-2)^{2}\alpha\zeta^{4}+\left(9\alpha^{5}-19\alpha^{4}+25\alpha^{3}-2\alpha^{2}+4\alpha-8\right)\zeta+4\alpha\left(\alpha^{2}-3\bar{\alpha}\right)\right]\nonumber \\
 & +\frac{2}{r^{4}}\left[(\alpha-2)^{2}\alpha\zeta^{4}-\alpha\left(7\alpha^{4}-33\alpha^{3}+45\alpha^{2}-28\alpha+16\right)\zeta^{3}+2\alpha\left(16\alpha^{4}-52\alpha^{3}+63\alpha^{2}-43\alpha+4\right)\zeta^{2}\right.\nonumber \\
 & \qquad-\left.\left(25\alpha^{5}-37\alpha^{4}+37\alpha^{3}+8\alpha^{2}+4\alpha-8\right)\zeta-\alpha\left(19\alpha^{2}+6\alpha-12\right)\right]\nonumber \\
 & +\frac{1}{r^{6}}\left[-2(\alpha-2)^{2}\alpha\zeta^{4}+\left(9\alpha^{4}-32\alpha^{3}+28\alpha^{2}-16\right)\zeta^{3}-2\left(37\alpha^{4}-90\alpha^{3}+56\alpha^{2}-32\alpha+8\right)\zeta^{2}\right.+\nonumber \\
 & \qquad\left.+\left(101\alpha^{4}-104\alpha^{3}+100\alpha^{2}+40\alpha-16\right)\zeta+2\left(5\alpha^{3}+40\alpha^{2}-4\alpha-8\right)\right]\nonumber \\
 & +\frac{4\left((2-\alpha)\left(\alpha^{2}-\alpha+2\right)\zeta^{3}+\left(11\alpha^{3}-20\alpha^{2}-2\alpha+12\right)\zeta^{2}-\left(24\alpha^{3}-9\alpha^{2}+10\alpha-4\right)\zeta-2(\alpha+2)(4\alpha-1)\right)}{r^{8}}+\nonumber \\
 & \left.-\frac{8\left(2\alpha^{2}-5\alpha+4\right)\zeta^{2}-4\left(13\alpha^{2}-8\right)\zeta-24\alpha}{r^{10}}+\frac{16\bar{\alpha}\zeta}{r^{12}}\right\} +\left(\mathinner{\color{white}\frac{}{}}\zeta\to-\zeta\right)\nonumber 
\end{align}
\begin{align}
d_{\gamma\chi_{c2}}^{(+-,+1)} & \left(r,\,\alpha,\,\zeta=x/\xi\right)=\frac{\zeta\sqrt{\left(1-\alpha\right)\left(\alpha r^{2}-1\right)}\left(\alpha\bar{\alpha}+\alpha r^{2}-1\right)}{2\bar{\alpha}\alpha^{2}\left(\zeta+1-i0\right)\left(\zeta-1+i0\right)r^{2}\left(r^{2}-1\right)^{2}}+\left(\mathinner{\color{white}\frac{}{}}\zeta\to-\zeta\right)=0
\end{align}
\begin{align}
c_{\gamma\chi_{c2}}^{(++,+2)} & \left(r,\,\alpha,\,\zeta=x/\xi\right)=\frac{4\mathfrak{C}_{\chi_{c}}\left(1-\alpha\right)\left(\alpha\,r^{2}-1\right)r^{2}}{\alpha^{2}\left(1+\bar{\alpha}\right)^{2}\left(\bar{\alpha}+1/r^{2}\right)\left(r^{2}-1\right)^{2}\left(\zeta+1-i0\right)\left(\zeta-1+i0\right)\left(\zeta-\kappa+i0\right)^{2}}\times\\
\times & \left\{ \mathinner{\color{white}\frac{\frac{}{}}{\frac{}{}}}\bar{\alpha}\alpha^{2}\left[-(2-\alpha)\alpha\zeta^{4}+\left(2\alpha^{4}-11\alpha^{3}+20\alpha^{2}-16\alpha+4\right)\zeta^{3}-2(2-\alpha)\bar{\alpha}\alpha(2\alpha-1)\zeta^{2}+\right.\right.\nonumber \\
 & \qquad+\left.\left(2\alpha^{4}-3\alpha^{3}+4\alpha-4\right)\zeta-(2-\alpha)\alpha\right]\nonumber \\
 & +\frac{1}{r^{2}}\left[(\alpha-2)\alpha\left(\alpha^{3}+\alpha^{2}+2\alpha-8\right)\zeta^{4}+\alpha\left(2\alpha^{6}-11\alpha^{5}+23\alpha^{4}-37\alpha^{3}+44\alpha^{2}-44\alpha+32\right)\zeta^{3}+\alpha^{5}-\alpha^{4}-4\alpha^{3}\right.\nonumber \\
 & \qquad+\left.2(2-\alpha)\alpha\left(2\alpha^{5}+\alpha^{4}-6\alpha^{3}+2\alpha^{2}+4\alpha+2\right)\zeta^{2}+\alpha^{2}\left(2\alpha^{5}+5\alpha^{4}-17\alpha^{3}+15\alpha^{2}-4\alpha+8\right)\zeta+16\alpha^{2}-8\alpha\right]\nonumber \\
 & +\frac{1}{r^{4}}\left[2(2-\alpha)\alpha\left(\alpha^{2}-2\right)\zeta^{4}+-2\left(\alpha^{6}-8\alpha^{5}+20\alpha^{4}-34\alpha^{3}+36\alpha^{2}-28\alpha+24\right)\zeta^{3}+4\left(3\alpha^{6}-13\alpha^{5}+20\alpha^{4}\right)\zeta^{2}\right.\nonumber \\
 & \qquad+\left.4\left(22\alpha^{2}\bar{\alpha}-10\alpha-12\right)\zeta^{2}-2\alpha\left(5\alpha^{5}-6\alpha^{4}+6\alpha^{2}-16\alpha+36\right)\zeta+2(2-\alpha)\left(\alpha^{3}-10\alpha^{2}-6\alpha+4\right)\right]\nonumber \\
 & +\frac{1}{r^{6}}\left[2(\alpha-2)^{2}\alpha\zeta^{4}+\left(-7\alpha^{4}+16\alpha^{3}-4\alpha^{2}-24\alpha+32\right)\zeta^{3}-2\left(4\alpha^{5}-31\alpha^{4}+52\alpha^{3}-42\alpha^{2}+52\alpha-64\right)\zeta^{2}\right.+\nonumber \\
 & \qquad\left.+\left(16\alpha^{5}-51\alpha^{4}+40\alpha^{3}-28\alpha^{2}+24\alpha+96\right)\zeta-2\alpha\left(3\alpha^{2}+30\alpha-56\right)\right]\nonumber \\
 & \qquad+\frac{4\left[(\alpha-2)\left(\alpha^{2}-\alpha+2\right)\zeta^{3}-\left(9\alpha^{3}-14\alpha^{2}-8\alpha+20\right)\zeta^{2}-\left(2\alpha^{4}-16\alpha^{3}+7\alpha^{2}-4\alpha+28\right)\zeta+2\left(3\alpha^{2}+2\alpha-10\right)\right]}{r^{8}}\nonumber \\
 & \left.+\frac{4\left[2\left(2\alpha^{2}-5\alpha+4\right)\zeta^{2}-\left(11\alpha^{2}-16\right)\zeta-2(3\alpha-4)\right]}{r^{10}}-\frac{16\bar{\alpha}\zeta}{r^{12}}\right\} +\left(\mathinner{\color{white}\frac{}{}}\zeta\to-\zeta\right)\nonumber 
\end{align}
\begin{align}
d_{\gamma\chi_{c2}}^{(++,+2)} & \left(r,\,\alpha,\,\zeta=x/\xi\right)=-\frac{4\mathfrak{C}_{\chi_{c}}\bar{\alpha}}{\alpha^{2}\left(1+\bar{\alpha}\right)^{2}\left(\bar{\alpha}+1/r^{2}\right)^{2}\left(r^{2}-1\right)^{2}\left(\zeta+1-i0\right)\left(\zeta-1+i0\right)\left(\zeta-\kappa+i0\right)^{2}}\times\\
\times & \left\{ \mathinner{\color{white}\frac{\frac{}{}}{\frac{}{}}}\frac{\bar{\alpha}\alpha}{4}\left[(\alpha-2)^{2}\zeta^{4}-2(2-\alpha)\left(\alpha^{3}-3\alpha^{2}+3\alpha-3\right)\zeta^{3}-2\left(2\alpha^{4}-6\alpha^{3}+6\alpha^{2}-\alpha-4\right)\zeta^{2}\right.\right.\nonumber \\
 & \qquad\left.+2\left(\alpha^{4}-\alpha^{3}-\alpha^{2}+5\alpha-2\right)\zeta+\left(3\alpha^{2}+2\alpha-4\right)\right]+\nonumber \\
 & +\frac{1}{2r^{2}}\left[(\alpha-2)\left(2\alpha^{4}-7\alpha^{3}+7\alpha^{2}-5\alpha+2\right)\zeta^{3}+2\left(-4\alpha^{5}+15\alpha^{4}-21\alpha^{3}+15\alpha^{2}-2\alpha-2\right)\zeta^{2}+\right.\nonumber \\
 & \qquad+\left.\alpha\left(6\alpha^{4}-11\alpha^{3}+5\alpha^{2}+13\alpha-12\right)\zeta-4\bar{\alpha}(\alpha+1)(2\alpha-1)\right]\nonumber \\
 & +\frac{1}{4r^{4}}\left[(\alpha-2)^{2}\alpha\zeta^{4}+2(2-\alpha)\bar{\alpha}^{2}\alpha\zeta^{3}+2(\alpha-2)\left(10\alpha^{3}-13\alpha^{2}+11\alpha-4\right)\zeta^{2}\right.+\nonumber \\
 & \qquad\left.2\left(-13\alpha^{4}+20\alpha^{3}-15\alpha^{2}-6\alpha+8\right)\zeta+\alpha\left(\alpha^{2}-46\alpha+40\right)\right]\nonumber \\
 & +\frac{(\alpha-2)\left(\alpha^{2}-\alpha+2\right)\zeta^{3}+2\bar{\alpha}\left(3\alpha^{2}-4\alpha-2\right)\zeta^{2}+\left(13\alpha^{3}-11\alpha^{2}+8\alpha-4\right)\zeta+4\left(\alpha^{2}+4\alpha-3\right)}{2r^{6}}+\nonumber \\
 & \left.\frac{\left(2\alpha^{2}-5\alpha+4\right)\zeta^{2}+4\bar{\alpha}(\alpha+1)\zeta-3\alpha}{r^{8}}-\frac{2\bar{\alpha}\zeta}{r^{10}}\right\} +\left(\mathinner{\color{white}\frac{}{}}\zeta\to-\zeta\right)\nonumber 
\end{align}
\begin{align}
c_{\gamma\chi_{c2}}^{(+-,+2)} & \left(r,\,\alpha,\,\zeta=x/\xi\right)=\frac{4\mathfrak{C}_{\chi_{c}}\left(1-\alpha\right)\left(\alpha\,r^{2}-1\right)r^{2}}{\alpha^{2}\left(1+\bar{\alpha}\right)^{2}\left(\bar{\alpha}+1/r^{2}\right)\left(r^{2}-1\right)^{2}\left(\zeta+1-i0\right)\left(\zeta-1+i0\right)\left(\zeta-\kappa+i0\right)^{2}}\times\\
\times & \left\{ \mathinner{\color{white}\frac{\frac{}{}}{\frac{}{}}}\bar{\alpha}\alpha^{2}(2\alpha-1)\left[\alpha^{2}\zeta+2(2-\alpha)\alpha\zeta^{2}+(2-\alpha)^{2}\zeta^{3}\right]\right.\nonumber \\
 & +\frac{1}{r^{2}}\left[-5(\alpha-2)^{2}\alpha^{2}\zeta^{4}-\alpha\left(6\alpha^{5}-37\alpha^{4}+87\alpha^{3}-115\alpha^{2}+80\alpha-12\right)\zeta^{3}\alpha^{2}\left(5\alpha^{2}+8\alpha-8\right)\right.\nonumber \\
 & \qquad+\left.2\alpha^{2}\left(6\alpha^{4}-29\alpha^{3}+56\alpha^{2}-45\alpha+8\right)\zeta^{2}-\alpha^{2}(2\alpha-1)\left(3\alpha^{3}-9\alpha^{2}+7\alpha+8\right)\zeta\right]\nonumber \\
 & +\frac{1}{r^{4}}\left[-(\alpha-2)^{2}\bar{\alpha}\alpha\zeta^{4}+2\alpha\left(4\alpha^{4}-20\alpha^{3}+26\alpha^{2}-19\alpha+18\right)\zeta^{3}-2\alpha^{2}\left(20\alpha^{3}-73\alpha^{2}+109\alpha-78\right)\zeta^{2}\right.\nonumber \\
 & \qquad+\left.2\left(16\alpha^{5}-34\alpha^{4}+38\alpha^{3}+9\alpha^{2}+2\alpha-4\right)\zeta+\alpha\left(\alpha^{3}+35\alpha^{2}+4\alpha-12\right)\right]\nonumber \\
 & +\frac{1}{r^{6}}\left[(\alpha-2)^{2}\alpha\zeta^{4}+\left(-4\alpha^{4}+15\alpha^{3}-8\alpha^{2}-8\alpha+8\right)\zeta^{3}+4\left(10\alpha^{4}-25\alpha^{3}+22\alpha^{2}-20\alpha+2\right)\zeta^{2}\right.+\nonumber \\
 & \qquad\left.+\left(-60\alpha^{4}+83\alpha^{3}-104\alpha^{2}-40\alpha+8\right)\zeta-11\alpha^{3}-64\alpha^{2}+4\alpha+8\right]\nonumber \\
 & +\frac{2(\alpha-2)\left(\alpha^{2}-\alpha+2\right)\zeta^{3}-4\left(5\alpha^{3}-7\alpha^{2}-5\alpha+6\right)\zeta^{2}+2\left(25\alpha^{3}-9\alpha^{2}+28\alpha-4\right)\zeta+4\left(7\alpha^{2}+11\alpha-2\right)}{r^{8}}\nonumber \\
 & \left.+\frac{4\left[\left(2\alpha^{2}-5\alpha+4\right)\zeta^{2}-\left(6\alpha^{2}+3\alpha-4\right)\zeta-5\alpha\right]}{r^{10}}-\frac{8\bar{\alpha}\zeta}{r^{12}}\right\} +\left(\mathinner{\color{white}\frac{}{}}\zeta\to-\zeta\right)\nonumber 
\end{align}
\begin{align}
d_{\gamma\chi_{c2}}^{(+-,+2)} & \left(r,\,\alpha,\,\zeta=x/\xi\right)=-\frac{4\mathfrak{C}_{\chi_{c}}\bar{\alpha}}{\alpha^{2}\left(1+\bar{\alpha}\right)^{2}\left(\bar{\alpha}+1/r^{2}\right)^{2}\left(r^{2}-1\right)^{2}\left(\zeta+1-i0\right)\left(\zeta-1+i0\right)\left(\zeta-\kappa+i0\right)^{2}}\times\\
 & \times\left\{ \mathinner{\color{white}\frac{\frac{}{}}{\frac{}{}}}\frac{\alpha}{4}\left[(\alpha-2)^{2}\alpha\zeta^{4}+2\left(\alpha^{2}-2\alpha+2\right)\left[(\alpha-2)\left(\alpha^{2}-3\alpha+1\right)\zeta^{3}+\left(\alpha^{3}-\alpha^{2}-\alpha+2\right)\zeta\right]\right.\right.\nonumber \\
 & \qquad\,\,\,\left.-2\left(2\alpha^{4}-10\alpha^{3}+19\alpha^{2}-14\alpha+2\right)\alpha\zeta^{2}+\alpha^{3}\right]+\nonumber \\
 & -\frac{1}{2r^{2}}\left[(\alpha-2)^{2}\alpha\zeta^{4}-(2-\alpha)\left(2\alpha^{4}-9\alpha^{3}+13\alpha^{2}-11\alpha+1\right)\zeta^{3}-\alpha\left(8\alpha^{4}-38\alpha^{3}+70\alpha^{2}-49\alpha+5\right)\zeta^{2}\right.\nonumber \\
 & \qquad\left.+\left(6\alpha^{5}-17\alpha^{4}+19\alpha^{3}-3\alpha^{2}-3\alpha+2\right)\zeta+\alpha\left(3\alpha^{2}+\bar{\alpha}\right)\right]\nonumber \\
 & +\frac{(\alpha-2)\left(\alpha^{3}+4\alpha\bar{\alpha}-4\right)\zeta^{3}-2\alpha\left(5\alpha^{3}-22\alpha^{2}+39\alpha-24\right)\zeta^{2}+\alpha\left(13\alpha^{3}-34\alpha^{2}+36\alpha-4\right)\zeta+6\alpha^{2}}{2r^{4}}\nonumber \\
 & \left.\frac{(2-\alpha)\zeta^{3}+\left(4\alpha^{3}-16\alpha^{2}+31\alpha-16\right)\zeta^{2}-\left(12\alpha^{3}-28\alpha^{2}+27\alpha+2\right)\zeta-7\alpha}{2r^{6}}+\frac{2\left[\left(\alpha^{2}+2\bar{\alpha}\right)\zeta-\zeta^{2}+1\right]}{r^{8}}\right\} +\left(\mathinner{\color{white}\frac{}{}}\zeta\to-\zeta\right)\nonumber 
\end{align}

 \end{document}